\documentclass{emulateapj}
\slugcomment{{\sc Accepted to ApJ:} February 2, 2009} 
\usepackage{color}
\usepackage{xspace}

\begin{document}

\title{Characterizing the properties of Clusters of Galaxies as a function of luminosity and redshift}

\author{ K.~Andersson$^{1,2,3}$, J.R.~Peterson$^4$, G.~Madejski$^{3,5}$, A.~Goobar$^{2,6}$}

\affil{$^1$ MKI, Massachusetts Institute of Technology, Cambridge, MA 02139, USA}

\affil{$^2$ Department of Physics, Stockholm University, Albanova University Center, SE-106 91, Stockholm, Sweden}

\affil{$^3$ Stanford Linear Accelerator Center, Menlo Park, CA 94025, USA}

\email{kanderss@space.mit.edu}

\affil{$^4$ Department of Physics, Purdue University, West Lafayette, IN 47907, USA}

\affil{$^5$ KIPAC, Stanford University, PO Box 20450, Stanford, CA 94309, USA}

\affil{$^6$ The Oskar Klein Centre for Cosmo Particle Physics, AlbaNova, SE-106 91 Stockholm, Sweden}


\begin{abstract}  

We report the application of the new Monte Carlo method, Smoothed Particle Inference 
(SPI, described in a pair of companion papers), towards analysis and interpretation of X-ray observations 
of clusters of galaxies with the XMM-Newton satellite.  
Our sample consists of publicly available well-exposed observations of clusters at redshifts $z > 0.069$, 
totaling 101 objects.  
We determine the luminosity and temperature structure of the X-ray emitting gas, 
with the goal to 
quantify the scatter and the evolution of the $L_X$ - $T$ relation, as well as 
to investigate the dependence on cluster substructure with redshift.
This work is important for the establishment of the potential robustness of 
mass estimates from X-ray data 
which in turn is essential towards the use of clusters for measurements of cosmological parameters.  
We use the luminosity and temperature maps derived via the SPI technique to determine 
the presence of cooling cores, via measurements of luminosity and temperature contrast.  
The $L_{X} - T$ relation is investigated, and we confirm that $L_{X} \propto T^{3}$.  
We find a weak redshift dependence ($\propto (1+z)^{\beta_{LT}}, \beta_{LT}=0.50 \pm 0.34$), 
in contrast to some Chandra results.
The level of dynamical activity is established using the ``power ratios'' method, and we compare our 
results to previous application of this method to Chandra data for clusters.  
We find signs of evolution in the $P_3/P_0$ power ratio.
A new method, the ``temperature two-point correlation function,'' is proposed. This method is used to 
determine the ``power spectrum'' of temperature fluctuations in the X-ray emitting gas as a function of spatial 
scale. We show how this method can be fruitfully used to identify cooling core clusters as well as those 
with disturbed structures, presumably due to on-going or recent merger activity.

\end{abstract}


\section{Introduction}

Clusters of galaxies are the largest gravitationally 
bound structures in the Universe, and thus should provide 
a fair sample of its matter content.  This makes clusters 
good candidates for cosmological studies.  In particular, 
the gravitational growth of initial density perturbations 
can be used to constrain cosmological parameters via determination of the 
mass function of clusters of galaxies \citep[e.g.\ ][]{voit05}, but 
this requires good knowledge of cluster masses.  

One of the most promising avenues towards the measurement of the mass 
function of clusters is based on estimates of the cluster temperature 
and luminosity from X-ray observations of a large number of objects.
Specifically, the number density for clusters of different masses can 
be estimated using mass-observable relations, calibrated using 
nearby clusters where spatially resolved spectroscopy is available.  
The mass-temperature ($M$-$T$) \citep[e.g.\ ][]{arnaud05} and 
luminosity-temperature ($L_X$-$T$) \citep[e.g.\ ][]{arnaud99} 
scaling relations are of particular importance since temperature 
is often used as a proxy for mass and the relation 
to luminosity is needed to understand the sample selection function 
since X-ray selected samples generally are flux limited. 
More recently, a new proxy for cluster mass, $Y_X$, has been 
proposed \citep{kravtsov06}. This quantity is a simple product 
of the gas mass of the intra cluster medium and its temperature and 
has shown to exhibit a low amount of scatter.

Clusters are formed hierarchically through mergers of smaller 
clusters and groups. This merging activity is observed as 
distortions of their X-ray surface brightness profiles.  
Substructures and mergers affect the mass determinations 
and increase the scatter in the scaling relations.  Measurements 
using only ``relaxed'' clusters have achieved high precision 
\citep{vikh06} and selective studies, such as those of the 
gas mass fraction ($f_{gas}$) in clusters focus only on the 
largest clusters with minimal amount of substructure \citep{allen07}. 
However, substructure was found to be present in $\sim 50\%$ 
of clusters in a ROSAT study \citep{schuecker01} and studies of XMM 
data in a REFLEX-DXL study find substructures present in all 
clusters in that sample \citep{fino05}. Understanding, and 
assessing the effect of substructure on the robustness of mass 
determination is thus crucial.  

Besides the complications associated with the merger activity, evolution 
of clusters can affect the applicability of scaling relations.  
The high density environments in galaxy cluster cores cause them 
to cool radiatively and this is observed in undisturbed clusters 
as a decrease of the average projected temperature towards the 
center \citep[e.g.\ ][]{peterson06}. The undisturbed ``cooling 
clusters'' have sharply peaked luminosity profiles, a feature 
that is not observed in their non-cooling counterparts. Cool-core 
clusters deviate from the $L_X$-$T$ relation since the core has 
higher luminosity and lower temperature than the cluster population 
on average.
Observational 
evidence exists supporting the argument that cool cores can survive 
to some extent during a cluster merger.  These core remnants are 
then observed as sharp contact discontinuities or ``cold fronts'' 
\citep[e.g.\ ][]{markevitch07}, possibly affecting the $M$-$T$ scaling.  

All this indicates that it is important to assess and quantify 
the dynamical state of a cluster 
when the cluster data are used for the determination of cosmological 
parameters.  
While this has been successfully attempted in the past 
\citep[see e.g][]{maughan07,chen07}, 
statistical studies of cluster substructure have so far only used 
the information from the spatial distribution of X-ray counts on the 
sky, hence only mapping the luminosity structure. 
Ideally, the knowledge about the temperature structure should 
also be included in these searches since it holds important information 
about the dynamical history of the cluster.

The Smoothed Particle Inference (SPI) technique, developed recently 
by some of us \citep{peterson07} as an alternative to standard analysis 
techniques, is well suited for detecting the effects of 
cooling cores and substructure from both the projected temperature 
and luminosity distributions of clusters \citep[see\ ][]{andersson07}. 
This method relies on a description of a cluster as a 
large set of smoothed particles (two-dimensional, spatial Gaussians), 
each of which is described by a luminosity, spatial position, 
Gaussian width, temperature, redshift, and a set of elemental abundances.  
A large set of these particles is propagated through an instrument model, 
and the model parameters are adjusted using Markov Chain methods.  The 
resulting distribution becomes a kilo-parametric description of the cluster.  

In this paper, the SPI method is used on a large number of cluster 
observations available through the XMM public archives.  A cluster model is 
built using the the imaging spectroscopy XMM data.  The output of the modeling 
is used to separate clusters with cooling cores from more disturbed clusters 
and to study their properties separately. Specifically, we aim to assess 
the effects of cooling cores and substructure on the luminosity-temperature 
relation 
and study any possible redshift dependence of these effects.  
We also apply a new statistic, including the spatial distribution 
of both luminosity and temperature to quantify the level of dynamical 
instability present in the clusters. This statistic is designed to 
distinguish cool core clusters and isothermal clusters from 
those with more temperature structure indicating a recent or ongoing 
merger event.

In Section 2 we describe the construction and properties of the cluster sample 
and outline the data reduction scheme, 
in Section 3 we explain the different methods used to analyze the data, 
and in Section 4 
we describe the processing performed on the output cluster models and the 
methods used to quantify cluster properties. 
In Section 5 we display the statistical results of our modeling 
and in Section 6 we discuss the possible systematic effects associated 
with this. We finally conclude in Section 7 with a discussion on 
the results, problems and possible improvements. 

In all calculations we have assumed a concordance cosmology with 
$\Omega_M = 0.3$, $\Omega_\Lambda = 0.7$ and 
$H_0 = 70~$km$~$s$^{-1}~$Mpc$^{-1}$.

\section{Cluster sample and data preparation}

The cluster sample was compiled by cross-correlating 
the NASA Extragalactic Database (NED) with the public XMM-Newton 
observation archive as of May 10, 2006. The requirement for selection 
in NED was 
that the cluster should be a known X-ray source and a known 
galaxy cluster (2005 clusters) or group (120 groups). 
The XMM pointing was required to be within $3.5'$ of the source 
position in NED.  This resulted in 278 matches for clusters and 
34 for groups, some of which were multiple matches of the same cluster. 
After multiples and sources not visible in the X-ray data were 
removed the total number of clusters and groups was 201.
We further removed clusters where we could not get a reliable spectrum, 
i.e. clusters with a fluence (time-integrated energy flux)  below 
$10^{-8}$ erg cm$^{-2}$, as well as nearby sources where we could not 
fit 1 Mpc within a 13$\arcmin$ radius.  This selection left us with 
101 sources (see Table \ref{spectab}).  We note that although this sample 
is by no means complete, it represents a broad range of cluster properties 
over a large range of redshift.  

\begin{figure*}[!htb]
  \begin{center}
    \begin{tabular}{c}
      \includegraphics[width=4in,angle=-90]{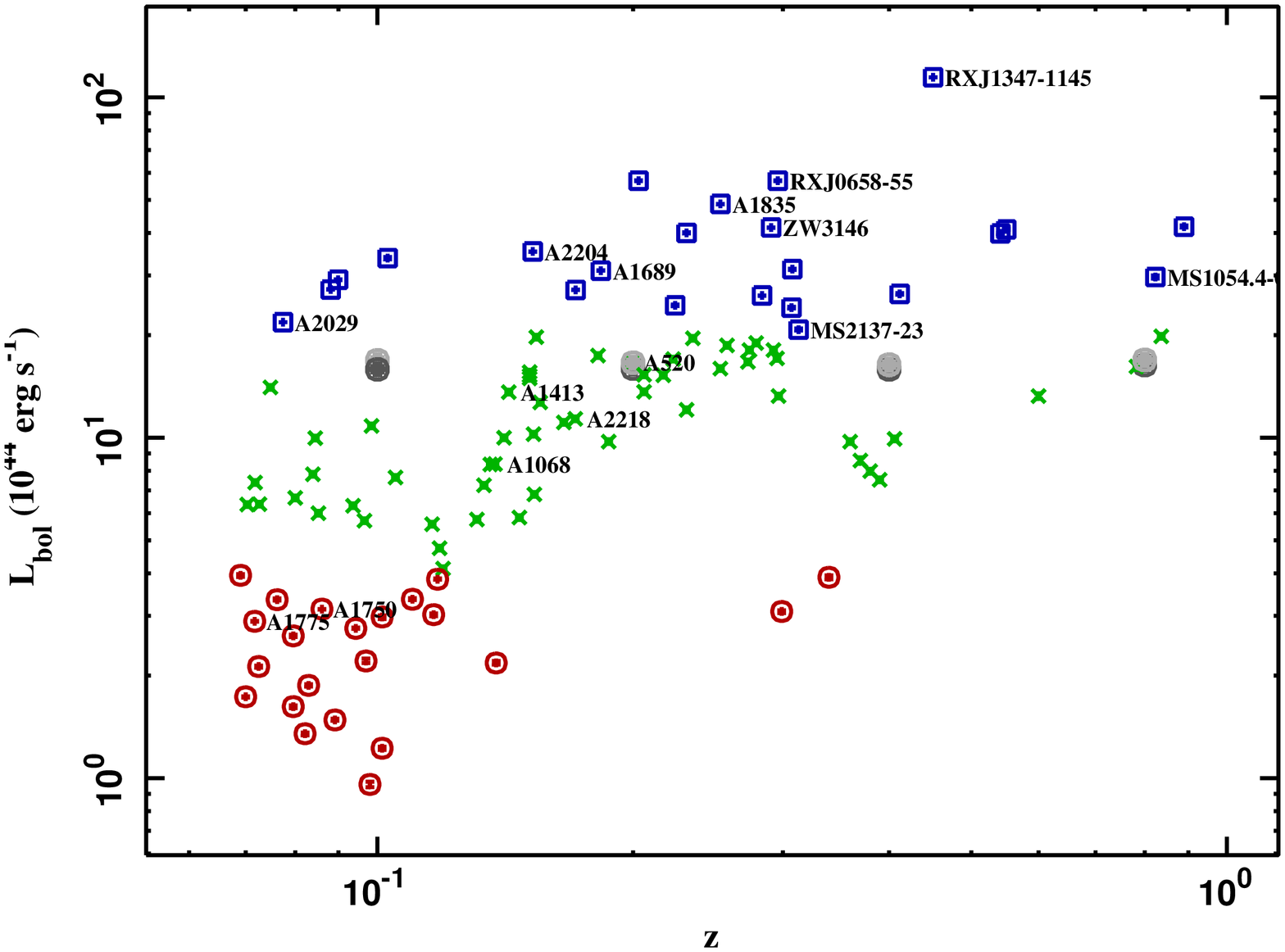} \\
    \end{tabular}
  \end{center}
  \caption{The distribution of bolometric luminosities for the sample. Low $L_X$ clusters ($< 4~10^{44} $erg s$^{-1}$) are shown as red circles, intermediate $L_X$ clusters ($4~10^{44}~$erg s$^{-1} \le L_X < 2~10^{45}~$erg s$^{-1}$) as green stars and high $L_X$ clusters ($\ge 2~10^{45}~$erg s$^{-1}$) as 
blue squares. Simulated cool core clusters are shown as gray filled circles.\label{L-z}}
\end{figure*}

The sample encompasses clusters with luminosities from 
$9 \times 10^{43}~$erg$~$s$^{-1}$ 
to $1.1 \times 10^{46}~$erg$~$s$^{-1}$, average temperatures from 
$2.2$ to $11.6~$keV (See Section \ref{standardana}) 
and redshifts from $z=0.069$ to $z=0.89$. 
The distribution of luminosities with 
redshift for the sample is shown in Figure \ref{L-z} where the names of a 
subsample of well-known clusters are printed in the plot.
Luminosities were calculated from the observed flux, using tabulated 
redshifts (as listed in NED), assuming a concordance cosmology  
(mentioned above), and applying a bolometric correction.

\subsection{Data reduction}

The data were reduced using standard pipeline processing and 
the calibration implementation as of XMM Science Analysis Software 
(SAS) version 6.5. 
Background flares from soft protons were removed using lightcurve 
filtering in both soft (MOS: 0.3-10 keV, PN: 0.3-12 keV) and 
hard (MOS: 10-12 keV, PN: 12-14 keV) X-ray bands. For the soft band, 
light curves were binned in 10 s intervals while we used 100 s bins 
for the hard band. In both cases the data were not included  
during the time when the count rate exceeded $3 \sigma$ above 
the quiescent count rate, indicating a proton flare. 

The event files were also filtered for non- X-ray events by selecting 
only single and double pixel events for PN and single to quadruple 
events for MOS. Bad pixels and pixel columns were removed by applying 
the standard keywords in event selection; FLAG=0 and \#XMMEA\_EM / \#XMMEA\_EP. 
The data reduction follows that described in \citet{andersson07}. 

\begin{deluxetable*}{lccccc}
\tabletypesize{\scriptsize}
\tablecaption{Spectral fit parameters \label{spectab}}
\tablewidth{0pt}
\tablehead{
\colhead{Name} &
\colhead{z} &
\colhead{$T~$[keV]} &
\colhead{Abundance wrt. Solar} &
\colhead{$L_{bol}$ [erg s$^{-1}$]} &
\colhead{Exposure [ks]} \\
}
\tablecomments{Redshifts, temperatures, metal abundances (wrt. Solar), luminosities and exposure times for the sample. Redshifts are taken from the NED listing. Temperatures and metal abundances were derived using spectral fits (Section \ref{standardana}). Luminosities within a 1 Mpc aperture were derived using the SPI modeling (Section \ref{spi}).}

\startdata
CIZAJ1645.4-7334 & 0.069 & $ 3.86^{+0.24}_{-0.22} $ & $ 0.65^{+0.12}_{-0.11} $ & $ 3.9 ~ 10^{44} $ & 14 \\
A1837 & 0.070 & $ 3.85^{+0.08}_{-0.08} $ & $ 0.38^{+0.03}_{-0.03} $ & $ 1.7 ~ 10^{44} $ & 45  \\
A3112 & 0.070 & $ 4.32^{+0.05}_{-0.04} $ & $ 0.50^{+0.02}_{-0.02} $ & $ 6.4 ~ 10^{44} $ & 21  \\
A1775 & 0.072 & $ 3.70^{+0.10}_{-0.06} $ & $ 0.49^{+0.03}_{-0.03} $ & $ 2.9 ~ 10^{44} $ & 23  \\
A399 & 0.072 & $ 6.92^{+0.23}_{-0.23} $ & $ 0.28^{+0.04}_{-0.04} $ & $ 7.4 ~ 10^{44} $ & 11  \\
A1589 & 0.072 & $ 4.81^{+0.20}_{-0.18} $ & $ 0.30^{+0.06}_{-0.05} $ & $ 2.1 ~ 10^{44} $ & 15  \\
A2065 & 0.073 & $ 5.36^{+0.11}_{-0.10} $ & $ 0.27^{+0.02}_{-0.03} $ & $ 6.4 ~ 10^{44} $ & 18  \\
A401 & 0.075 & $ 7.79^{+0.22}_{-0.23} $ & $ 0.27^{+0.03}_{-0.03} $ & $ 1.4 ~ 10^{45} $ & 11  \\
A2670 & 0.076 & $ 4.02^{+0.12}_{-0.11} $ & $ 0.40^{+0.04}_{-0.04} $ & $ 3.3 ~ 10^{44} $ & 17  \\
A2029 & 0.077 & $ 6.88^{+0.11}_{-0.08} $ & $ 0.41^{+0.02}_{-0.02} $ & $ 2.2 ~ 10^{45} $ & 11  \\
RXCJ1236.7-3354 & 0.080 & $ 2.80^{+0.11}_{-0.10} $ & $ 0.46^{+0.08}_{-0.07} $ & $ 1.6 ~ 10^{44} $ & 12  \\
RXCJ2129.8-5048 & 0.080 & $ 4.55^{+0.26}_{-0.19} $ & $ 0.27^{+0.06}_{-0.06} $ & $ 2.6 ~ 10^{44} $ & 21  \\
A2255 & 0.080 & $ 6.98^{+0.35}_{-0.28} $ & $ 0.28^{+0.07}_{-0.05} $ & $ 6.7 ~ 10^{44} $ & 10  \\
RXCJ0821.8+0112 & 0.082 & $ 3.69^{+0.29}_{-0.24} $ & $ 0.28^{+0.10}_{-0.09} $ & $ 1.3 ~ 10^{44} $ & 9  \\
RXCJ1302.8-0230 & 0.083 & $ 3.52^{+0.09}_{-0.09} $ & $ 0.49^{+0.06}_{-0.05} $ & $ 1.9 ~ 10^{44} $ & 22  \\
A1650 & 0.084 & $ 5.53^{+0.07}_{-0.08} $ & $ 0.36^{+0.02}_{-0.02} $ & $ 7.8 ~ 10^{44} $ & 37  \\
A1651 & 0.084 & $ 6.20^{+0.18}_{-0.16} $ & $ 0.34^{+0.03}_{-0.03} $ & $ 10.0 ~ 10^{44} $ & 10  \\
A2597 & 0.085 & $ 3.46^{+0.03}_{-0.03} $ & $ 0.39^{+0.01}_{-0.01} $ & $ 6.0 ~ 10^{44} $ & 56  \\
A1750 & 0.086 & $ 4.45^{+0.13}_{-0.13} $ & $ 0.31^{+0.04}_{-0.04} $ & $ 3.1 ~ 10^{44} $ & 29  \\
A478 & 0.088 & $ 6.04^{+0.04}_{-0.04} $ & $ 0.37^{+0.01}_{-0.01} $ & $ 2.7 ~ 10^{45} $ & 96  \\
A278 & 0.089 & $ 3.39^{+0.13}_{-0.14} $ & $ 0.26^{+0.05}_{-0.05} $ & $ 1.5 ~ 10^{44} $ & 27  \\
A2142 & 0.090 & $ 8.15^{+0.24}_{-0.28} $ & $ 0.30^{+0.04}_{-0.04} $ & $ 2.9 ~ 10^{45} $ & 6  \\
A3921 & 0.094 & $ 5.65^{+0.15}_{-0.14} $ & $ 0.33^{+0.04}_{-0.04} $ & $ 6.3 ~ 10^{44} $ & 29  \\
A13 & 0.094 & $ 5.00^{+0.17}_{-0.17} $ & $ 0.27^{+0.04}_{-0.04} $ & $ 2.8 ~ 10^{44} $ & 31 \\
A3911 & 0.097 & $ 5.94^{+0.19}_{-0.14} $ & $ 0.26^{+0.03}_{-0.03} $ & $ 5.7 ~ 10^{44} $ & 25 \\
RXCJ2319.6-7313 & 0.097 & $ 2.27^{+0.09}_{-0.07} $ & $ 0.32^{+0.05}_{-0.05} $ & $ 2.2 ~ 10^{44} $ & 8 \\
CL0852+1618 & 0.098 & $ 2.76^{+0.25}_{-0.21} $ & $ 0.85^{+0.24}_{-0.19} $ & $ 9.6 ~ 10^{43} $ & 31 \\
A3827 & 0.098 & $ 6.93^{+0.15}_{-0.13} $ & $ 0.27^{+0.02}_{-0.02} $ & $ 1.1 ~ 10^{45} $ & 21 \\
RXCJ0211.4-4017 & 0.101 & $ 2.22^{+0.07}_{-0.07} $ & $ 0.43^{+0.07}_{-0.06} $ & $ 1.2 ~ 10^{44} $ & 26 \\
A2241 & 0.101 & $ 3.66^{+0.45}_{-0.28} $ & $ 0.81^{+0.26}_{-0.18} $ & $ 3.0 ~ 10^{44} $ & 5 \\
PKS0745-19 & 0.103 & $ 6.44^{+0.07}_{-0.07} $ & $ 0.34^{+0.01}_{-0.01} $ & $ 3.4 ~ 10^{45} $ & 20 \\
RXCJ0645.4-5413 & 0.105 & $ 7.39^{+0.33}_{-0.26} $ & $ 0.24^{+0.04}_{-0.04} $ & $ 7.7 ~ 10^{44} $ & 13 \\
RXCJ0049.4-2931 & 0.110 & $ 4.02^{+0.19}_{-0.18} $ & $ 0.40^{+0.07}_{-0.06} $ & $ 3.4 ~ 10^{44} $ & 19 \\
A1302 & 0.116 & $ 6.59^{+0.46}_{-0.42} $ & $ 0.53^{+0.10}_{-0.09} $ & $ 5.6 ~ 10^{44} $ & 16 \\
RXCJ0616.8-4748 & 0.116 & $ 5.05^{+0.45}_{-0.44} $ & $ 0.26^{+0.10}_{-0.10} $ & $ 3.0 ~ 10^{44} $ & 9 \\
RXCJ2149.1-3041 & 0.118 & $ 3.53^{+0.07}_{-0.07} $ & $ 0.47^{+0.04}_{-0.04} $ & $ 3.8 ~ 10^{44} $ & 23 \\
RXCJ1516.3+0005 & 0.118 & $ 5.25^{+0.16}_{-0.15} $ & $ 0.30^{+0.04}_{-0.04} $ & $ 4.7 ~ 10^{44} $ & 25 \\
RXCJ1141.4-1216 & 0.119 & $ 3.53^{+0.06}_{-0.06} $ & $ 0.53^{+0.04}_{-0.03} $ & $ 4.1 ~ 10^{44} $ & 26 \\
RXCJ0020.7-2542 & 0.131 & $ 6.47^{+0.27}_{-0.23} $ & $ 0.24^{+0.04}_{-0.04} $ & $ 5.8 ~ 10^{44} $ & 16 \\
RXCJ1044.5-0704 & 0.134 & $ 3.67^{+0.04}_{-0.06} $ & $ 0.36^{+0.03}_{-0.02} $ & $ 7.3 ~ 10^{44} $ & 25 \\
RXCJ0145.0-5300 & 0.136 & $ 6.80^{+0.42}_{-0.44} $ & $ 0.34^{+0.09}_{-0.08} $ & $ 8.3 ~ 10^{44} $ & 14 \\
A1068 & 0.138 & $ 3.81^{+0.08}_{-0.07} $ & $ 0.39^{+0.03}_{-0.03} $ & $ 8.4 ~ 10^{44} $ & 20 \\
RXJ1416.4+2315 & 0.138 & $ 3.58^{+0.38}_{-0.33} $ & $ 0.25^{+0.14}_{-0.13} $ & $ 2.2 ~ 10^{44} $ & 6 \\
RXCJ0605.8-3518 & 0.141 & $ 4.52^{+0.08}_{-0.08} $ & $ 0.39^{+0.03}_{-0.03} $ & $ 1.0 ~ 10^{45} $ & 21 \\
A1413 & 0.143 & $ 7.30^{+0.19}_{-0.19} $ & $ 0.36^{+0.04}_{-0.04} $ & $ 1.4 ~ 10^{45} $ & 24 \\
RXCJ2048.1-1750 & 0.147 & $ 5.92^{+0.28}_{-0.23} $ & $ 0.23^{+0.05}_{-0.05} $ & $ 5.8 ~ 10^{44} $ & 23 \\
A3888 & 0.151 & $ 9.31^{+0.69}_{-0.51} $ & $ 0.23^{+0.08}_{-0.07} $ & $ 1.5 ~ 10^{45} $ & 5 \\
A2034 & 0.151 & $ 7.41^{+0.27}_{-0.21} $ & $ 0.29^{+0.04}_{-0.04} $ & $ 1.6 ~ 10^{45} $ & 12 \\
RXCJ2234.5-3744 & 0.151 & $ 7.88^{+0.22}_{-0.22} $ & $ 0.19^{+0.03}_{-0.03} $ & $ 1.5 ~ 10^{45} $ & 23 \\
A2204 & 0.152 & $ 6.44^{+0.08}_{-0.09} $ & $ 0.37^{+0.02}_{-0.02} $ & $ 3.5 ~ 10^{45} $ & 19 \\
RXCJ0958.3-1103 & 0.153 & $ 5.20^{+0.19}_{-0.20} $ & $ 0.40^{+0.05}_{-0.05} $ & $ 1.0 ~ 10^{45} $ & 8 \\
\enddata
\end{deluxetable*}

\begin{deluxetable*}{lccccc}
\tabletypesize{\scriptsize}
\tablecaption{Spectral fit parameters, Contd. \label{spectab2}}
\tablewidth{0pt}
\tablehead{
\colhead{Name} &
\colhead{z} &
\colhead{$T~$[keV]} &
\colhead{Abundance wrt. Solar} &
\colhead{$L_{bol}$ [erg s$^{-1}$]} &
\colhead{Exposure [ks]} \\
}

\tablecomments{Redshifts, temperatures, metal abundances (wrt. Solar), luminosities and exposure times for the sample. Redshifts are taken from the NED listing. Temperatures and metal abundances were derived using spectral fits (Section \ref{standardana}). Luminosities within a 1 Mpc aperture were derived using the SPI modeling (Section \ref{spi}).}
\startdata
A868 & 0.153 & $ 5.86^{+0.54}_{-0.39} $ & $ 0.21^{+0.08}_{-0.08} $ & $ 6.8 ~ 10^{44} $ & 6 \\
RXCJ2014.8-2430 & 0.154 & $ 4.90^{+0.08}_{-0.07} $ & $ 0.37^{+0.02}_{-0.02} $ & $ 2.0 ~ 10^{45} $ & 23 \\
A2104 & 0.155 & $ 9.55^{+1.43}_{-0.97} $ & $ 0.38^{+0.14}_{-0.13} $ & $ 1.3 ~ 10^{45} $ & 5 \\
RXCJ0547.6-3152 & 0.166 & $ 6.92^{+0.22}_{-0.20} $ & $ 0.31^{+0.04}_{-0.03} $ & $ 1.1 ~ 10^{45} $ & 22 \\
A2218 & 0.171 & $ 7.17^{+0.21}_{-0.17} $ & $ 0.22^{+0.03}_{-0.03} $ & $ 1.1 ~ 10^{45} $ & 16 \\
A1914 & 0.171 & $ 9.49^{+0.28}_{-0.18} $ & $ 0.25^{+0.03}_{-0.03} $ & $ 2.7 ~ 10^{45} $ & 22 \\
A665 & 0.182 & $ 7.94^{+0.23}_{-0.24} $ & $ 0.25^{+0.03}_{-0.03} $ & $ 1.7 ~ 10^{45} $ & 57 \\
A1689 & 0.183 & $ 9.07^{+0.17}_{-0.12} $ & $ 0.28^{+0.02}_{-0.02} $ & $ 3.1 ~ 10^{45} $ & 35 \\
A383 & 0.187 & $ 4.21^{+0.09}_{-0.08} $ & $ 0.46^{+0.04}_{-0.03} $ & $ 9.7 ~ 10^{44} $ & 28 \\
A520 & 0.199 & $ 8.45^{+0.33}_{-0.26} $ & $ 0.24^{+0.04}_{-0.03} $ & $ 1.7 ~ 10^{45} $ & 38 \\
A2163 & 0.203 & $ 11.12^{+0.36}_{-0.39} $ & $ 0.21^{+0.04}_{-0.04} $ & $ 5.7 ~ 10^{45} $ & 10 \\
A209 & 0.206 & $ 7.21^{+0.27}_{-0.26} $ & $ 0.27^{+0.04}_{-0.04} $ & $ 1.5 ~ 10^{45} $ & 17 \\
A963 & 0.206 & $ 6.43^{+0.22}_{-0.19} $ & $ 0.31^{+0.04}_{-0.03} $ & $ 1.4 ~ 10^{45} $ & 23 \\
A773 & 0.217 & $ 7.41^{+0.33}_{-0.26} $ & $ 0.29^{+0.05}_{-0.05} $ & $ 1.5 ~ 10^{45} $ & 15 \\
A1763 & 0.223 & $ 7.67^{+0.34}_{-0.33} $ & $ 0.34^{+0.05}_{-0.06} $ & $ 1.7 ~ 10^{45} $ & 12 \\
A2261 & 0.224 & $ 8.66^{+0.71}_{-0.67} $ & $ 0.41^{+0.11}_{-0.11} $ & $ 2.5 ~ 10^{45} $ & 4 \\
A267 & 0.231 & $ 6.67^{+0.38}_{-0.37} $ & $ 0.34^{+0.07}_{-0.06} $ & $ 1.2 ~ 10^{45} $ & 15 \\
A2390 & 0.231 & $ 8.68^{+0.29}_{-0.27} $ & $ 0.35^{+0.04}_{-0.04} $ & $ 4.0 ~ 10^{45} $ & 12 \\
RXJ2129.6+0005 & 0.235 & $ 5.74^{+0.04}_{-0.10} $ & $ 0.38^{+0.03}_{-0.03} $ & $ 2.0 ~ 10^{45} $ & 43 \\
A1835 & 0.253 & $ 7.14^{+0.10}_{-0.11} $ & $ 0.30^{+0.02}_{-0.02} $ & $ 4.9 ~ 10^{45} $ & 37 \\
RXCJ0307.0-2840 & 0.253 & $ 6.47^{+0.38}_{-0.35} $ & $ 0.32^{+0.06}_{-0.06} $ & $ 1.6 ~ 10^{45} $ & 11 \\
E1455+2232 & 0.258 & $ 4.59^{+0.08}_{-0.09} $ & $ 0.35^{+0.03}_{-0.03} $ & $ 1.9 ~ 10^{45} $ & 33 \\
RXCJ2337.6+0016 & 0.273 & $ 7.74^{+0.66}_{-0.52} $ & $ 0.19^{+0.07}_{-0.07} $ & $ 1.7 ~ 10^{45} $ & 11 \\
RXCJ0303.8-7752 & 0.274 & $ 8.21^{+0.64}_{-0.62} $ & $ 0.26^{+0.08}_{-0.07} $ & $ 1.8 ~ 10^{45} $ & 10 \\
A1758 & 0.279 & $ 9.16^{+0.39}_{-0.43} $ & $ 0.29^{+0.06}_{-0.06} $ & $ 1.9 ~ 10^{45} $ & 44 \\
RXCJ0232.2-4420 & 0.284 & $ 7.13^{+0.31}_{-0.29} $ & $ 0.30^{+0.05}_{-0.05} $ & $ 2.6 ~ 10^{45} $ & 10 \\
ZW3146 & 0.291 & $ 6.21^{+0.14}_{-0.10} $ & $ 0.33^{+0.02}_{-0.02} $ & $ 4.1 ~ 10^{45} $ & 50 \\
RXCJ0043.4-2037 & 0.292 & $ 6.95^{+0.46}_{-0.42} $ & $ 0.29^{+0.08}_{-0.07} $ & $ 1.8 ~ 10^{45} $ & 10 \\
RXCJ0516.7-5430 & 0.295 & $ 8.33^{+0.84}_{-0.74} $ & $ 0.19^{+0.09}_{-0.09} $ & $ 1.7 ~ 10^{45} $ & 10 \\
RXJ0658-55 & 0.296 & $ 11.58^{+0.26}_{-0.35} $ & $ 0.23^{+0.03}_{-0.03} $ & $ 5.7 ~ 10^{45} $ & 29 \\
RXCJ2308.3-0211 & 0.297 & $ 7.22^{+0.91}_{-0.67} $ & $ 0.40^{+0.12}_{-0.11} $ & $ 1.3 ~ 10^{45} $ & 9 \\
RXJ2237.0-1516 & 0.299 & $ 3.46^{+0.44}_{-0.42} $ & $ 0.46^{+0.24}_{-0.20} $ & $ 3.1 ~ 10^{44} $ & 19 \\
RXCJ1131.9-1955 & 0.307 & $ 7.69^{+0.55}_{-0.40} $ & $ 0.28^{+0.06}_{-0.06} $ & $ 2.4 ~ 10^{45} $ & 11 \\
RXCJ0014.3-3022 & 0.308 & $ 8.36^{+0.46}_{-0.45} $ & $ 0.24^{+0.05}_{-0.05} $ & $ 3.1 ~ 10^{45} $ & 13 \\
MS2137-23 & 0.313 & $ 4.67^{+0.17}_{-0.19} $ & $ 0.36^{+0.05}_{-0.06} $ & $ 2.1 ~ 10^{45} $ & 11 \\
MS1208.7+3928 & 0.340 & $ 5.85^{+1.04}_{-0.80} $ & $ 0.79^{+0.39}_{-0.33} $ & $ 3.9 ~ 10^{44} $ & 11 \\
RXJ0256.5+0006 & 0.360 & $ 6.68^{+1.02}_{-0.80} $ & $ 0.47^{+0.18}_{-0.17} $ & $ 9.7 ~ 10^{44} $ & 11 \\
RXJ0318.2-0301 & 0.370 & $ 6.07^{+1.07}_{-0.81} $ & $ 0.22^{+0.15}_{-0.15} $ & $ 8.6 ~ 10^{44} $ & 16 \\
RXJ0426.1+1655 & 0.380 & $ 7.85^{+2.64}_{-1.73} $ & $ 0.54^{+0.33}_{-0.27} $ & $ 8.0 ~ 10^{44} $ & 10 \\
RXJ1241.5+3250 & 0.390 & $ 6.63^{+0.74}_{-0.72} $ & $ 0.32^{+0.15}_{-0.14} $ & $ 7.5 ~ 10^{44} $ & 17 \\
A851 & 0.406 & $ 6.25^{+0.41}_{-0.45} $ & $ 0.22^{+0.07}_{-0.07} $ & $ 9.9 ~ 10^{44} $ & 43 \\
RXCJ2228+2037 & 0.412 & $ 9.03^{+0.50}_{-0.49} $ & $ 0.23^{+0.06}_{-0.06} $ & $ 2.6 ~ 10^{45} $ & 23 \\
RXJ1347-1145 & 0.451 & $ 11.44^{+0.26}_{-0.29} $ & $ 0.27^{+0.03}_{-0.03} $ & $ 1.1 ~ 10^{46} $ & 32 \\
CL0016+16 & 0.541 & $ 9.20^{+0.50}_{-0.55} $ & $ 0.29^{+0.06}_{-0.06} $ & $ 4.0 ~ 10^{45} $ & 28 \\
MS0451.6-0305 & 0.550 & $ 10.02^{+0.80}_{-0.60} $ & $ 0.36^{+0.08}_{-0.07} $ & $ 4.1 ~ 10^{45} $ & 28 \\
RXJ1120.1+4318 & 0.600 & $ 6.09^{+0.89}_{-0.69} $ & $ 0.54^{+0.18}_{-0.16} $ & $ 1.3 ~ 10^{45} $ & 18 \\
MS1137.5+6625 & 0.782 & $ 8.58^{+2.33}_{-1.98} $ & $ 0.35^{+0.37}_{-0.26} $ & $ 1.6 ~ 10^{45} $ & 18 \\
MS1054.4-0321 & 0.823 & $ 9.20^{+1.26}_{-1.03} $ & $ 0.22^{+0.13}_{-0.13} $ & $ 3.0 ~ 10^{45} $ & 25 \\
WARPJ0152.7-1357 & 0.837 & $ 7.93^{+0.73}_{-0.45} $ & $ 0.29^{+0.11}_{-0.11} $ & $ 2.0 ~ 10^{45} $ & 48 \\
CLJ1226.9+3332 & 0.890 & $ 10.69^{+0.82}_{-0.81} $ & $ 0.15^{+0.08}_{-0.08} $ & $ 4.2 ~ 10^{45} $ & 69 \\
\enddata
\end{deluxetable*}

\section{Data analysis methods}
We analyze all objects using both simple spectral analysis, where a single 
spectrum is extracted and fitted, as well as the SPI analysis, using the 
Monte Carlo approach and modeling the clusters both spatially and spectrally.

\subsection{Standard analysis}
\label{standardana}
All clusters are first analyzed using ``standard'' spectral analysis.  
This was conducted by extracting X-ray counts from a circular 
region centered on the peak of X-ray emission. The extraction radius was 
determined by estimating the radius at which a circle encompasses 90\% of the 
background subtracted surface intensity. 

The background was estimated using the surface intensity at the edge 
of the field. Background spectra were extracted in regions outside 
the source extraction region. Point sources were detected using SAS routine 
\verb1emldetect1 using only sources measured with a likelihood above 100. 

The extracted spectra were fitted using XSPEC \citep{xspec} software, employing a MEKAL 
\citep{mekal1,mekal2,mekal3,mekal4} 
thermal plasma model with Solar abundances absorbed by a WABS \citep{wabs} model, 
which we allow to be fitted as a free 
parameter. In the fit, the redshift was 
fixed to the known optical value as listed in the NED 
database entry. 

The results of these ``standard'' spectral fits including plasma temperature and 
metal abundances w.r.t. Solar are listed in Table \ref{spectab}. 
The bolometric luminosities shown in the table are derived using 
the SPI analysis below. 
The redshift as given in NED is 
also shown as well as the average effective exposure time after filtering. 
All observations had usable data for all three EPIC detectors with the 
exception of A665, A1413, A2261, A2597 and A3921 where only the two MOS 
detectors were available.

\subsection{SPI based MCMC analysis} 
\label{spi}
The cluster event files are modeled using Smoothed Particle 
Inference (SPI) \citep{peterson07} with a Monte Carlo model of the 
XMM-Newton EPIC Camera \citep{andersson07}. 
Within the SPI analysis, clusters are modeled as conglomerations of 2D spatial 
Gaussians with individual spectral models. In this work we use the 
MEKAL model to describe the thermal plasma. 
Model photons are simulated and propagated through the detector model 
adding background Monte Carlo events representing internal fluorescent 
lines, electronic noise and soft proton signals. 
Data and model photons are binned in three dimensional adaptive bins 
and compared via a two-sample likelihood function. 
All parameters, spatial and spectral, are iterated in a Markov Chain 
with an adaptive step length. All model samples in the converged 
part of the chain are used to reconstruct the cluster. 

\subsubsection{SPI setup}

The number of SPI particles are determined by the number of photons in the 
data so that, on average, $N_\gamma/N_p=400$, where $N_p$ is the number 
of particles and $N_\gamma$ is the number of X-ray events in the data. 
This number is chosen based on the Bayesian evidence calculation 
in \citet{peterson07} where we find the evidence reaching a plateau 
near 400 SPI particles for a 155000 photon observation. 
We choose to scale the number of particles with the number of photons 
since that is what ultimately determines the complexity of the data and 
we want a corresponding complexity of the model. 
We constrain the minimum number particles to be 100 and the maximum 
to be 1000.

The oversimulate factor, the factor that determines the number of model 
photons, is set so that there are 10 times as many photons in the model 
compared to the data with a maximum of 4 million model photons. 
This number is motivated by the drastic improvement of the optimization 
when using a factor of 10 or above as shown in \citet{peterson07}.  The upper limit 
is set to minimize computing time for very well exposed clusters. 

The three dimensional (x,y,pulseheight) adaptive binning grid is created so 
that every bin with more than 20 photons is divided in 2. The spectral 
dimension is divided 10 times more often than the spatial dimensions on 
average in order to achieve appropriate spectral resolution.

\subsubsection{Model setup}
The setup of the instrumental background model, consisting of electronic 
noise, soft proton detections and fluorescent emission lines, is analogous to 
the setup in \citet{andersson07}. The fraction of photons going to the 
background model is variable from 0 to 1, as are the relative normalizations of 
the included components. 

We model the soft X-ray background originating from our Galaxy using a uniform emission
component consisting of a thermal plasma spectral model.  This emission consists of local
(and thus weakly absorbed) component plus an absorbed, more distant contribution from the 
Galactic halo;  it is
adequately described as several thermal components in the 0.05-0.5 keV range \citep[e.g.\ ][]{kuntz00}. 
Motivated by our analysis of blank sky data files, we find that unabsorbed MEKAL model at
temperature of 0.16 keV, with metal abundance of $0.3 Z_\sun$ at $z=0$ describes
the data well, and this is a model we adopt for the Galactic background: 
given the limited EPIC bandpass, our approximation is sufficient. We keep in mind
that the spectral accuracy of the method is limited in the regions of the clusters
where a large fraction of the photons come from the X-ray background, such as at
large radii. 
While there the cluster flux is low, the adaptive binning grid will allow only
gross spectral features to be detectable.

Similarly the Cosmic X-ray Background (CXB) - presumably due to 
superposition of unresolved AGN - is modeled using a 
powerlaw with photon index $\Gamma=1.47$ and absorption fixed at 
the galactic value at the coordinates as given by \citet{dickey90}.

The cluster model consists of spatial Gaussians described by 
an x and y position and a Gaussian $\sigma$. 
Each Gaussian is assigned a spectral MEKAL model with WABS 
absorption. The allowed ranges for the absorbing equivalent 
hydrogen column, $n_H$, the plasma temperature, $T$, the 
metallicity w.r.t. the Solar values of \citet{anders}, $Z$, and 
redshift $z$ are shown in Table \ref{modtab}. 
Absorption is assumed to be within 20\% of the $n_{H,gal}$ 
values of \citet{dickey90} at the cluster coordinates.

\begin{deluxetable}{lll}
\tabletypesize{\scriptsize}
\tablecaption{Cluster model parameter ranges used in the SPI analysis\label{modtab}}
\tablewidth{0pt}
\tablehead{
\colhead{Parameter} &
\colhead{min value} &
\colhead{max value} \\
}
\tablecomments{The parameter ranges used in the SPI modeling. $n_H$ is variable within 20 \% of the values of \citet{dickey90}. The $x$ and $y$ positions are variable within $12 \arcmin$ of the XMM nominal pointing.}
\startdata
Spectral model & \quad & \quad \\
$n_H$ & 0.8$~n_{H,gal}$ & 1.2$~n_{H,gal}$ \\
$T$ [keV] & $0.5$ & $15.0$ \\ 
$Z$ wrt. Solar & 0 & 2 \\
$z$ & $z_{NED}$ fixed & \quad \\
\hline

Spatial model & \quad & \quad \\
$x$ & $-12\arcmin$ & $+12\arcmin$ \\
$y$ & $-12\arcmin$ & $+12\arcmin$ \\
ln$~\sigma (\arcsec)$ & 0 & 6 \\
\enddata
\end{deluxetable}

\subsubsection{Convergence}
In \citet{peterson07} a criterion for convergence of the Markov chain is 
applied and it is found to converge within 200 iterations. 
Here, the more conservative limit of 750 iterations is used as 
the point of convergence.  All the results derived from the model 
samples are from the iterations from 750 to 2000. 
This range is chosen based on slight deviations in cluster properties 
when derived from iteration 200 and onward. 

\subsubsection{Simulated clusters}
In order to study the systematic effects induced by the X-ray mirror PSF 
and limited statistics on faint sources, we simulate a massive, spherically 
symmetric, weak cooling  
core cluster at $z = 0.1$, $0.2$, $0.4$ and $0.8$. 
The simulated cluster model consists of two superimposed beta models with 
core radii of $74$ and $370$ kpc, both with $\beta = 0.7$. 
The smaller, core component is assigned a spectral model with $kT = 4~$keV 
whereas the ambient component has $kT = 9~$keV. 
The normalization ratio of the cold to the hot component in 
terms of emission measure, $\int n_e n_p \mathrm{d}V$, is $11/9$ 
and the overall normalization is set so that the 
total bolometric luminosity of the cluster is $L_{Bol} = 1.7 ~ 10^{45}~$erg s$^{-1}$, 
a typical luminosity in our sample, present at all redshifts $0.1\le z <0.8$.
The model is based on observations of the cool core clusters A2029 and A2241. 
Bolometric luminosities for the cluster sample are estimated using the $0.01$ - $100~$ keV 
energy band as described in Section \ref{mapsec}.

These simulated clusters all have the same background level 
($5~10^{-3}$ s$^{-1}~(\arcmin)^{-2}$) and are assumed to have an effective 
exposure time of 20 ks in all 3 EPIC detectors.  The assumed luminosity results in a total 
number of 850058, 209294, 46358 and 9353 cluster photons respectively for the 
$z = 0.1$, $0.2$, $0.4$ and $0.8$ clusters 
in the adopted energy range (MOS:0.3-10 keV, pn:1.1-10 keV). As a reference, the same clusters 
are also simulated without background and reconstructed separately. 
 
\begin{figure*}[!htb]
  \begin{center}
    \begin{tabular}{cc}
      \includegraphics[width=1.7in]{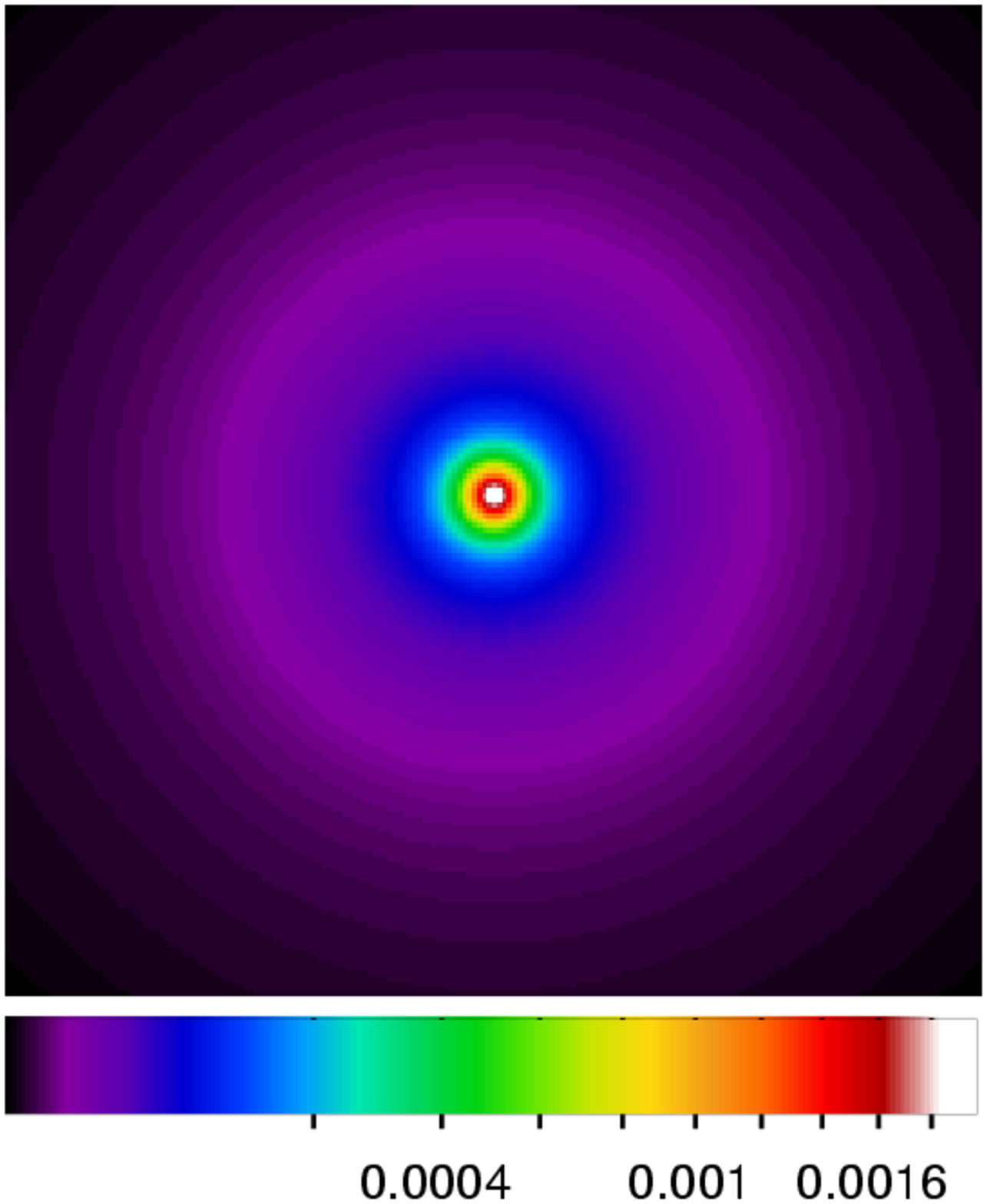} & 
      \includegraphics[width=1.7in]{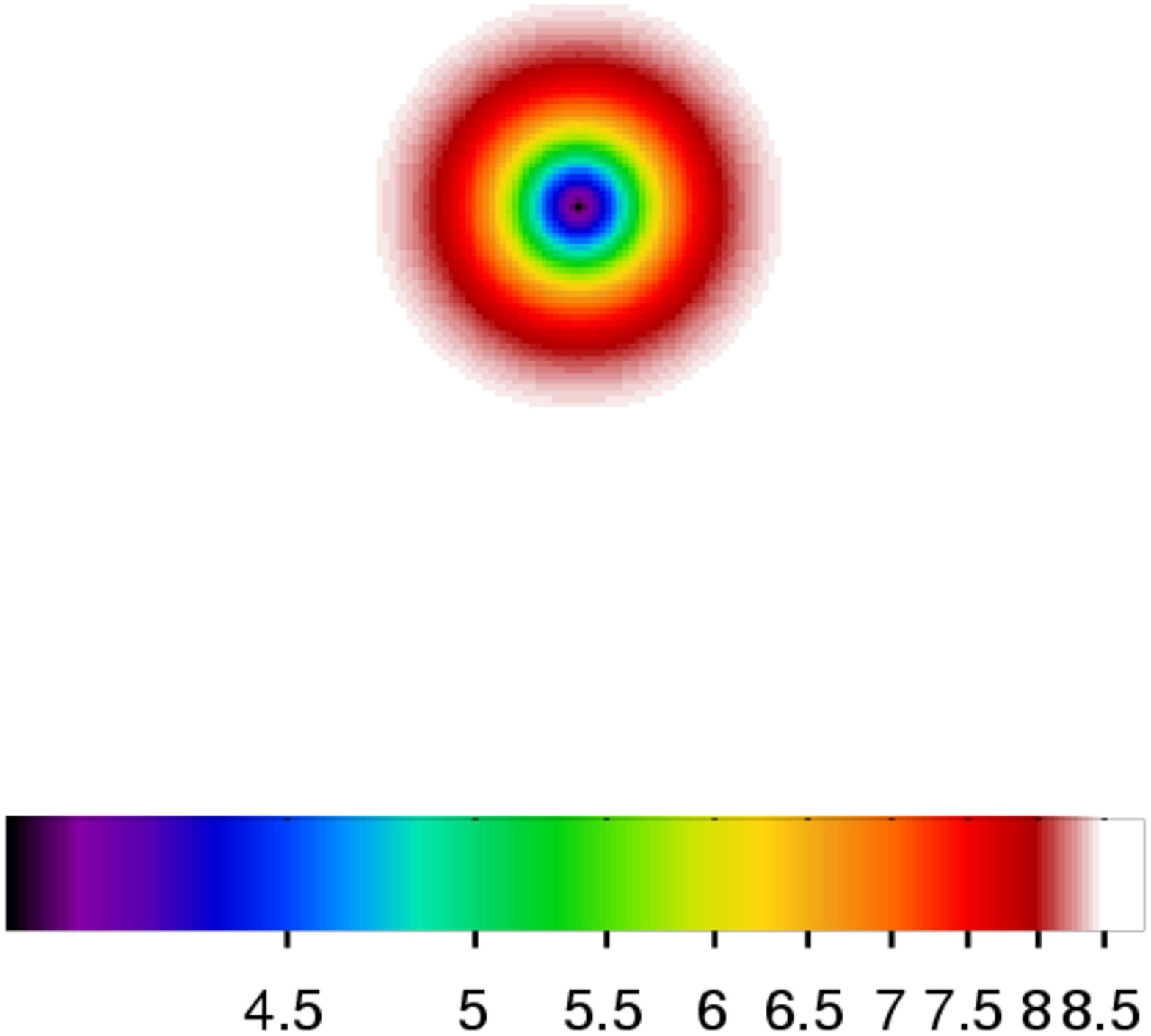} \\
    \end{tabular}
  \end{center}
  \caption{$2 \times 2$ Mpc images of the bolometric luminosity in units of $10^{44}~$erg s$^{-1}$ $(\arcsec)^{-2}$ (left panel) and temperature in keV (right panel) of the original simulated cool core cluster model. \label{simorig}}
\end{figure*}

The apparent evolution with $z$ in the derived properties for these identical 
clusters will be used to categorize the clusters in the sample.
Figure \ref{simorig} shows the original cluster model before propagated 
through the instrument model visualized with a luminosity and temperature map 
within the 1 Mpc radius. Luminosity per unit solid angle in this 
Figure is given in units of $10^{44}~$erg s$^{-1}$ $(\arcsec)^{-2}$ and temperature in keV. 

\begin{figure*}[!htb]
  \begin{center}
    \begin{tabular}{cc}
      \includegraphics[width=1.7in]{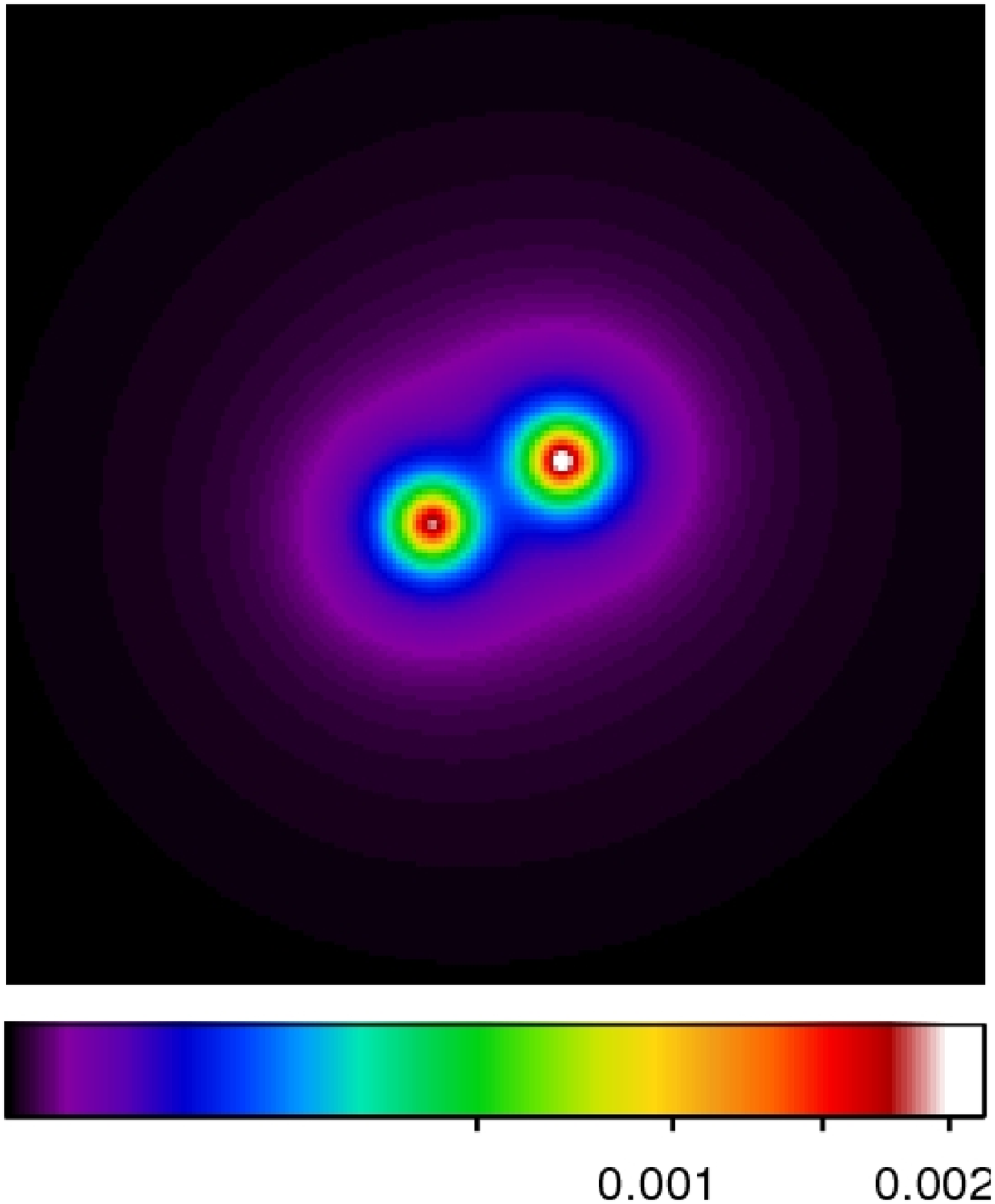} & 
      \includegraphics[width=1.7in]{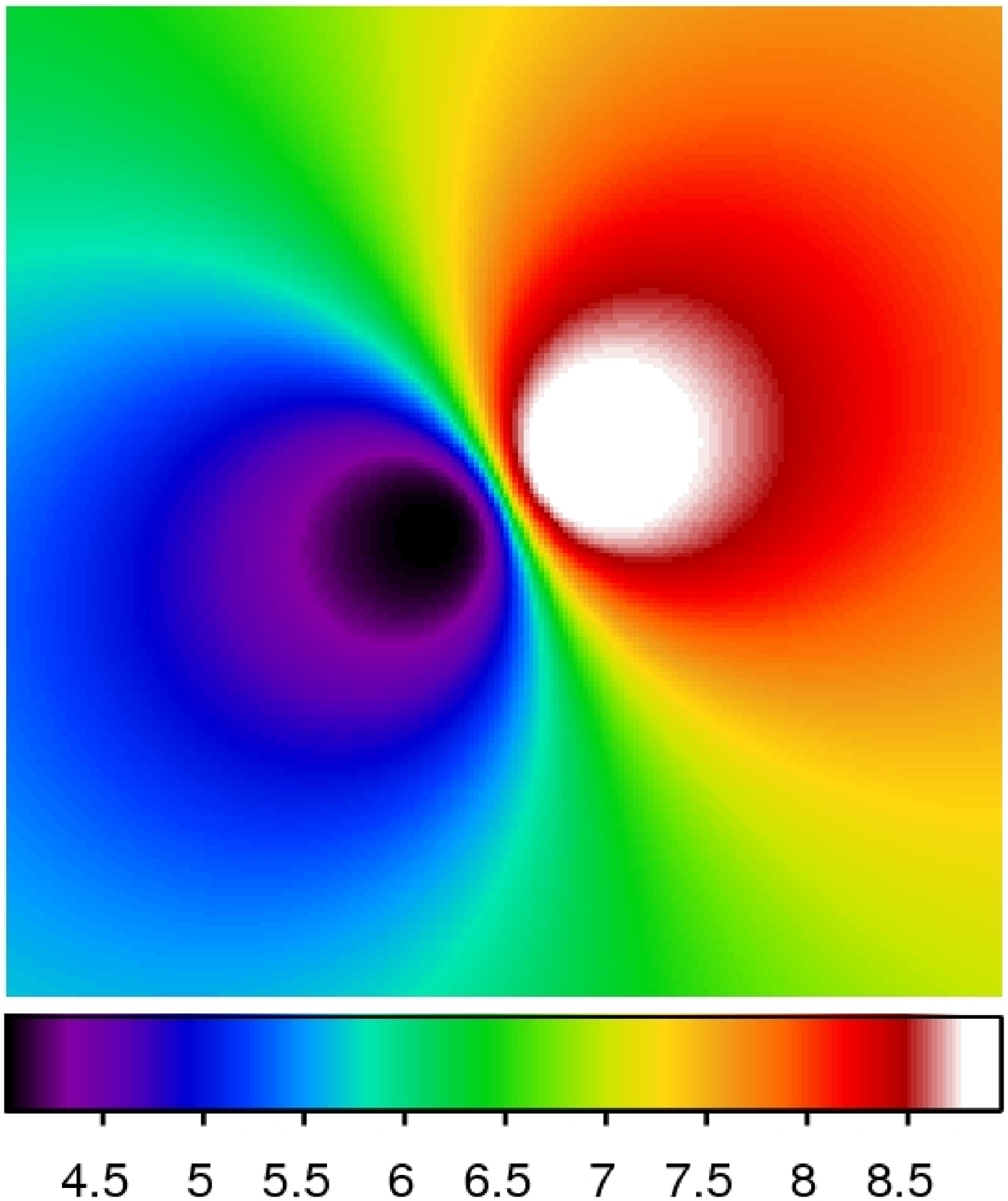} \\
    \end{tabular}
  \end{center}
  \caption{$2 \times 2$ Mpc images of the bolometric luminosity in units of $10^{44}~$erg s$^{-1}$ $(\arcsec)^{-2}$ (left panel) and temperature in keV (right panel) of the simulated two-subcluster model. \label{simorig2}}
\end{figure*}

\begin{figure*}[!htb]
  \begin{center}
    \begin{tabular}{cc}
      \includegraphics[width=1.7in]{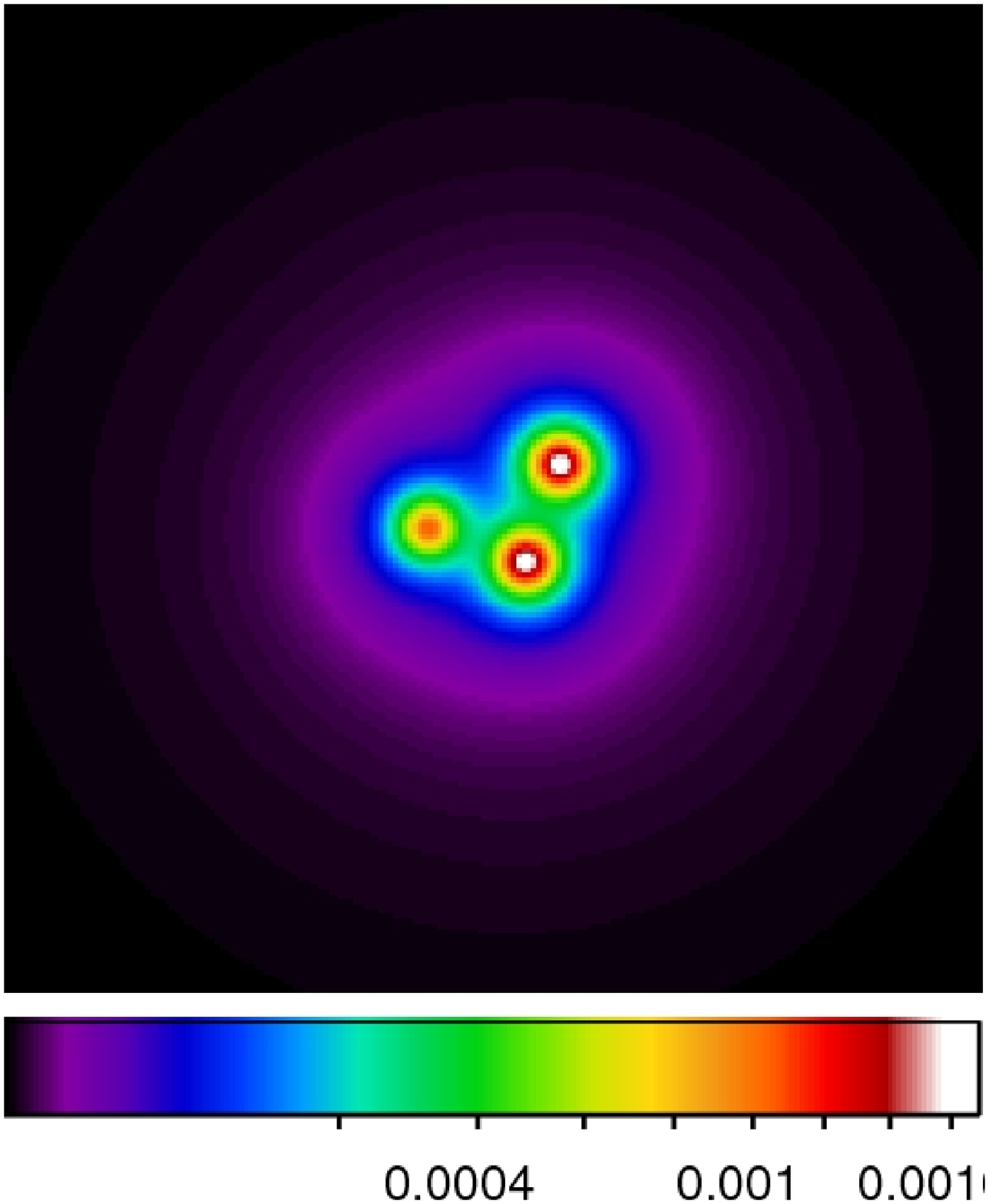} & 
      \includegraphics[width=1.7in]{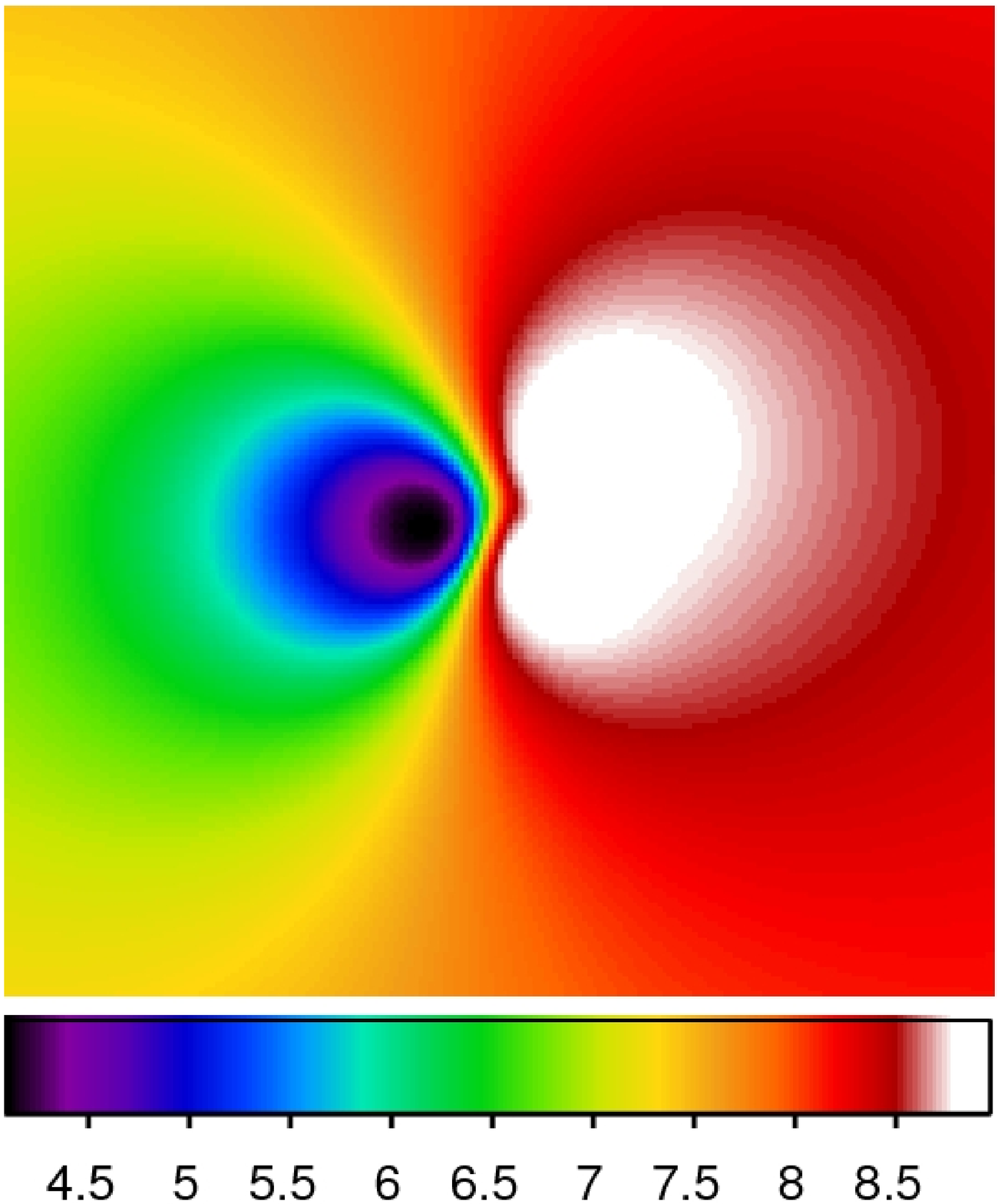} \\
    \end{tabular}
  \end{center}
  \caption{$2 \times 2$ Mpc images of the bolometric luminosity in units of $10^{44}~$erg s$^{-1}$ $(\arcsec)^{-2}$ (left panel) and temperature in keV (right panel) of the simulated three-subcluster model. \label{simorig3}}
\end{figure*}

In order to follow the evolution of substructure in clusters we also simulate 
two sets of clusters with irregular morphology, one using a two ``subcluster'' model and 
the other, a three ``subcluster'' model.
The two-subcluster model consists of two beta models, at 4 and 9 keV respectively, 
each with $r_c=74~$kpc and $\beta = 0.7$ separated by $300~$kpc. 
The total luminosity of the clusters is the same as for the original cool core cluster, 
$L_{Bol} = 1.7 ~ 10^{45}~$erg s$^{-1}$. The cluster model is shown in Figure \ref{simorig2}.
The three-subcluster model consists of three beta models, at 4, 9 and 9 keV respectively, 
with core radii of $r_c=74~$kpc and $\beta = 0.7$. The clusters are separated 
by a triangle with sides of $300$, $200$ and $200~$kpc. The total luminosity of the 
three clusters is $L_{Bol} = 1.7 ~ 10^{45}~$erg s$^{-1}$. This model is shown in 
Figure \ref{simorig3}.

\section{Post-processing}

Here, we describe the raw output of the SPI analysis - temperature 
and luminosity maps of the clusters considered here - and discuss 
the estimation of uncertainties.  We also 
outline the methods that will be applied in the subsequent sections 
towards determination of the luminosity and temperature structure 
of clusters in our sample.  First, we consider a luminosity and temperature 
contrast analysis with the primary goal to identify clusters containing 
cool cores.  The luminosity 
contrast analysis is similar to that of \citet{vikh06b}.  
We also consider the ``Power Ratios'' method, suggested by \citet{buote95} 
to quantify the cluster substructuring. 
Finally, we propose a new method, the ``Temperature two-point correlation'' 
which is specifically designed to quantify the temperature structure, 
and is enabled by the SPI analysis.  

\subsection{Temperature and luminosity maps}
\label{mapsec}
We create $2 \times 2~$Mpc cluster luminosity and emission weighted temperature maps 
centered on the peak of cluster emission with $10~$kpc bins. 
The method of creating median parameter maps is described in 
\citet{andersson07}.  Maps are created for each sample in the 
chain from iteration 750 to 2000 and are averaged by taking 
the median in each spatial point over the whole sample.

\begin{figure*}[!htb]
  \begin{center}
    \begin{tabular}{c}
      \includegraphics[width=4in]{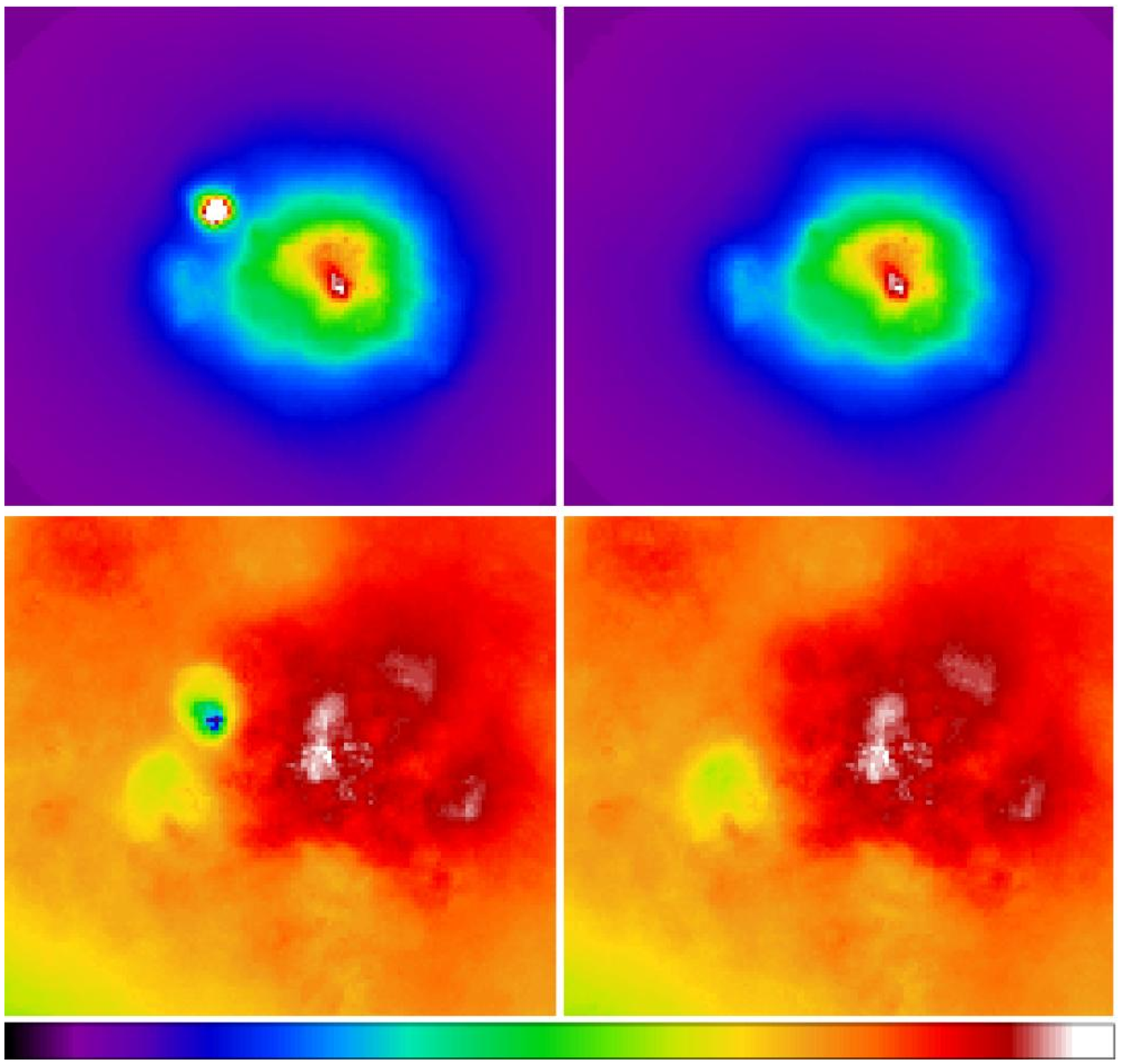} \\
    \end{tabular}
  \end{center}
  \caption{$10 \arcmin \times 10 \arcmin$ field showing luminosity (top) and temperature (bottom) maps of Abell 3888 before (left) and after (right) point source removal. The color scale in the luminosity map is set so that white corresponds to the maximum cluster flux. The point source is 100 times brighter than this level. The scale in the temperature map ranges from 2 to 10 keV.\label{psrmfig}}
\end{figure*}

Point sources are removed by filtering out all cluster particles 
within a radius of $-16.1 + \log_{10}(L_2) ~ 15~\arcsec$, from the point 
source as detected by \verb1emldetect1, where $L_2$ is the likelihood 
of detection.  We find that this method successfully removes any point source 
contamination without removing the cluster flux in the region of 
the point source. This method is effective because particles that represent 
the point source emission are generally smaller in size whereas the 
cluster particles are larger and fill the region where the point source 
was removed.  
An example of such point source removal is shown in Figure \ref{psrmfig} 
which shows maps of luminosity (top) and temperature (bottom) for Abell 3888 
before (left) and after (right) point source removal.

\begin{figure*}[!htb]
  \begin{center}
    \begin{tabular}{cccc}
      \includegraphics[width=1.6in]{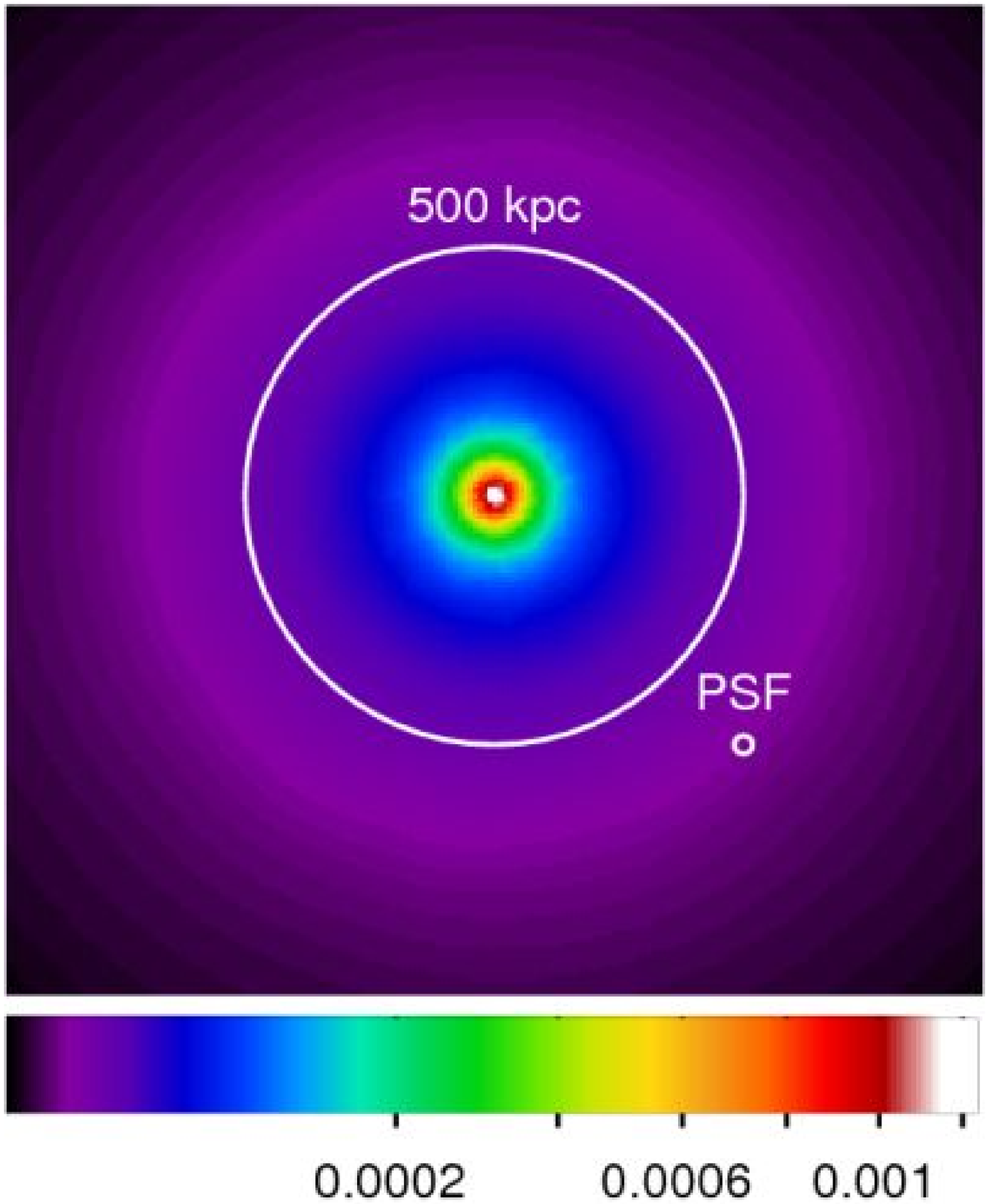} & 
      \includegraphics[width=1.6in]{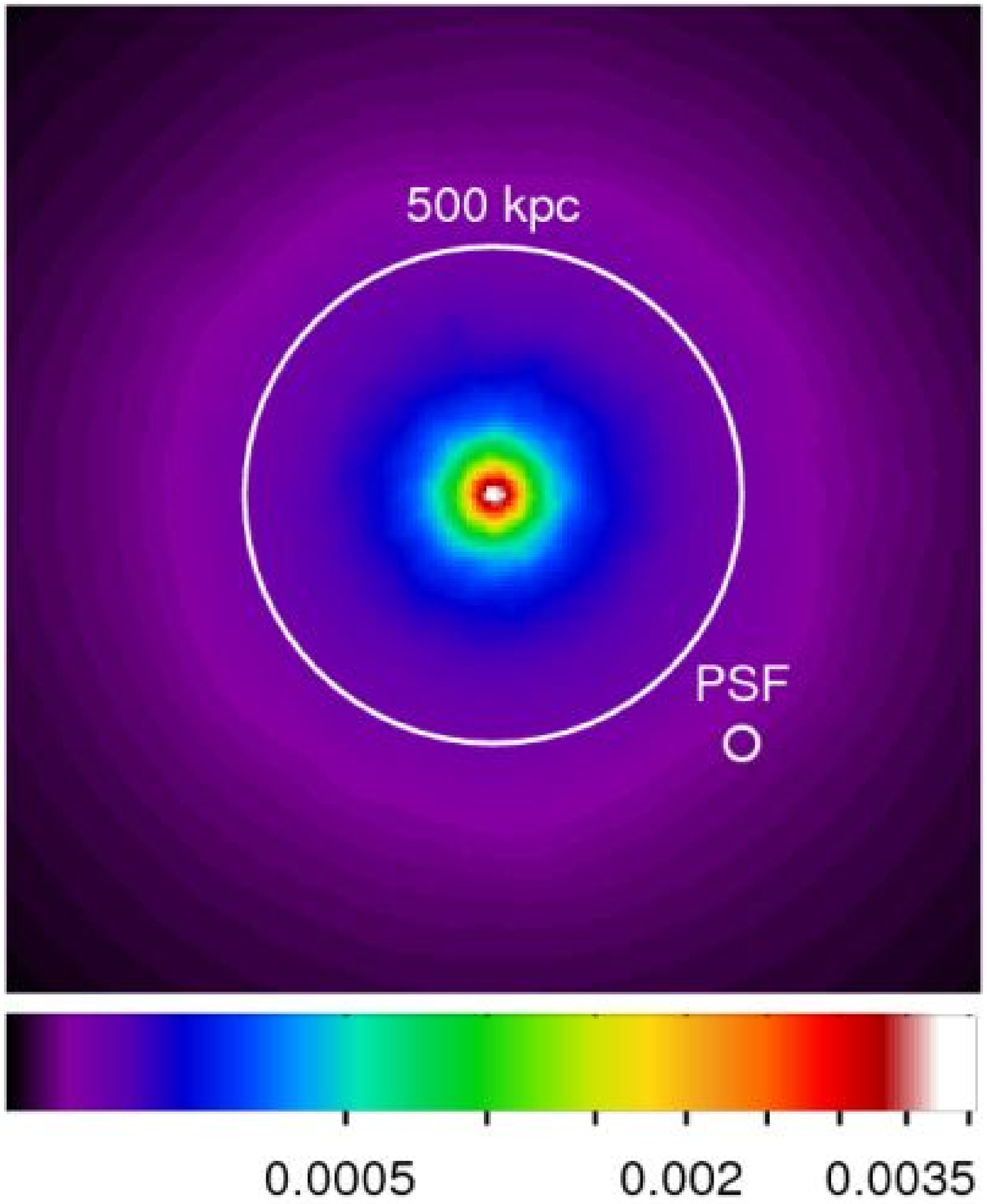} & 
      \includegraphics[width=1.6in]{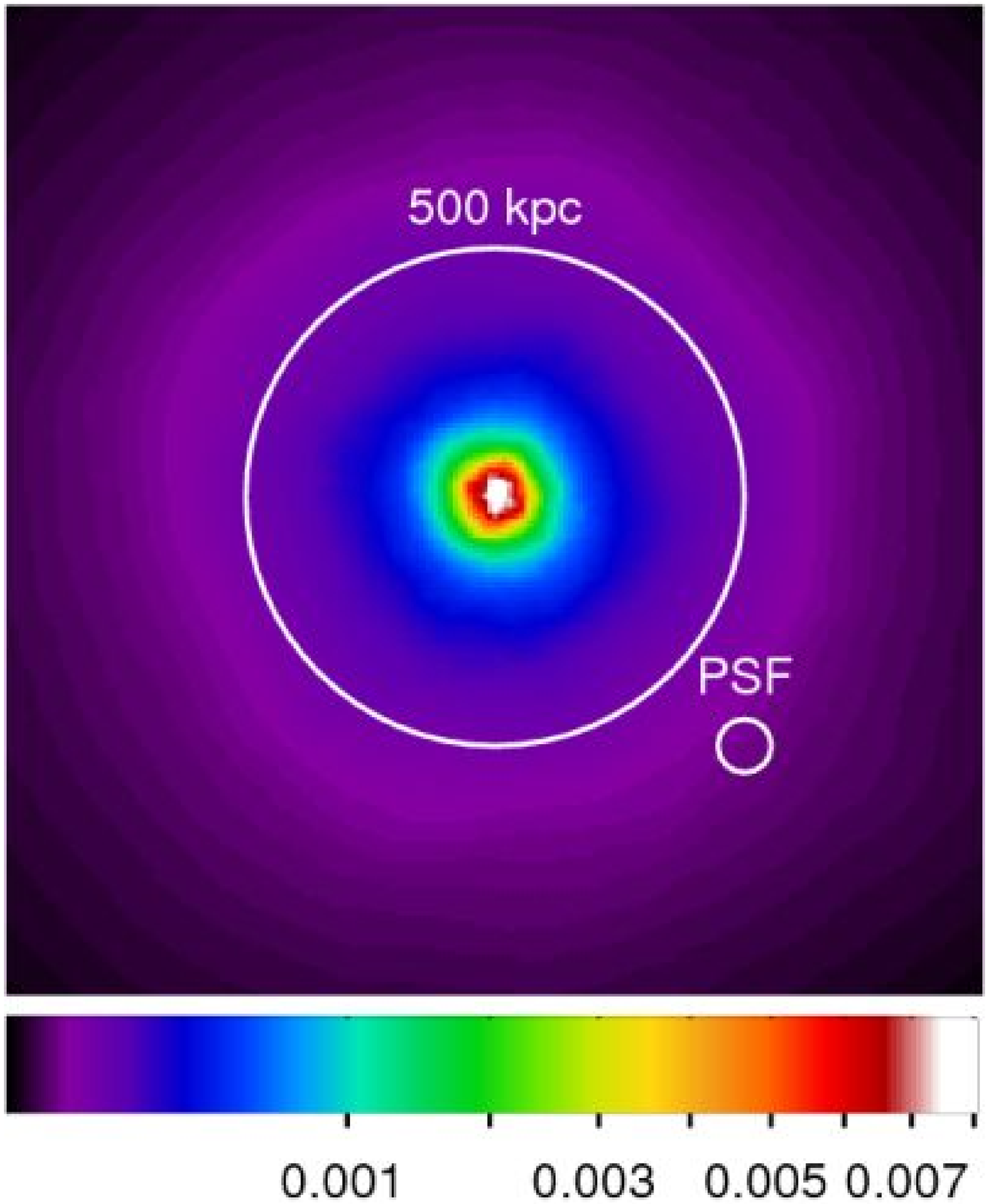} & 
      \includegraphics[width=1.6in]{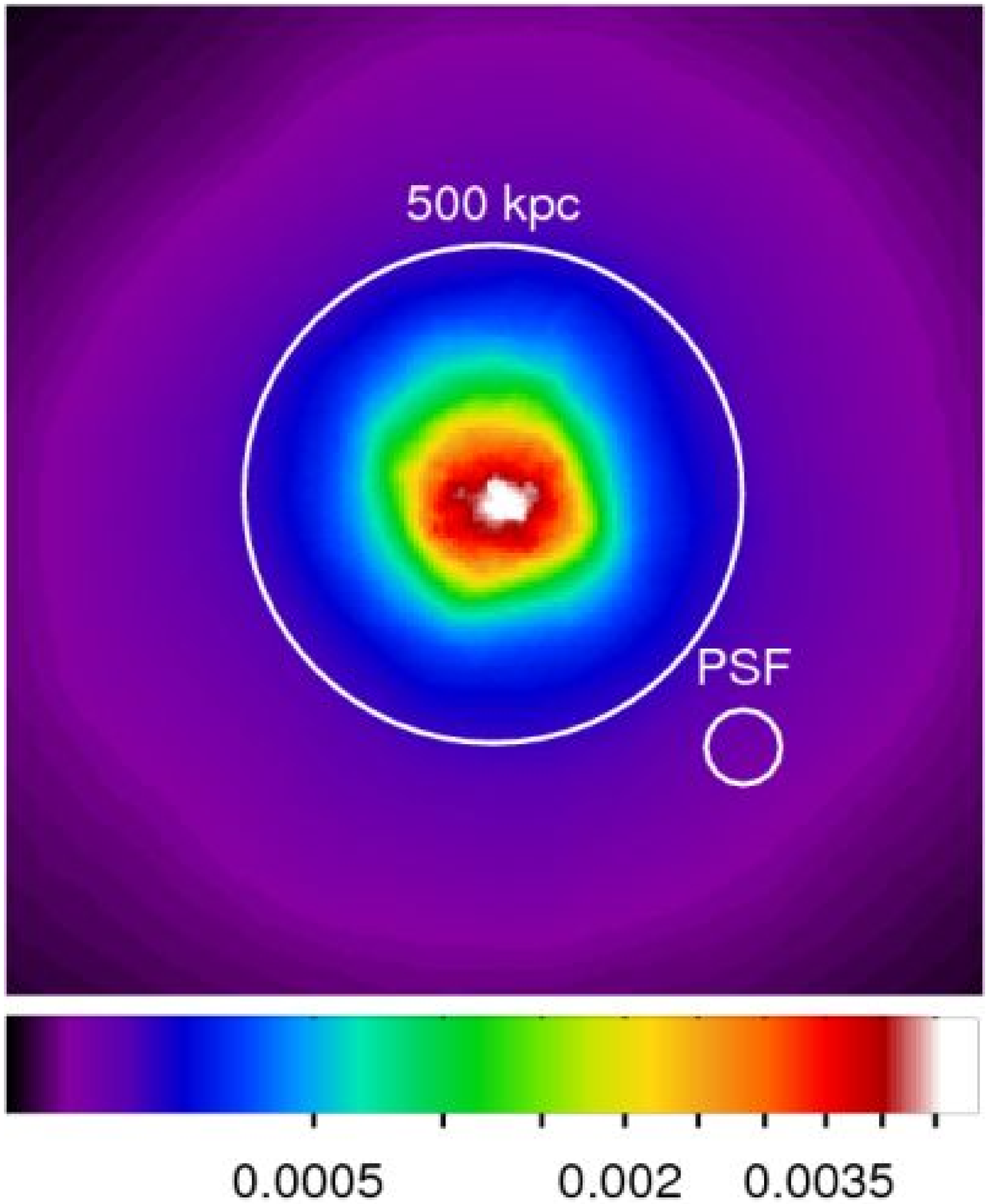} \\
      \includegraphics[width=1.6in]{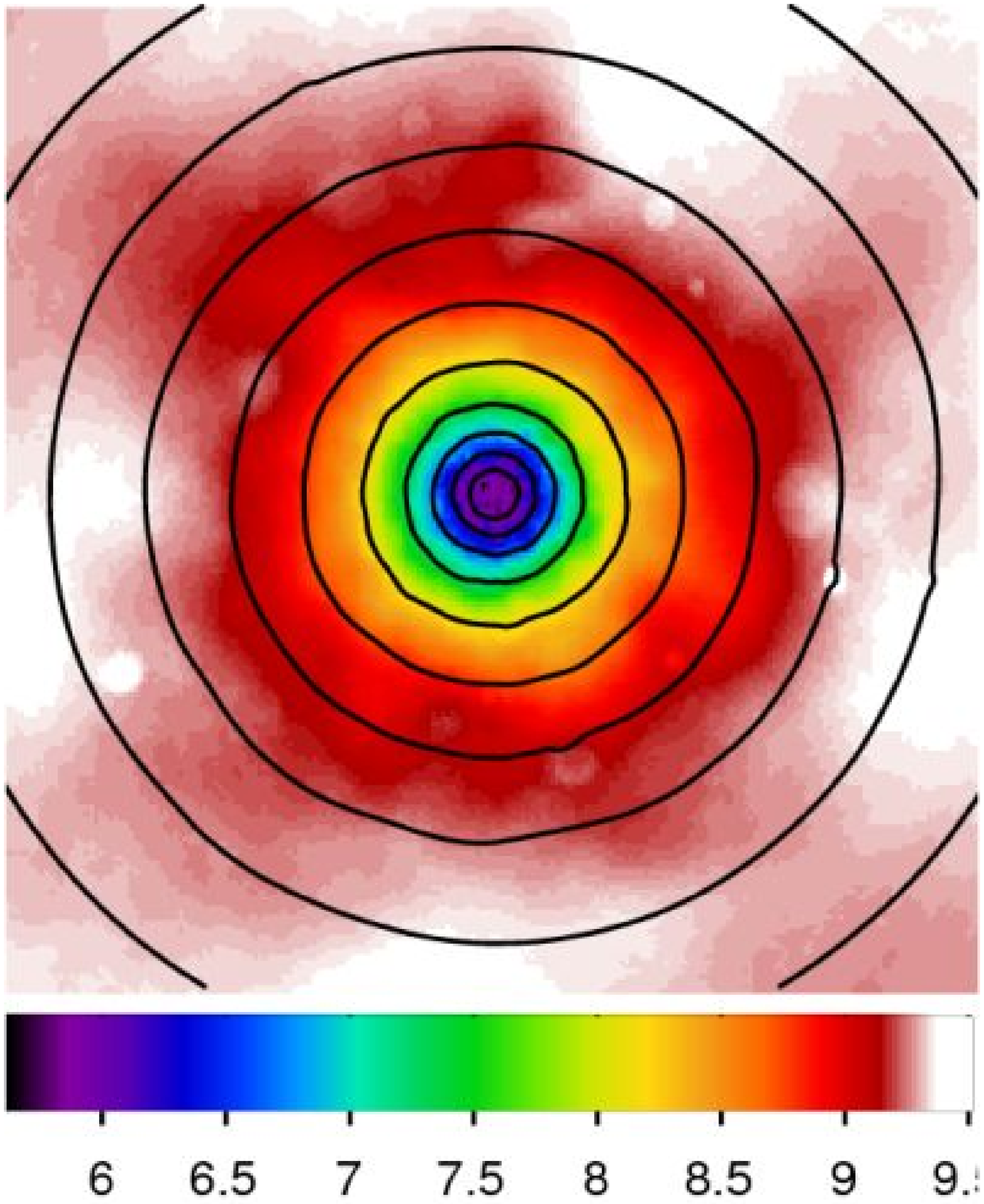} & 
      \includegraphics[width=1.6in]{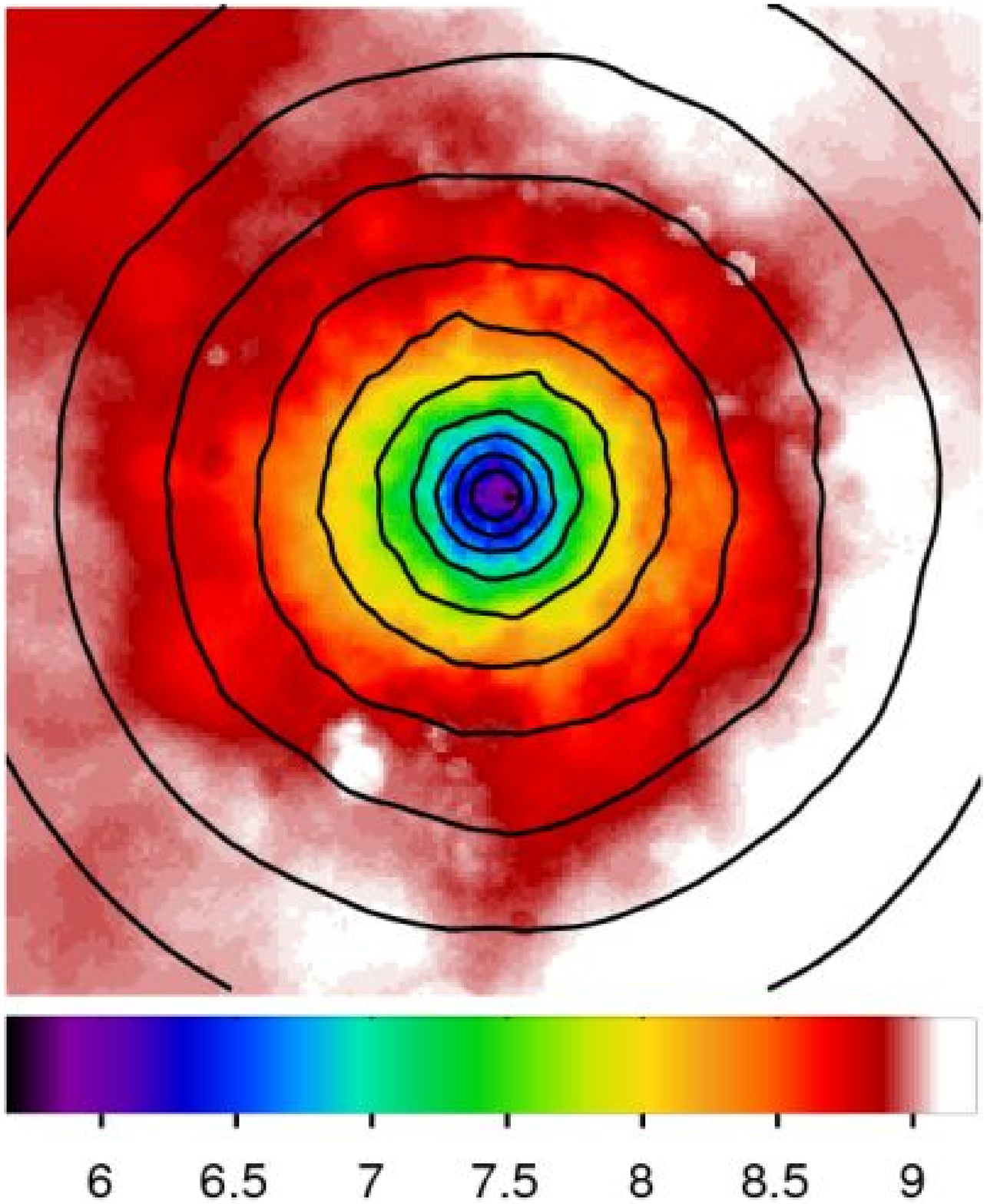} & 
      \includegraphics[width=1.6in]{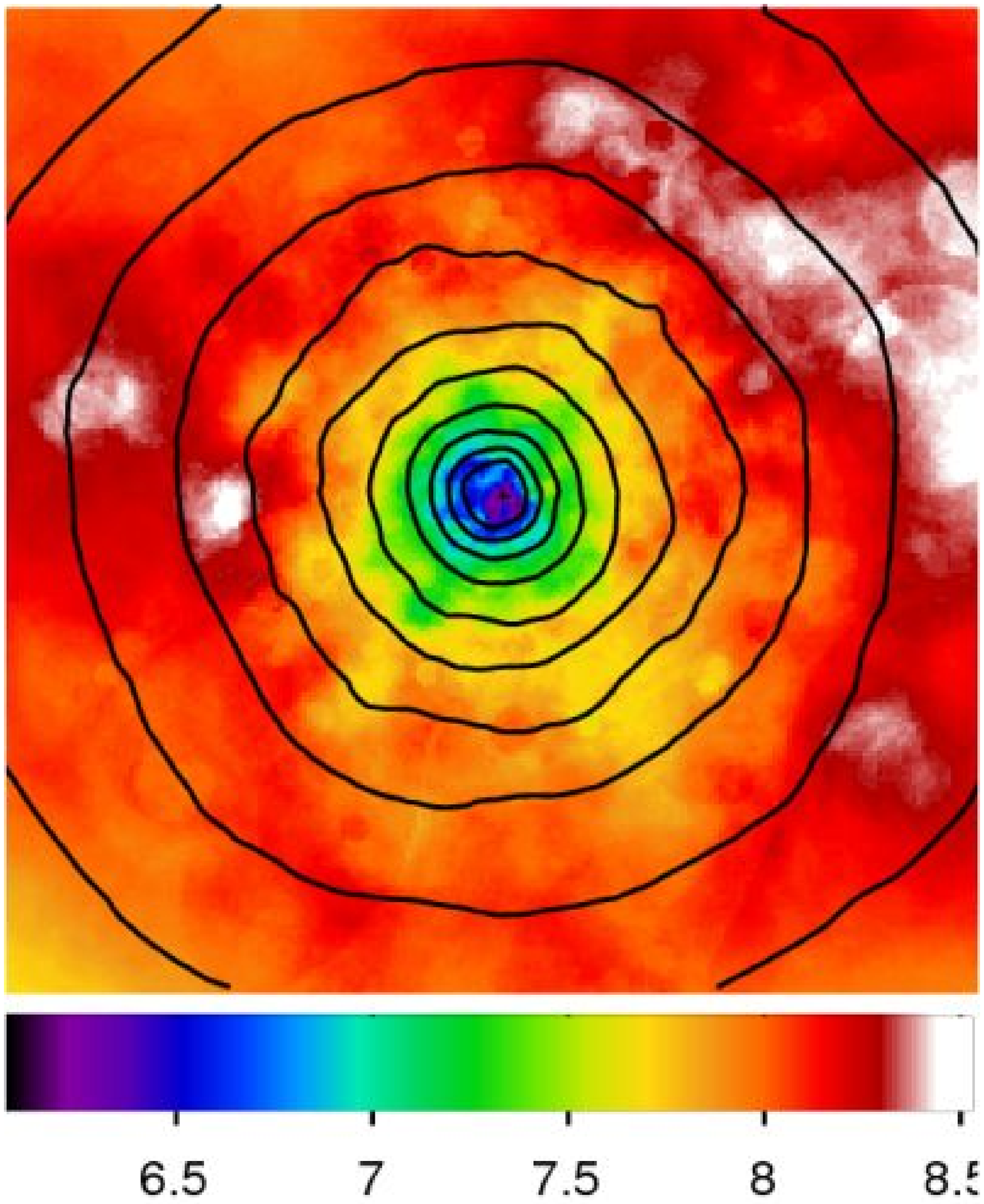} & 
      \includegraphics[width=1.6in]{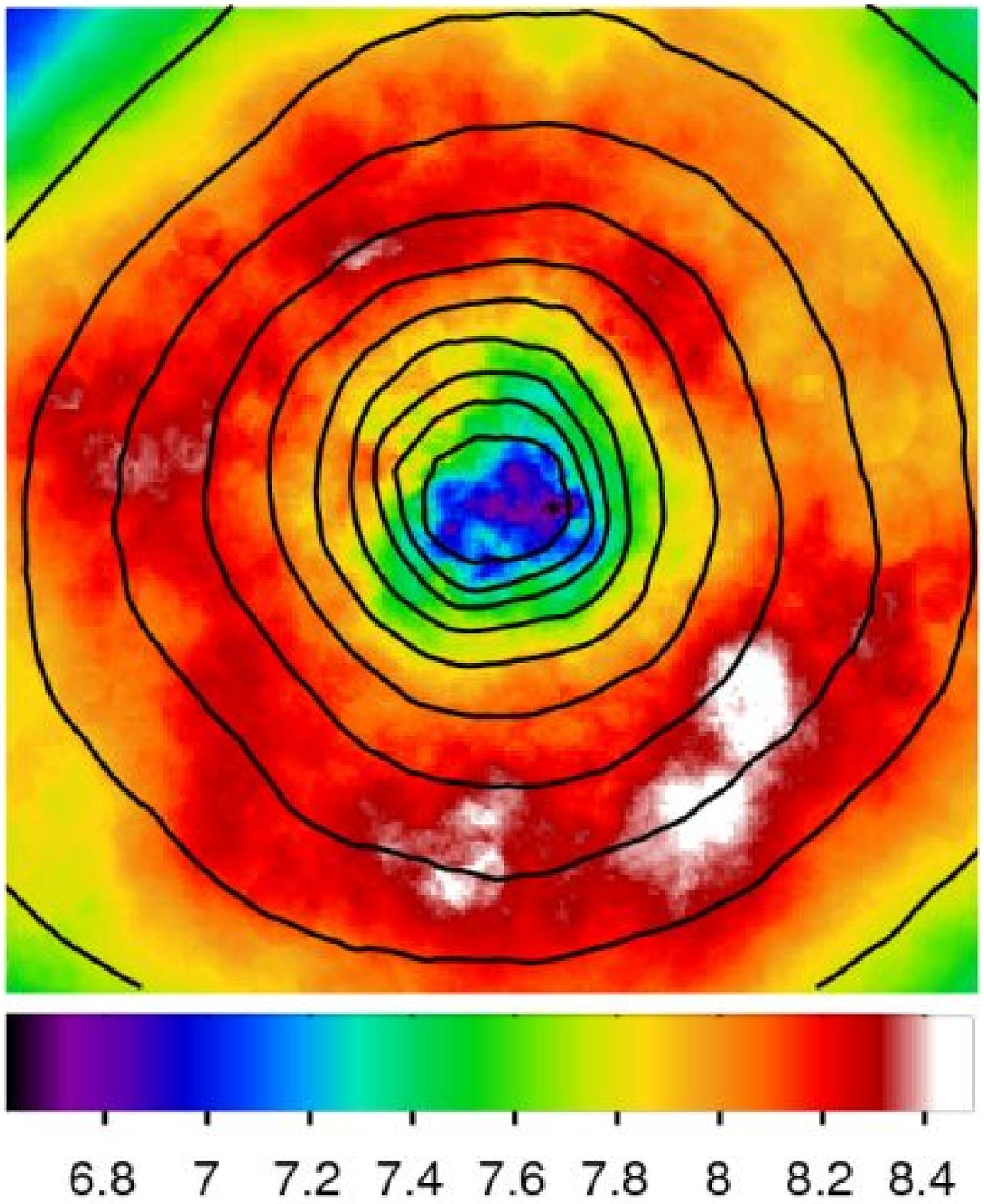} \\
    \end{tabular}
  \end{center}
  \caption{Luminosity (top) and temperature (bottom) maps for the reconstructed simulated cool core 
clusters, with background, at redshift 0.1, 0.2, 0.4 and 0.8. Luminosity per unit solid angle 
is in units of $10^{44}~$erg s$^{-1}$ $(\arcsec)^{-2}$ and temperatures in keV.
\label{figsim}}
\end{figure*}

The reconstruction of the four simulated cool core clusters within 
$2~\times~2~$Mpc is shown in Figure \ref{figsim} where 
luminosity per unit solid angle is given in units of 
$10^{44}~$erg s$^{-1}$ $(\arcsec)^{-2}$ and temperature in units of keV.
Here, the effects of the XMM point spread function (PSF) as 
well as the loss of photons with redshift is clearly seen as 
a distortion and flattening of the profile at high $z$. 

The maps for a subsample of clusters are shown in Figures \ref{A1068} 
through \ref{ZW3146} including the named clusters in Figure \ref{L-z}.
We comment briefly on these maps in Section \ref{clustercomments} and 
compare them to previously published results.
Note that the temperatures in Table \ref{spectab} do not necessarily 
agree with the scale shown in the figures. This is because these values 
are derived using two different methods. The tabulated values are 
determined using standard analysis (Section \ref{standardana}) requiring 
a single value of gas temperature for the whole cluster, whereas 
the parameter maps are created using a range of temperatures derived from 
the SPI runs (Section \ref{spi}). 
The source of the disagreement has to do with the fact that SPI 
bases the spectral modeling on superpositions of thermal spectra 
with many different temperatures.  The fact that the similarities between 
these spectra increase with higher temperature implies that the 
average temperature value of these also increases. It also becomes 
dependent on the prior range of temperatures (here: 0.5 - 15 keV). 
Whenever referring to the cluster average temperature, the tabulated 
values are used.  A bolometric correction is applied to the cluster 
particles individually, where the temperature, abundance and emission 
measure of the particle are used to calculate a bolometric luminosity 
in the 0.01 eV to 100 keV range using the MEKAL spectral model. 
This is the luminosity that is listed in Table \ref{spectab}.

\subsection{Estimation of uncertainties} \label{uncertain}

In all cases, except for the temperature and elemental abundance values 
used in Table \ref{spectab}, uncertainties are estimated by identifying 
5 independent sets of samples in the Markov chain posterior. The samples 
are taken sequentially in ranges containing 250 iterations each, starting 
at iteration 750. The luminosity and temperature maps and the quantities 
derived from these are then evaluated separately and the uncertainty on 
these quantities is determined from the rms scatter within these 5 sets.  
This is a valid approximation of statistical uncertainty 
because each subset can be seen as an independent reconstruction of the 
cluster. The motions of particles in the parameter space are large and 
within a few iterations 
there is no memory about earlier states. This is largely due to the high 
dimensionality of the overall parameter space and the lack of local likelihood 
maxima. Of course, it would be more accurate to use more than 5 sets for this 
calculation but at least 250 iterations are needed in order to describe 
a cluster model and there is a limitation in CPU time to produce more iterations. 

\subsection{Luminosity and temperature contrast methods and identification of 
cooling clusters \label{contrast}}
A common definition of a ``cooling flow'' cluster regards the central 
cooling time, $t_{cool}=T/\mathrm{d}(ln(T))$, being much less than a 
``Hubble time'', $t_{cool} << t_H$. The calculation of the cooling time 
requires high resolution spatially resolved spectroscopy which is not 
available for most of the clusters in our sample. Instead we use the 
information about the steepness of the luminosity profile as well as 
the gradient of temperature in the cluster core to empirically select 
the clusters with cool cores. 

The luminosity contrast is calculated as 

\begin{equation}
C_L = \frac{L\left(r \le 0.02 r_{500}\right)}{L\left(0.08 r_{500} \le r < 0.12 r_{500}\right)}
\end{equation}

\noindent where $L$ is luminosity per solid angle. We use the luminosity based 
definition of $r_{500}$;  

\begin{equation}
r_{500} = 909~E(z) ~ \left( \frac{L_{bol}}{10^{44} \mathrm{erg~s}^{-1}}\right)^{0.172} \mathrm{kpc}.
\end{equation}

\noindent where 

\begin{eqnarray}
E(z)=H(z)/H_0 = \nonumber \\ 
\sqrt{(1+z)^2 ~ (1+\Omega_M~ z) - z~(2+z)~\Omega_\Lambda}.
\end{eqnarray}

This value of $r_{500}$ is derived by combining the $M$-$T$ 
relation from \citet{arnaud05} and the $L$-$T$ relation from 
\citet{arnaud99}. The fractions of $r_{500}$ are used to 
account for the difference in size for clusters of various luminosities. 

Similarly we define the temperature contrast as 

\begin{equation}
C_T = \frac{T\left(r \le 0.02 r_{500}\right) }{T\left(0.45 r_{500} \le r < 0.55 r_{500}\right)}
\end{equation}

\noindent where we choose to estimate the gradient farther from the core ($0.5 r_{500}$) than 
for the luminosity contrast above because it is less sensitive to smearing by 
redshift dependent effects. 
We find that the radius of $0.5 r_{500}$ is where we could 
get the most distinguishing power. We discuss the use of 
$C_L$ and $C_T$ towards identifying cooling clusters in Section \ref{ccident}.  

\subsection{Power Ratios method \label{powsec}}

One of the successful approaches towards assessing the amount of 
substructure in clusters is the power ratios method \citep{buote95}.  
It was used in \citet{jeltema05} for a sample of Chandra-observed 
clusters to establish that clusters are more dynamically active at high $z$. 
This method is based on the multipole expansion of the surface brightness $\Sigma(R,\phi)$ 
around the cluster centroid where the multipole moments $a_m$ and $b_m$ are 

\begin{eqnarray}
a_m(R) = \int_{R' \le R} \Sigma(R,\phi) (R')^m \cos m \phi' \mathrm{d}^2 x' \\
b_m(R) = \int_{R' \le R} \Sigma(R,\phi) (R')^m \sin m \phi' \mathrm{d}^2 x'
\end{eqnarray}

\noindent so that the powers in the multipole $m$ can be written 

\begin{eqnarray}
P_0 = \left( a_0 \mathrm{ln} (R) \right)^2 \\ 
P_m = \frac{1}{2 m^2 R^{2m}} \left( a^2_m + b^2_m \right).
\end{eqnarray}

Here, we define the location of the cluster centroid by the requirement 
that $P_1$ should be zero at this location.  The above method is applied 
to the luminosity maps described in the previous section and the ratios 
$P_2/P_0$ and $P_3/P_0$ are calculated within a radius of 500 kpc.  We expect 
that $P_2/P_0$ will be larger for elongated clusters whereas $P_3/P_0$ will 
be large for clusters with much substructure in the luminosity map.

\subsection{Temperature two-point correlation}

We suggest that the variation of temperature inside the clusters can be 
estimated and characterized by taking a (non-standard) two point correlation of 
the temperature difference, weighted by products of luminosity 
so as to enhance temperature differences among the regions with the 
highest brightness.
This is accomplished by taking the following sum over all 
10 kpc $\times$ 10 kpc pixels in the 
generated maps:  
 
\begin{equation}
A(r_k)=\sqrt{ \sum^{i,j} \sqrt{L_i ~ L_j} ~ (T_i - T_j)^2 / L_T } \\
\end{equation}

where $L_i$ is the luminosity of pixel $i$, $T_i$ is the temperature and 
$L_T=\sum^{i,j} \sqrt{L_i ~ L_j}$.
$i$ and $j$ are such that both $(x_i,y_i)$ and $(x_j,y_j)$ cover 
all points within 500 kpc of the centroid calculated in Section \ref{powsec}.
$A$ is binned based on the distance between $i$ and $j$ as 
$10 k~$kpc$ < r_k \le 10 (k+1)~$kpc, $k=0\ldots99$, where  
$r_k=\sqrt{(x_i-x_j)^2 + (y_i-y_j)^2}$ is the pixel distance and  
$(x_i,y_i)$ is the spatial position of pixel $i$.

This statistic is designed to quantify the amount of distortion in the 
intra cluster medium, specifically regarding temperature features that 
can result from cluster mergers. Basically, it gives a ``power spectrum'' 
of strong temperature gradients over different scales, where 
the ``strength'' of the gradient is determined by $\sqrt{L_i ~ L_j}$. 
In integrating this quantity over small or large distances it should be 
possible to distinguish small scale features, such as cooling cores, from 
larger scale disturbances. We discuss the application of this method 
in Section \ref{tcorrref}.

\section{Results}

\subsection{Identification of cooling core clusters}
\label{ccident}

\begin{figure*}[!htb]
  \begin{center}
    \begin{tabular}{cc}
      \includegraphics[width=2.5in,angle=-90]{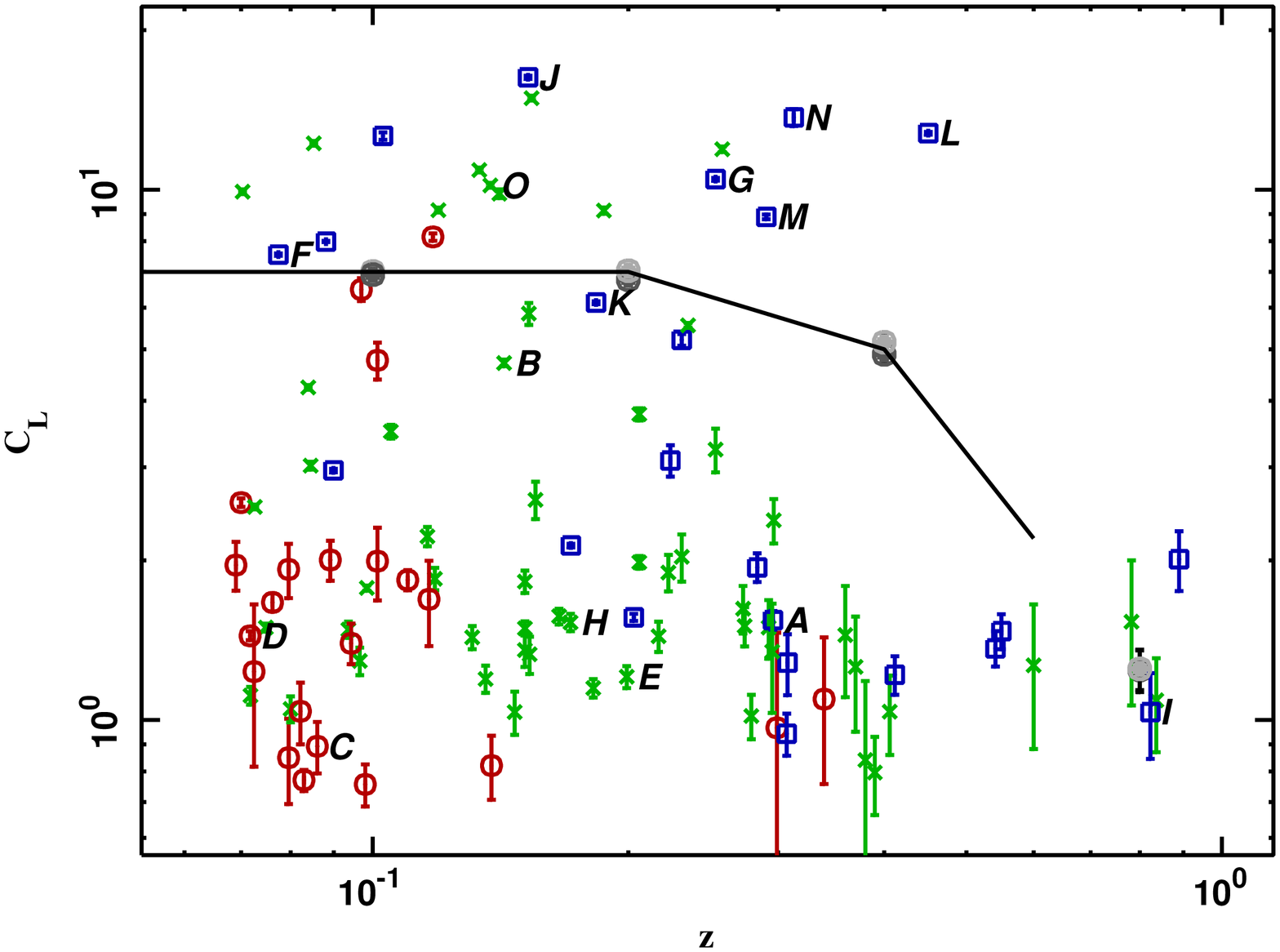} & 
      \includegraphics[width=2.5in,angle=-90]{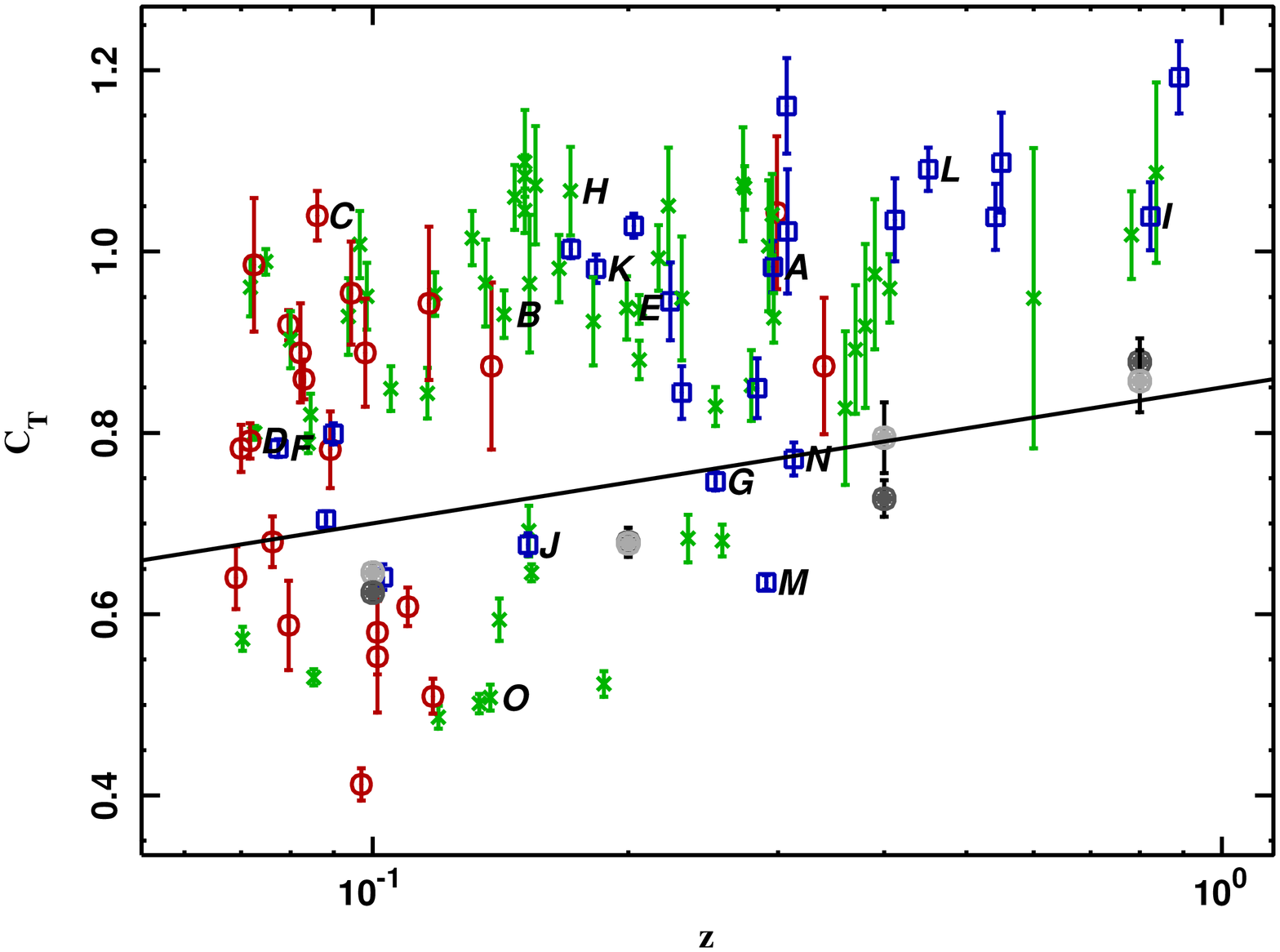} \\
    \end{tabular}
  \end{center}
  \caption{Luminosity contrast (left) and temperature contrast (right) as a function of redshift. 
The different symbols represent luminosity (red circles 
$\ge 10^{43}~$erg s$^{-1}$, green stars $\ge 4~10^{44}~$erg s$^{-1}$, and blue squares $\ge 2~10^{45}~$erg s$^{-1}$). 
The cuts on $C_L$ and $C_T$ are shown as solid lines. The cooling clusters are above and below the line respectively. The simulated cool core clusters without (with) background are shown as light (dark) gray filled circles, occasionally overlapping. A few selected clusters are denoted using letters from A to O as follows: 
(A) RXJ0658-55, (B) A1413, (C) A1750, (D) A1775, (E) A520, (F) A2029, (G) A1835, (H) A2218, (I) MS1054.4-0321, (J) A2204, (K) A1689, (L) RXJ1347-1145, (M) ZW3146, (N) MS2137-23, (O) A1068.
\label{Lc}}
\end{figure*}

The first step is to identify clusters with cooling cores. We use the approach 
described in Section \ref{contrast}, and plot the luminosity contrast vs 
redshift in Figure \ref{Lc} (left panel). Here, 
we also show the names for some well known clusters. 
Low $L_X$ 
clusters ($< 4~10^{44}~$erg s$^{-1}$) are shown as red circles, intermediate $L_X$  
clusters ($4~10^{44}~$erg s$^{-1} \le L_X < 2~10^{45}~$erg s$^{-1}$) as green stars and 
high $L_X$ clusters ($\ge 2~10^{45}~$erg s$^{-1}$) as 
blue squares. 
Likewise, the temperature contrast vs redshift is shown in Figure \ref{Lc} 
(right panel).

In Figure \ref{Lc} we also show the calculated values for 
our simulated cool core cluster at redshift $z=0.1$, $0.2$, $0.4$ and $0.8$. 
It is easily seen that the effects of PSF smoothing and loss of photons 
due to increased distance severely affects both quantities. 
We show both the simulated clusters with (dark gray) and without (light gray) 
simulated background, totaling 8 simulated clusters. 

$C_L$ is estimated at $0.1~r_{500}$ ($\sim 10''$ at $z=0.8$) and it 
becomes increasingly difficult to estimate at high redshifts.  
This can be seen in the sharp drop in $C_L$ in the trend of the 
simulated clusters above $z=0.4$. At $z=0.8$, the detection of cooling 
cores using $C_L$ is no longer sensitive and thus we decide to 
rely on the $C_T$ statistic alone for identification purposes above
 $z=0.6$. We identify those clusters with a $C_L$ above 
a line interpolated between the points set by the simulated clusters 
as cool core clusters. 
This cut corresponds to a linear interpolation of the points 
$C_L=(7,7,5,2.2)$ at $z=(0.1,0.2,0.4,0.6)$.
This comfortably separates known cool core clusters (e.g. A2029) 
from disturbed and intermediate core clusters (e.g. A1689). 

For $C_T$, the simulated clusters can be seen to have a trend that is more 
linear and not as dramatic as the trend in $C_L$. This is largely due to the 
larger radius ($0.5~r_{500}$) used when calculating $C_T$. 
Using $C_T$ to identify cool core clusters we make a cut corresponding 
to a straight line approximately following the trend of the simulated clusters 
so that clusters below the line, $C_T \le 0.85 + 0.15~\log{z}$, are selected. 
These cuts select well known cooling clusters like A1068, A1835 and A2204 but 
fail to select clusters that are known not to have pronounced cool cores such as 
A1689 and A1413.

\begin{figure*}[!htb]
  \begin{center}
    \begin{tabular}{c}
      \includegraphics[width=3in,angle=-90]{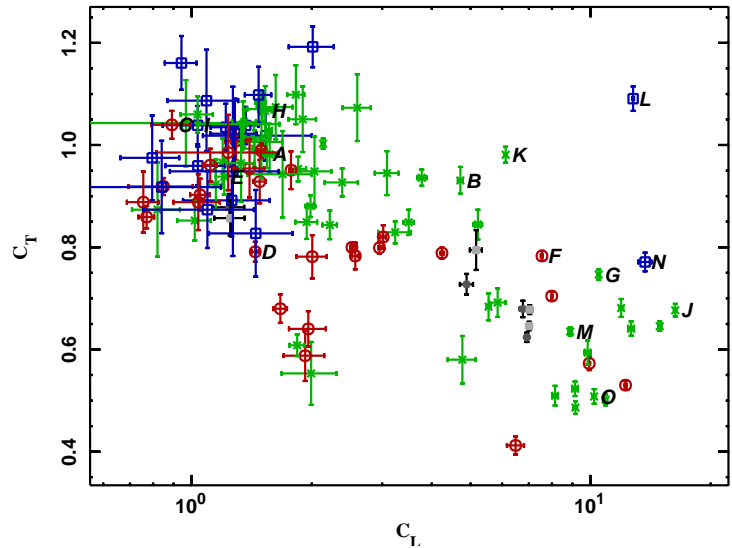} \\ 
    \end{tabular}
  \end{center}
  \caption{Temperature contrast plotted versus luminosity contrast. Here the different symbols represent redshift (red circles $0.069\le z<0.1$, green stars $0.1\le z<0.3$, and blue squares $z\ge 0.3$). Simulated cool core clusters are shown as filled gray circles. Letters from A to O denote a few selected clusters (see Figure \ref{Lc}).
\label{TcLc}}
\end{figure*}

In subsequent plots we show the values for the 
reconstructions of the simulated clusters as filled circles and 
use this to measure how accurately a weak cooling core cluster at 
different redshifts can be resolved. 
The cuts are shown in Figure \ref{Lc}.
In Figure \ref{TcLc} we also show $C_T$ plotted against $C_L$, clearly showing 
the separation of cooling clusters like A1835 and A2204 (G, J - shown bottom right), nearly isothermal 
clusters such as A1413 and A1689 (B, K - shown center) and disturbed clusters like A2218 and A520 (H, E - shown top left).

Out of 101 clusters, 31 are identified as cooling clusters.

\subsection{Dynamical activity}

\subsubsection{Power ratios}
\label{pratsec}
\begin{deluxetable*}{lll}
\tabletypesize{\scriptsize}
\tablecaption{Power ratios \label{prattab}}
\tablewidth{0pt}
\tablehead{
\colhead{Name} &
\colhead{$P_2/P_0 (\times 10^{-7})$} &
\colhead{$P_3/P_0 (\times 10^{-7})$} \\
}
\tablecomments{Results of the Power Ratio analysis (Section \ref{powsec}) of the clusters.}
\startdata

CIZAJ1645.4-7334 & $ 4.72 \pm 0.95 $ & $ 0.41 \pm 0.49 $ \\ 
A1837 & $ 112 \pm 7 $ & $ 2.37 \pm 0.31 $ \\ 
A3112 & $ 39.5 \pm 1.6 $ & $ 0.085 \pm 0.031 $ \\ 
A1775 & $ 0.5 \pm 0.33 $ & $ 3.86 \pm 0.65 $ \\ 
A399 & $ 62.0 \pm 3.2 $ & $ 3.55 \pm 0.31 $ \\ 
A1589 & $ 298 \pm 17 $ & $ 0.88 \pm 0.51 $ \\ 
A2065 & $ 208 \pm 6 $ & $ 0.089 \pm 0.032 $ \\ 
A401 & $ 104 \pm 3 $ & $ 0.92 \pm 0.07 $ \\ 
A2670 & $ 4.96 \pm 0.91 $ & $ 4.07 \pm 0.43 $ \\ 
A2029 & $ 55.9 \pm 2.1 $ & $ 0.051 \pm 0.022 $ \\ 
RXCJ1236.7-3354 & $ 21.6 \pm 8.6 $ & $ 8.97 \pm 2.72 $ \\ 
RXCJ2129.8-5048 & $ 78.0 \pm 7.8 $ & $ 27.1 \pm 1.8 $ \\ 
A2255 & $ 64.5 \pm 5.0 $ & $ 0.58 \pm 0.42 $ \\ 
RXCJ0821.8+0112 & $ 106 \pm 16 $ & $ 0.87 \pm 0.58 $ \\ 
RXCJ1302.8-0230 & $ 45.0 \pm 11.7 $ & $ 0.59 \pm 0.63 $ \\ 
A1650 & $ 87.9 \pm 2.3 $ & $ 0.24 \pm 0.06 $ \\ 
A1651 & $ 35.6 \pm 3.1 $ & $ 0.12 \pm 0.08 $ \\ 
A2597 & $ 26.5 \pm 0.6 $ & $ 0.058 \pm 0.026 $ \\ 
A1750 & $ 39.9 \pm 5.7 $ & $ 1.84 \pm 0.73 $ \\ 
A478 & $ 46.4 \pm 0.5 $ & $ 0.054 \pm 0.004 $ \\ 
A278 & $ 16.2 \pm 2.1 $ & $ 1.23 \pm 0.57 $ \\ 
A2142 & $ 247 \pm 9 $ & $ 0.15 \pm 0.05 $ \\ 
A3921 & $ 228 \pm 10 $ & $ 2.2 \pm 0.48 $ \\ 
A13 & $ 88.2 \pm 10.1 $ & $ 0.19 \pm 0.12 $ \\ 
A3911 & $ 344 \pm 11 $ & $ 2.45 \pm 0.51 $ \\ 
RXCJ2319.6-7313 & $ 128 \pm 18 $ & $ 0.22 \pm 0.09 $ \\ 
CL0852+1618 & $ 12.5 \pm 9.6 $ & $ 22.9 \pm 6.5 $ \\ 
A3827 & $ 17.0 \pm 1.7 $ & $ 0.96 \pm 0.15 $ \\ 
RXCJ0211.4-4017 & $ 18.3 \pm 1.2 $ & $ 1.48 \pm 0.55 $ \\ 
A2241 & $ 7.01 \pm 1.39 $ & $ 0.48 \pm 0.36 $ \\ 
PKS0745-19 & $ 36.0 \pm 1.7 $ & $ 0.018 \pm 0.01 $ \\ 
RXCJ0645.4-5413 & $ 126 \pm 6 $ & $ 2.11 \pm 0.67 $ \\ 
RXCJ0049.4-2931 & $ 13.3 \pm 2.3 $ & $ 0.19 \pm 0.14 $ \\ 
A1302 & $ 28.9 \pm 3.0 $ & $ 0.26 \pm 0.23 $ \\ 
RXCJ0616.8-4748 & $ 112 \pm 18 $ & $ 3.0 \pm 1.7 $ \\ 
RXCJ2149.1-3041 & $ 29.3 \pm 3.7 $ & $ 0.028 \pm 0.015 $ \\ 
RXCJ1516.3+0005 & $ 82.9 \pm 4.0 $ & $ 0.63 \pm 0.41 $ \\ 
RXCJ1141.4-1216 & $ 19.4 \pm 1.5 $ & $ 0.12 \pm 0.11 $ \\ 
RXCJ0020.7-2542 & $ 118 \pm 6 $ & $ 1.36 \pm 0.31 $ \\ 
RXCJ1044.5-0704 & $ 47.0 \pm 3.0 $ & $ 0.77 \pm 0.09 $ \\ 
RXCJ0145.0-5300 & $ 342 \pm 29 $ & $ 2.63 \pm 0.93 $ \\ 
A1068 & $ 63.0 \pm 4.4 $ & $ 1.21 \pm 0.32 $ \\ 
RXJ1416.4+2315 & $ 180 \pm 20 $ & $ 3.37 \pm 3.16 $ \\ 
RXCJ0605.8-3518 & $ 47.0 \pm 3.0 $ & $ 0.52 \pm 0.17 $ \\ 
A1413 & $ 192 \pm 3 $ & $ 0.19 \pm 0.1 $ \\ 
RXCJ2048.1-1750 & $ 34.4 \pm 4.4 $ & $ 3.92 \pm 0.34 $ \\ 
A3888 & $ 117 \pm 7 $ & $ 6.33 \pm 0.8 $ \\ 
RXCJ2234.5-3744 & $ 119 \pm 4 $ & $ 4.06 \pm 0.49 $ \\ 
A2034 & $ 84.5 \pm 8.4 $ & $ 0.41 \pm 0.21 $ \\ 
A2204 & $ 3.81 \pm 0.33 $ & $ 0.077 \pm 0.045 $ \\ 
RXCJ0958.3-1103 & $ 79.0 \pm 9.1 $ & $ 0.14 \pm 0.12 $ \\ 
\enddata
\end{deluxetable*}
%
\begin{deluxetable*}{lll}
\tabletypesize{\scriptsize}
\tablecaption{Power ratios \label{prattab2}}
\tablewidth{0pt}
\tablehead{
\colhead{Name} &
\colhead{$P_2/P_0 (\times 10^{-7})$} &
\colhead{$P_3/P_0 (\times 10^{-7})$} \\
}
\tablecomments{Results of the Power Ratio analysis (Section \ref{powsec}) of the clusters.}
\startdata
A868 & $ 105 \pm 6 $ & $ 0.92 \pm 0.98 $ \\ 
RXCJ2014.8-2430 & $ 15.8 \pm 1.2 $ & $ 0.37 \pm 0.13 $ \\ 
A2104 & $ 68.0 \pm 16.7 $ & $ 2.83 \pm 2.54 $ \\ 
RXCJ0547.6-3152 & $ 29.4 \pm 3.1 $ & $ 0.33 \pm 0.14 $ \\ 
A2218 & $ 72.8 \pm 5.6 $ & $ 1.14 \pm 0.36 $ \\ 
A1914 & $ 32.4 \pm 1.0 $ & $ 2.45 \pm 0.26 $ \\ 
A665 & $ 44.8 \pm 3.5 $ & $ 7.75 \pm 0.77 $ \\ 
A1689 & $ 27.6 \pm 2.0 $ & $ 0.64 \pm 0.12 $ \\ 
A383 & $ 1.63 \pm 0.26 $ & $ 0.62 \pm 0.17 $ \\ 
A520 & $ 65.3 \pm 11.2 $ & $ 3.87 \pm 0.46 $ \\ 
A2163 & $ 45.6 \pm 5.9 $ & $ 10.7 \pm 1.7 $ \\ 
A209 & $ 98.8 \pm 5.5 $ & $ 1.18 \pm 0.83 $ \\ 
A963 & $ 8.14 \pm 1.32 $ & $ 1.34 \pm 0.67 $ \\ 
A773 & $ 83.2 \pm 11.2 $ & $ 0.82 \pm 0.39 $ \\ 
A1763 & $ 218 \pm 8 $ & $ 1.36 \pm 1.3 $ \\ 
A2261 & $ 12.9 \pm 11.4 $ & $ 2.79 \pm 1.7 $ \\ 
A267 & $ 95.8 \pm 9.2 $ & $ 0.92 \pm 0.64 $ \\ 
A2390 & $ 149 \pm 7 $ & $ 2.97 \pm 0.86 $ \\ 
RXJ2129.6+0005 & $ 66.1 \pm 1.4 $ & $ 0.21 \pm 0.03 $ \\ 
A1835 & $ 10.2 \pm 0.4 $ & $ 0.36 \pm 0.09 $ \\ 
RXCJ0307.0-2840 & $ 14.6 \pm 3.8 $ & $ 2.23 \pm 0.51 $ \\ 
E1455+2232 & $ 16.1 \pm 0.9 $ & $ 0.12 \pm 0.08 $ \\ 
RXCJ2337.6+0016 & $ 282 \pm 29 $ & $ 1.16 \pm 0.42 $ \\ 
RXCJ0303.8-7752 & $ 37.4 \pm 13.7 $ & $ 2.01 \pm 0.87 $ \\ 
A1758 & $ 502 \pm 13 $ & $ 0.78 \pm 0.32 $ \\ 
RXCJ0232.2-4420 & $ 50.6 \pm 4.8 $ & $ 2.5 \pm 0.62 $ \\ 
ZW3146 & $ 11.6 \pm 1.3 $ & $ 0.37 \pm 0.08 $ \\ 
RXCJ0043.4-2037 & $ 58.5 \pm 12.9 $ & $ 2.31 \pm 2.39 $ \\ 
RXCJ0516.7-5430 & $ 296 \pm 54 $ & $ 8.28 \pm 5.55 $ \\ 
RXJ0658-55 & $ 116 \pm 3 $ & $ 8.73 \pm 0.95 $ \\ 
RXCJ2308.3-0211 & $ 23.3 \pm 4.2 $ & $ 1.44 \pm 0.74 $ \\ 
RXJ2237.0-1516 & $ 198 \pm 41 $ & $ 8.29 \pm 7.19 $ \\ 
RXCJ1131.9-1955 & $ 142 \pm 18 $ & $ 2.03 \pm 1.29 $ \\ 
RXCJ0014.3-3022 & $ 5.54 \pm 4.01 $ & $ 7.16 \pm 1.1 $ \\ 
MS2137-23 & $ 1.47 \pm 0.96 $ & $ 0.37 \pm 0.49 $ \\ 
MS1208.7+3928 & $ 322 \pm 293 $ & $ 23.4 \pm 26.9 $ \\ 
RXJ0256.5+0006 & $ 6.12 \pm 2.42 $ & $ 64.0 \pm 15.3 $ \\ 
RXJ0318.2-0301 & $ 39.7 \pm 20.2 $ & $ 4.11 \pm 3.27 $ \\ 
RXJ0426.1+1655 & $ 35.9 \pm 37.5 $ & $ 4.66 \pm 5.68 $ \\ 
RXJ1241.5+3250 & $ 27.4 \pm 17.9 $ & $ 12.6 \pm 9.9 $ \\ 
A851 & $ 304 \pm 30 $ & $ 6.49 \pm 2.24 $ \\ 
RXCJ2228+2037 & $ 144 \pm 27 $ & $ 8.23 \pm 3.18 $ \\ 
RXJ1347-1145 & $ 32.6 \pm 1.0 $ & $ 0.87 \pm 0.24 $ \\ 
CL0016+16 & $ 114 \pm 17 $ & $ 5.54 \pm 2.87 $ \\ 
MS0451.6-0305 & $ 82.3 \pm 5.1 $ & $ 7.01 \pm 2.19 $ \\ 
RXJ1120.1+4318 & $ 137 \pm 49 $ & $ 9.82 \pm 7.75 $ \\ 
MS1137.5+6625 & $ 11.5 \pm 11.9 $ & $ 10.5 \pm 7.8 $ \\ 
MS1054.4-0321 & $ 220 \pm 73 $ & $ 17.3 \pm 14.6 $ \\ 
WARPJ0152.7-1357 & $ 2880 \pm 260 $ & $ 27.6 \pm 19.4 $ \\ 
CLJ1226.9+3332 & $ 11.1 \pm 3.2 $ & $ 0.87 \pm 0.4 $ \\ 
\enddata

\end{deluxetable*}

In assessing the dynamical activity in the clusters in our sample we perform 
a multipole expansion of our luminosity maps as described in Section \ref{powsec}. 
The power ratios $P_2/P_0$ and $P_3/P_0$ are calculated to a radius of 500 kpc 
in order to quantify the amount of substructure 
in the sample. These are listed in Table \ref{prattab}.

Here we divide the sample in 3 redshift bins; a local $z$ bin ($z < 0.1$), a low-$z$ 
bin ($0.1 \le z < 0.3$) and a high-$z$ bin ($z > 0.3$).
The results of the multipole expansion are shown in Table \ref{prtab} where we 
show the average values of $P_2/P_0$ and $P_3/P_0$ for the 3 samples. We also show 
the result obtained when using only the cooling core clusters identified in previous 
sections. These are not excluded from the other samples. 

\begin{figure*}[!htb]
  \begin{center}
    \begin{tabular}{c}
      \includegraphics[width=3in,angle=-90]{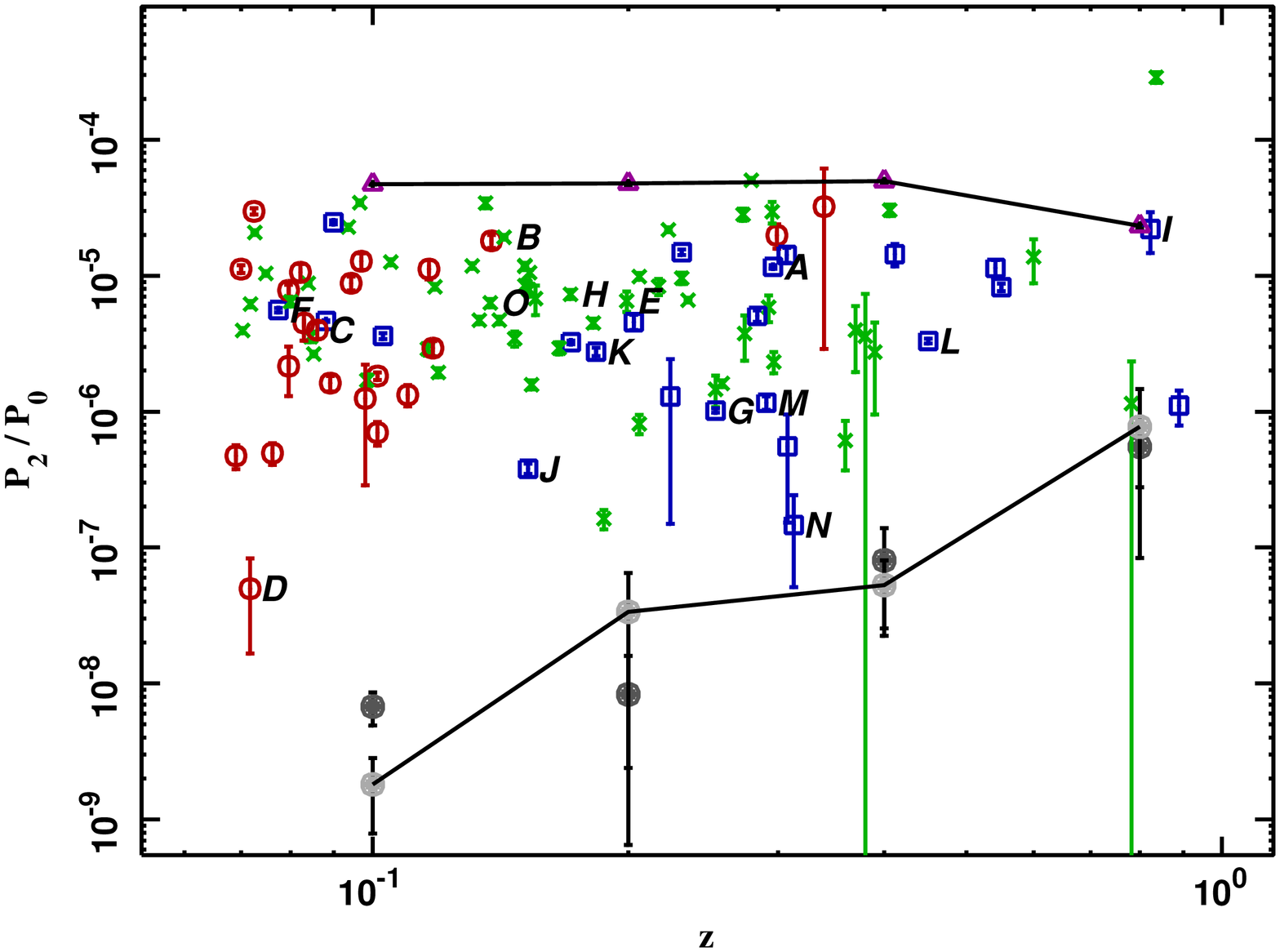} \\
    \end{tabular}
  \end{center}
  \caption{Power ratios $P_2/P_0$ vs redshift for all clusters. Low luminosity clusters are shown as red circles, intermediate luminosity as green stars and high luminosity as blue squares, cf. Figure \ref{Lc}. Simulated cool core clusters are shown as filled gray circles connected by black lines. The simulated two-subcluster model is shown as purple triangles connected by black lines. Letters from A to O denote a few selected clusters (see Figure \ref{Lc}).
\label{Prat2}}
\end{figure*}

\begin{figure*}[!htb]
  \begin{center}
    \begin{tabular}{c}
      \includegraphics[width=3in,angle=-90]{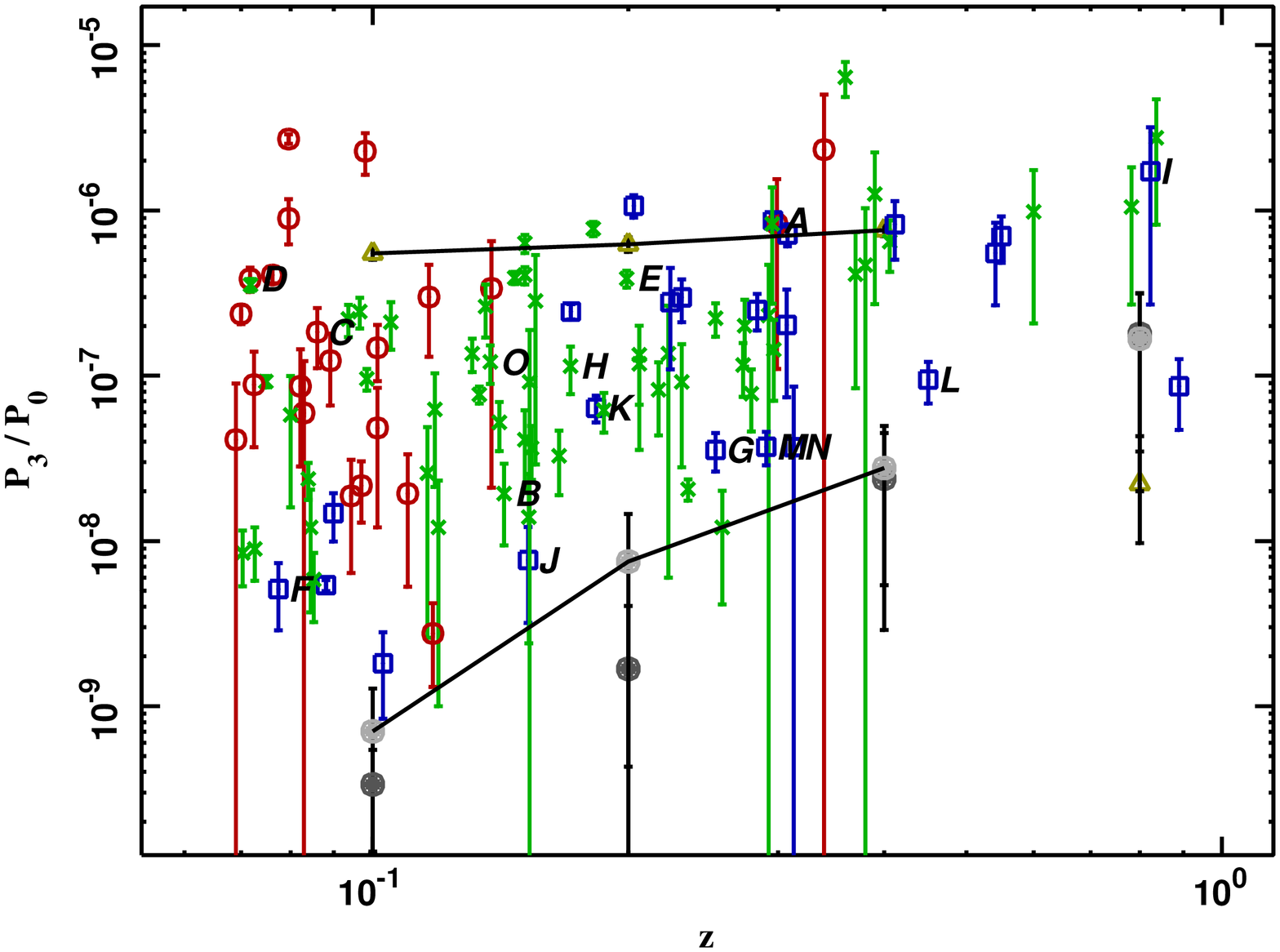} \\
    \end{tabular}
  \end{center}
  \caption{Power ratios $P_3/P_0$ vs redshift for all clusters. Low luminosity clusters are shown as red circles, intermediate luminosity as green stars and high luminosity as blue squares, cf. Figure \ref{Lc}. Simulated cool core clusters are shown as filled gray circles connected by black lines. The simulated two-subcluster model is shown as yellow triangles connected by black lines. Letters from A to O denote a few selected clusters (see Figure \ref{Lc}).
\label{Prat3}}
\end{figure*}

The values of $P_2/P_0$ and $P_3/P_0$ are plotted against $z$ in Figures \ref{Prat2} and 
\ref{Prat3} respectively where the simulated cool core clusters are shown 
as filled circles connected by black lines. 
In Figure \ref{Prat2} we also show the simulated two-component clusters as purple triangles 
connected by black lines and in Figure \ref{Prat3} we show the simulated three-component 
clusters as yellow triangles connected by black lines. 

The simulated cool core clusters represent the minimum amount of substructure that 
can be measured as they are simulated with perfect circular symmetry. This is true 
for clusters of similar luminosity and exposure. However, the $P_3/P_0$ values 
in Table \ref{prtab} are higher than those of the simulated cool core clusters and 
above $z = 0.3$ most clusters have $P_3/P_0$ values similar to those of the simulated 
three-component clusters. 
This is most clearly seen around $z = 0.4$ where the the datapoints cluster around 
the value of the three-component simulation.
Towards $z = 0.8$ the simulated irregular cluster shows a decreasing value with redshift
due to the smoothing effect introduced by PSF blurring and lack of photons. 
The datapoints around $z=0.6$ and $z=0.8$ show large $P_3/P_0$ values around $10^{-6}$, 
larger than for the simulated clusters.  

\begin{figure*}[!htb]
  \begin{center}
    \begin{tabular}{c}
      \includegraphics[width=3in,angle=-90]{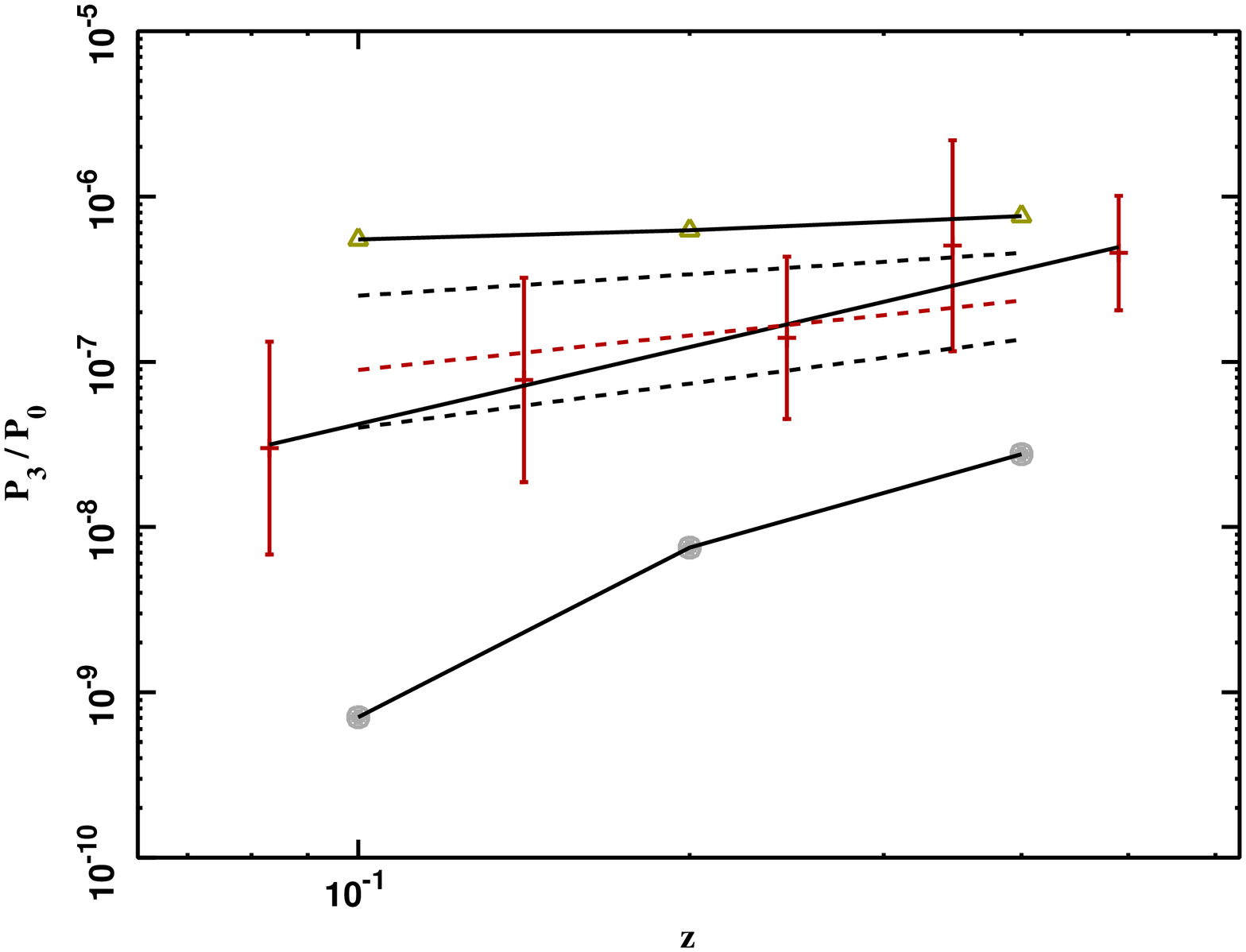} \\
    \end{tabular}
  \end{center}
  \caption{$P_3/P_0$ values for $L_{bol} \ge 4~10^{44}~$erg s$^{-1}$ clusters up to $z=0.6$ averaged in 
5 redshift bins. Best fit line (solid line) is shown along with interpolated trends from 
the simulated clusters (dashed lines, red dashed line has lowest $\chi^2$). 
Simulated cool core clusters are shown as filled gray circles and 
simulated irregular (3-core) clusters are shown as yellow stars.
\label{P3fit}}
\end{figure*}

To investigate further the proposed evolution of $P_3/P_0$ we restrict the analysis 
to clusters below $z=0.6$ due to the apparent breakdown in the analysis as seen 
in the simulated clusters at $z=0.8$. We also exclude the low-luminosity clusters 
($L_{bol} < 4~10^{44}~$erg s$^{-1}$) from the sample since these are not present 
at all $z$. We bin the $P_3/P_0$ values for the remaining clusters in 5 bins of 
redshift ($z=0$,$0.1$,$0.2$,$0.3$,$0.4$,$0.6$) and fit these datapoints to 
a straight line in $\log{z}$-$\log{P_3/P_0}$ space. The binned data is shown 
along with the best fit line in Figure \ref{P3fit}. In order to distinguish 
between real and apparent evolution we interpolate the slope linearly for 
values of $\log{P_3/P_0}$ from the simulated cool core clusters (gray circles) 
to the simulated clusters with irregular morphology (yellow stars). 
These interpolated lines are shown as dashed lines in Figure \ref{P3fit} 
where the red dashed line has the lowest $\chi^2$ when compared to the data. 
Compared to the best line-fit (solid line) the dashed line models have 
$\Delta \chi^2=2.6, 1.1$ and $3.3$ respectively (from bottom to top). 
We conclude that the evolution in $P_3/P_0$ in addition to the redshift-dependent bias 
is only slightly more than $1$-$\sigma$ significant.

\begin{deluxetable}{llll}
\tabletypesize{\scriptsize}
\tablecaption{Power ratio results \label{prtab}}
\tablewidth{0pt}
\tablehead{
\colhead{Sample} &
\colhead{N.o. objects} &
\colhead{$P_2/P_0 (\times 10^{-7})$} & 
\colhead{$P_3/P_0 (\times 10^{-7})$} \\
}
\tablecomments{Average values of the Power Ratios derived in different redshift bins.}
\startdata
$0.069 \le z < 0.1$ & 28 &  $71 \pm 3$ & $1.7 \pm 0.3$ \\
$0.1 \le z < 0.3$   & 55 &  $89 \pm 5$ & $2.2 \pm 0.5$ \\
$z \ge 0.3$        & 18 & $65 \pm 7$  & $7.0 \pm 2.8$ \\
cooling clusters   & 31 & $39 \pm 2$  & $0.29 \pm 0.07$ \\
\enddata
\end{deluxetable}

To compare our results with previous work we list our results for the 
clusters also found in \citet{jeltema05} along with their results in 
Table \ref{prtabJ}. 
These values were derived using Chandra data only. 
We find our values to be consistently about a factor 2 higher with a few 
exceptions but that they follow the same trend of low to high. This 
discrepancy could be related to the limited number of photons in the 
Chandra data causing structures to appear smoother but could also be due 
to artifacts in our modeling described further in following sections. 

\begin{deluxetable*}{lllll}
\tabletypesize{\footnotesize}
\tablecaption{Power ratio comparison \label{prtabJ}}
\tablewidth{0pt}
\tablehead{
\colhead{Name} &
\colhead{$P_2/P_0 (\times 10^{-7})$ \tablenotemark{a}} & 
\colhead{$P_2/P_0 (\times 10^{-7})$ } & 
\colhead{$P_3/P_0 (\times 10^{-7})$ \tablenotemark{a}} &
\colhead{$P_3/P_0 (\times 10^{-7})$ } \\
}
\tablecomments{Comparison of the Power Ratios in our analysis to the results using Chandra data by \citet{jeltema05}}
\startdata
A1413 & $ 58.0 $ & $ 192 \pm 3 $ & $ 0.0525 $ & $ 0.19 \pm 0.1 $ \\ 
A2034 & $ 14.1 $ & $ 84.5 \pm 8.4 $ & $ 0.476 $ & $ 0.41 \pm 0.21 $ \\ 
A2218 & $ 27.4 $ & $ 72.8 \pm 5.6 $ & $ 0.233 $ & $ 1.14 \pm 0.36 $ \\ 
A1914 & $ 15.7 $ & $ 32.4 \pm 1.0 $ & $ 0.709 $ & $ 2.45 \pm 0.26 $ \\ 
A665 & $ 19.4 $ & $ 44.8 \pm 3.5 $ & $ 2.43 $ & $ 7.75 \pm 0.77 $ \\ 
A520 & $ 40.7 $ & $ 65.3 \pm 11.2 $ & $ 2.43 $ & $ 3.87 \pm 0.46 $ \\ 
A963 & $ 5.03 $ & $ 8.14 \pm 1.32 $ & $ 0.342 $ & $ 1.34 \pm 0.67 $ \\ 
A773 & $ 46.9 $ & $ 83.2 \pm 11.2 $ & $ -0.125 $ & $ 0.82 \pm 0.39 $ \\ 
A2261 & $ 4.57 $ & $ 12.9 \pm 11.4 $ & $ 0.201 $ & $ 2.79 \pm 1.7 $ \\ 
A2390 & $ 58.0 $ & $ 149 \pm 7 $ & $ 0.291 $ & $ 2.97 \pm 0.86 $ \\ 
A267 & $ 61.4 $ & $ 95.8 \pm 9.2 $ & $ -0.306 $ & $ 0.92 \pm 0.64 $ \\ 
RXJ2129.6+0005 & $ 17.6 $ & $ 66.1 \pm 1.4 $ & $ -0.0814 $ & $ 0.21 \pm 0.03 $ \\ 
A1758 & $ 188.0 $ & $ 502 \pm 13 $ & $ 1.06 $ & $ 0.78 \pm 0.32 $ \\ 
ZW3146 & $ 4.42 $ & $ 11.6 \pm 1.3 $ & $ 0.078 $ & $ 0.37 \pm 0.08 $ \\ 
MS2137-23 & $ 1.81 $ & $ 1.47 \pm 0.96 $ & $ 0.00772 $ & $ 0.37 \pm 0.49 $ \\ 
CL0016+16 & $ 46.4 $ & $ 114 \pm 17 $ & $ 0.316 $ & $ 5.54 \pm 2.87 $ \\ 
MS0451.6-0305 & $ 66.1 $ & $ 82.3 \pm 5.1 $ & $ 2.19 $ & $ 7.01 \pm 2.19 $ \\ 
MS1137.5+6625 & $ 5.24 $ & $ 11.5 \pm 11.9 $ & $ 0.115 $ & $ 10.5 \pm 7.8 $ \\ 
MS1054.4-0321 & $ 150.0 $ & $ 220 \pm 73 $ & $ 10.3 $ & $ 17.3 \pm 14.6 $ \\ 
WARPJ0152.7-1357 & $ 264.0 $ & $ 2880 \pm 260 $ & $ 12.4 $ & $ 27.6 \pm 19.4 $ \\ 
CLJ1226.9+3332 & $ -0.821 $ & $ 11.1 \pm 3.2 $ & $ 0.77 $ & $ 0.87 \pm 0.4 $ \\ 
\enddata
\tablenotetext{a}{Derived by \citet{jeltema05} from Chandra data.}
\end{deluxetable*}

\subsection{Luminosity-Temperature relation}
Using our definition of a cooling cluster, we treat these 
clusters separately in the subsequent analysis. 
The remaining clusters are divided into 2 redshift bins; $0.069 \le z < 0.2$ 
and $z \ge 0.2$.  This results in 38 low redshift clusters with $\bar{z}=0.11$ 
and 32 high redshift clusters with $\bar{z}=0.39$. 

\begin{figure*}[!htb]
  \begin{center}
    \begin{tabular}{c}
      \includegraphics[width=4in,angle=-90]{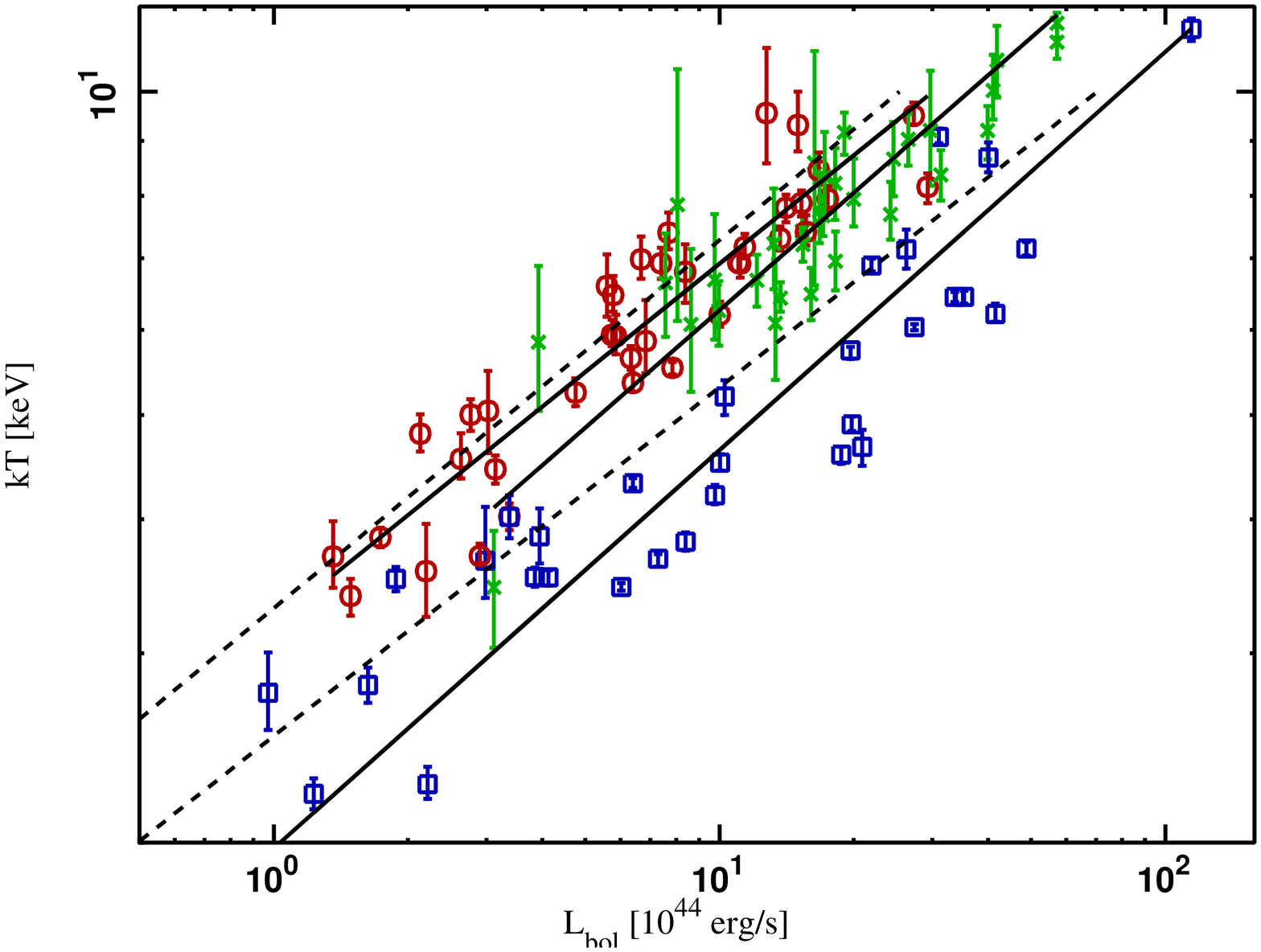} \\
    \end{tabular}
  \end{center}
  \caption{Luminosity temperature relation for low redshift clusters ($z<0.2$, red circles), high redshift clusters ($z\ge 0.2$, green stars) and cooling core clusters (blue squares). Best fit $L_X$-$T$ relations are shown as solid lines and the non-cooling and cooling samples at $z=0$ of \citet{allen98} are shown as dashed lines.
\label{figlt}}
\end{figure*}

The X-ray luminosity-temperature relations for these 3 samples are shown 
in Figure \ref{figlt}. We have used the bolometric luminosity as calculated 
within a 1 Mpc radius along with the temperatures given in Table \ref{spectab}.
The low redshift sample (red circles), the high redshift sample (green stars) 
and the cooling cluster sample (blue squares) are shown along with their best 
fit $L_X - T$ relations (solid lines). The best fit relations from the non-cooling 
and cooling samples of \citet{allen98}, at $z=0$, are shown as dashed lines for comparison. 

Our best fit values of $L_6$ and 
$\alpha_{LT}$ are shown in Table \ref{lttab}, where the bolometric X-ray 
luminosity is $L_X = L_6(kT/6~$keV$)^{\alpha_{LT}}$. We have added a systematic 
scatter in $\mathrm{log}(T)$ in order to achieve $\chi^2/dof. = 1$. 

\begin{deluxetable}{llll}
\tabletypesize{\scriptsize}
\tablecaption{Luminosity-temperature relation \label{lttab}}
\tablewidth{0pt}
\tablehead{
\colhead{Sample} &
\colhead{$L_6$} &
\colhead{$\alpha_{LT}$} & 
\colhead{$\sigma_{syst}(\log(T))$} \\
}
\tablecomments{Best fit parameters for the $L_X$-$T$ relation. In addition to the measurement errors, a systematic scatter in $\log (T)$ has been added in order to achieve $\chi^2/dof. = 1$.  The scatter in temperature in the non-cooling sample corresponds to $\sim 0.7~$ keV at 6 keV.}
\startdata
$0.069 \le z < 0.2$ & $6.5 \pm 0.4$ & $2.98 \pm 0.19$ & 0.046 \\
$z \ge 0.2$         & $8.9 \pm 0.7$ & $2.62 \pm 0.21$ & 0.050 \\
cooling clusters    & $20.3 \pm 2.0$  & $2.69 \pm 0.18$ & 0.075 \\
\enddata
\end{deluxetable}
 
It is obvious in this plot that it is of great importance to exclude cooling core 
clusters since these clusters have much 
higher $L_X$ for a given $T$ compared to non-cooling clusters. 
It is of particular importance to select these clusters 
using a method unbiased in redshift. 
A selection bias at high redshift leading to the failure to identify cooling 
clusters properly could easily be interpreted as evolution. 
We have based our selection of cooling clusters on simulated identical clusters 
at various redshifts, 
propagating their photons through our detector model to best account for 
any distance dependent systematic effect. 

Fitting the data to a generalized $L_X$-$T$ relation with a $z$ dependence; 
\begin{equation}
\label{ltbeq}
L_X = L_{6} ~ \left( \frac{T}{6~\mathrm{keV}} \right) ^{\alpha_{LT}} ~ (1+z)^{\beta_{LT}}, 
\end{equation}
weak evolution is found with the best fit values for the non-cooling clusters 
shown in Table \ref{lttab2} along with fits for high luminosity ($L_{bol} \ge 10^{45} $erg s$^{-1}$) 
non-cooling clusters, high $P_3/P_0$ clusters ($P_3/P_0 \ge 10^{-7}$) and for the cooling clusters separately. 

The results for the non-cooling clusters show weak evolution whereas 
the cooling cluster sample is consistent with no evolution. 
There is however large intrinsic scatter within the cooling sample. 
Interestingly, the sample containing the clusters with the highest luminosity 
substructure (high $P_3/P_0$, see Section \ref{pratsec}) shows significant 
evolution with $z$. Slightly higher than that of the non-cooling sample.

It is possible that this has to do with the incompleteness of the sample. 
A cut on $P_3/P_0$ selects a larger fraction of high-$z$ clusters, since 
$P_3/P_0$ has a weak positive trend with $z$. High redshift clusters tend 
to be more luminous by selection and this could cause a small bias in the 
observed evolution.
We find that our results for the clusters without cool cores are in good agreement
with theoretical models of cluster evolution that are based on balancing
gas cooling with feedback at the entropy set by the
cooling threshold \citep{voit02,voit05}.
This threshold $K_c(T,t)$ is the entropy at which constant-
entropy gas at temperature $T$ radiates an energy equivalent to its thermal
energy in time $t$.
These models are also successful in explaining the observed $L_X \propto T^3$ behavior
for clusters, contrary to early assumptions of pure gravitational self-similar
collapse, where $L_X \propto T^2$ was originally expected.

\begin{deluxetable*}{lllll}
\tabletypesize{\scriptsize}
\tablecaption{Redshift dependent luminosity-temperature relation \label{lttab2}}
\tablewidth{0pt}
\tablehead{
\colhead{Sample} &
\colhead{$L_6$} &
\colhead{$\alpha_{LT}$} & 
\colhead{$\beta_{LT}$} &
\colhead{$\sigma_{syst}(\log(T))$} \\
}
\tablecomments{Best fit parameters for the $L_X$-$T$ relation in Eq. \ref{ltbeq}.  In addition to the measurement errors, a systematic scatter in $\log (T)$ has been added in order to achieve $\chi^2/dof. = 1$. The scatter in temperature in the non-cooling sample corresponds to $\sim 0.5~$ keV at 6 keV. }
\startdata
non-cooling        & $6.8 \pm 0.5$ & $2.87 \pm 0.14$ & $0.50 \pm 0.34$ & 0.037 \\
cooling clusters   & $18.8 \pm 5.9$  & $2.79 \pm 0.26$ & $0.62 \pm 1.47$ & 0.067 \\
high $L_{bol}$ NC  & $7.6 \pm 0.7$ & $2.81 \pm 0.26$ & $0.48 \pm 0.32$ & 0.029 \\
high $P_3/P_0$ NC      & $6.4 \pm 0.6$ & $3.02 \pm 0.18$ & $0.82 \pm 0.42$ & 0.035 \\
\enddata
\end{deluxetable*}

\subsubsection{Temperature correlation}
\label{tcorrref}
Since the temperature two-point correlation function is new, we illustrate its 
power for four representative clusters that clearly show specific 
characteristics. $A(r_k)$ is calculated for all spatial points, with 
10 kpc resolution, within radius of 500 kpc of the cluster centroid.  

\begin{figure*}[!htb]
  \begin{center}
    \begin{tabular}{cc}
      \includegraphics[width=2.5in,angle=-90]{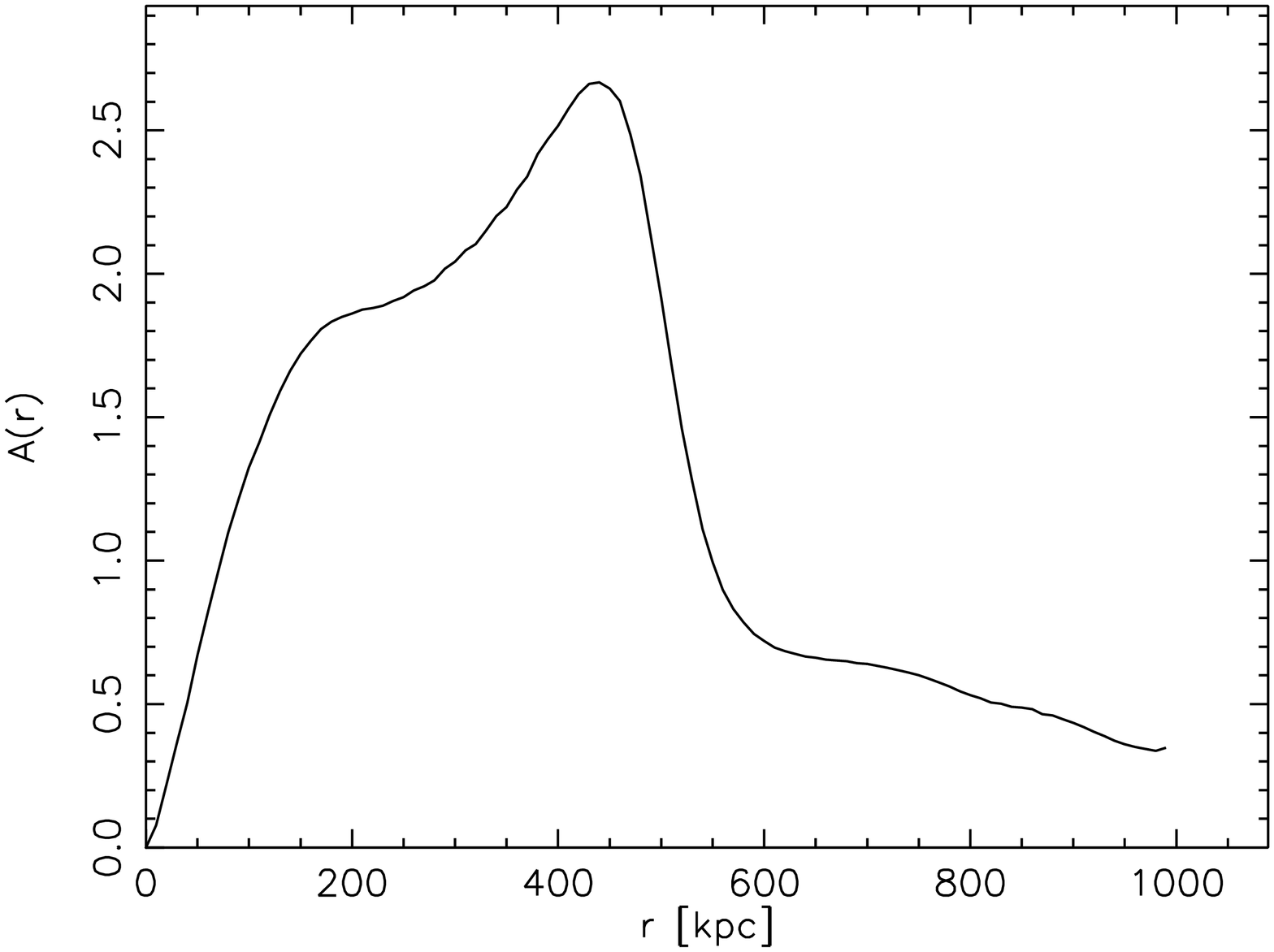} & 
      \includegraphics[width=2.5in,angle=-90]{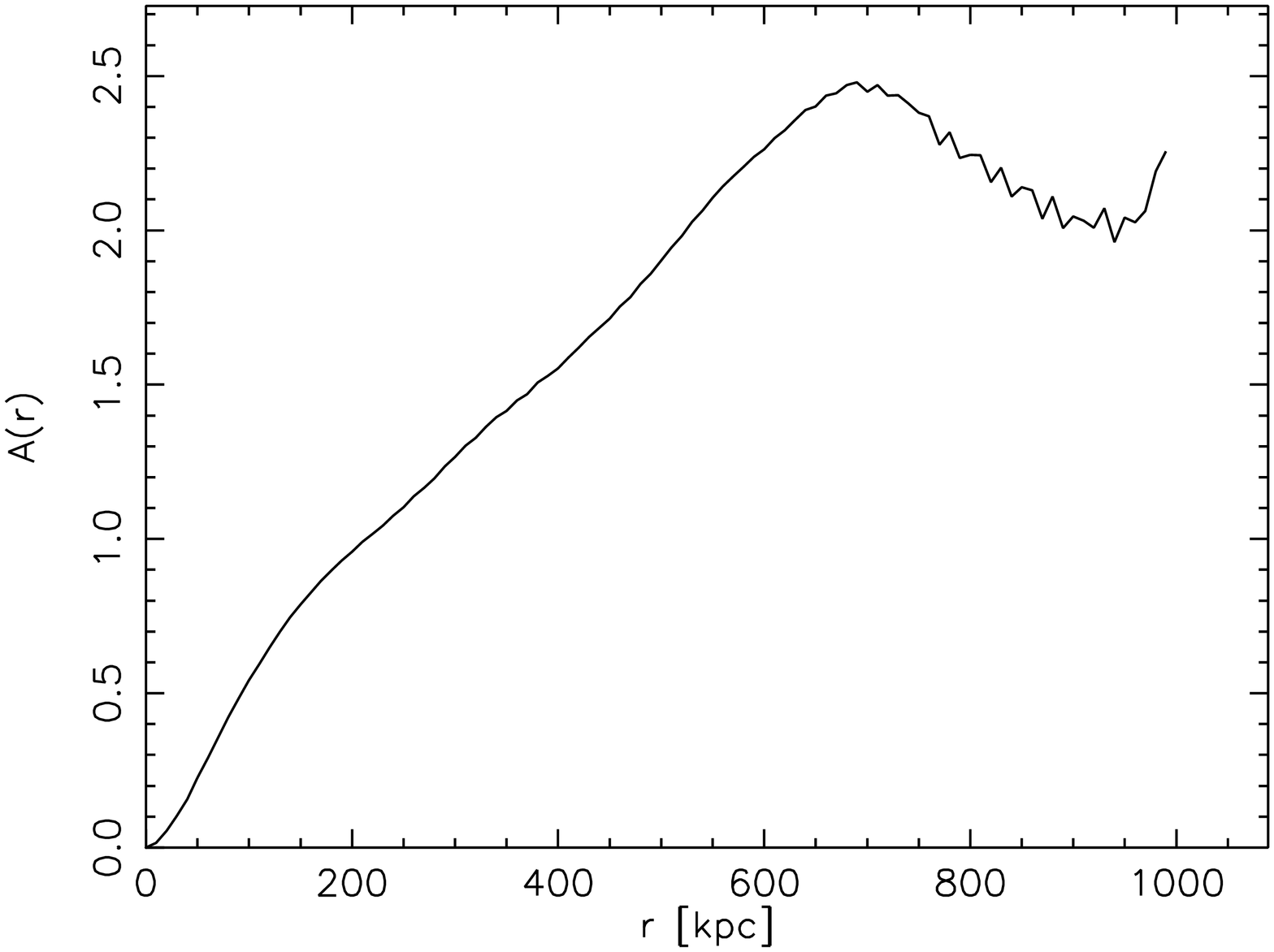} \\
      \includegraphics[width=2.5in,angle=-90]{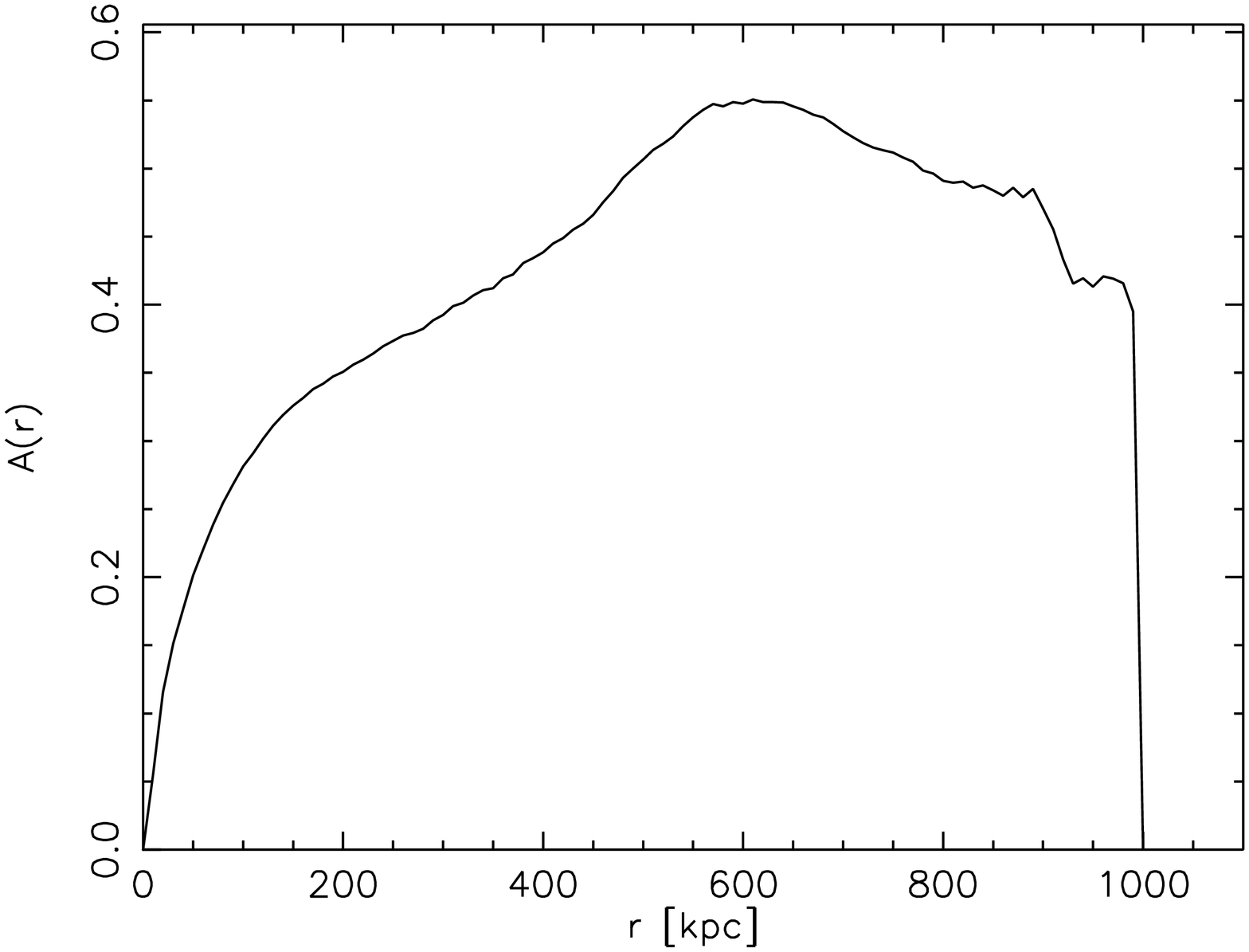} &
      \includegraphics[width=2.5in,angle=-90]{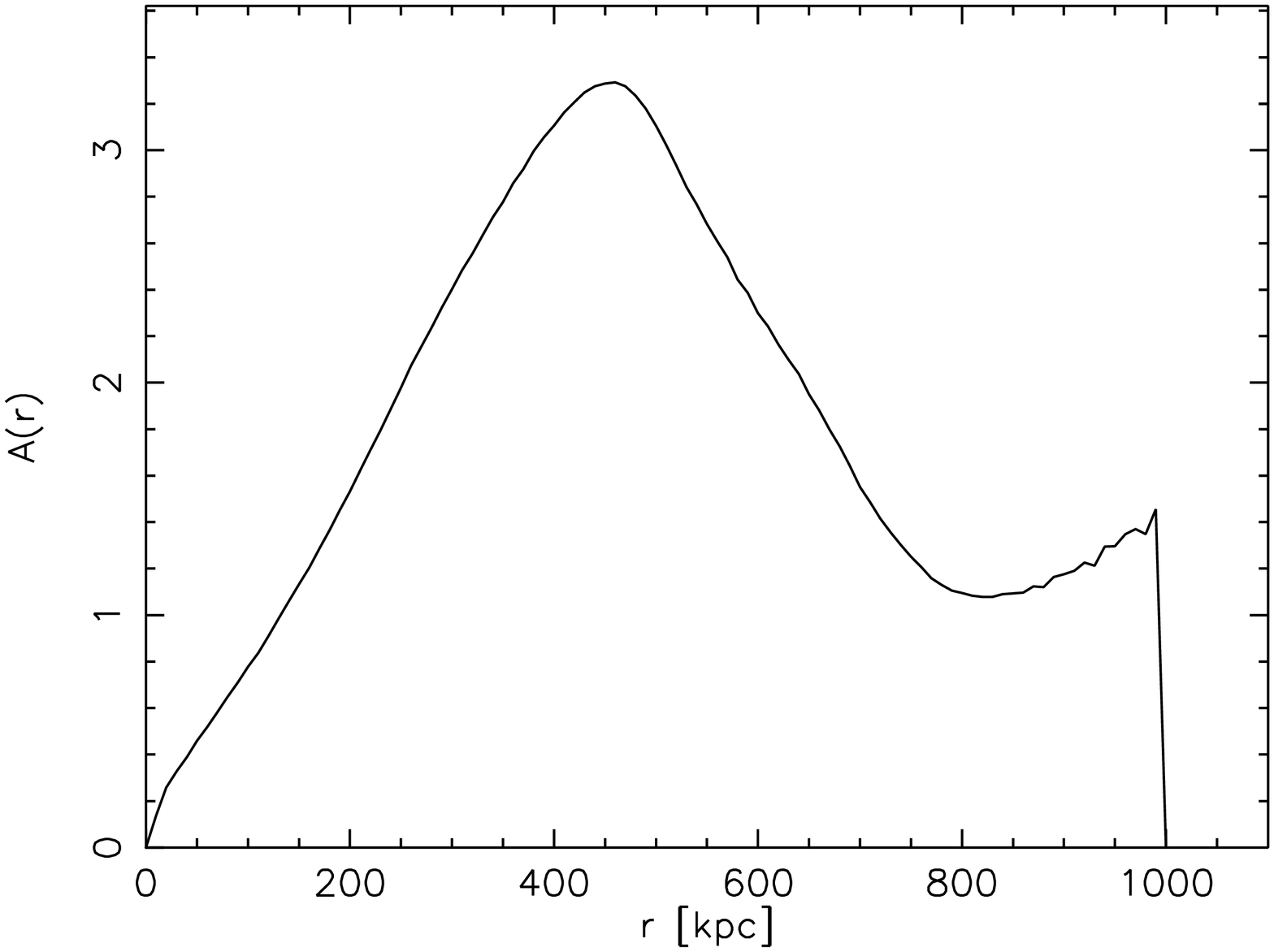} \\
      
    \end{tabular}
  \end{center}
  \caption{The temperature correlation function, $A(r_k)$ plotted for clusters A1835 (top left), MS1054.4-0321 (top right), A1689 (bottom left) and A2142 (bottom right).
\label{sTbins}}
\end{figure*}

First, we consider a well-known cooling core cluster, Abell 1835 
(see Fig. \ref{sTbins}, first panel).  Here, $A(r_k)$ is shown as a function 
of distance scale, $r$.
As expected, there is considerable power at 
small spatial scales, implying a compact structure with temperature clearly different 
from the rest of the cluster. There is a sharp drop after 500 kpc where the core 
is no longer visible. 
Another extreme example is MS1054.4-0321 (Fig. \ref{sTbins}, second panel), 
a highly substructured cluster consisting of multiple components. 
Here, the amplitude is high, distributed more uniformly over a large 
range of spatial scales, suggesting multiple high-luminosity components with 
different temperatures. 
As a third example, we also show the correlation function for a 
nearly isothermal cluster, A1689 (Fig. \ref{sTbins}, third panel), 
where small temperature fluctuations lead to lower values across the 
range of spatial scales. 
Finally we show the correlation function for A2142 (Fig. \ref{sTbins}, fourth panel), 
where an offset of the cluster core causes the sharp drop seen in A1835 
to be absent. The drop is gradual, implying an overall asymmetry in the 
temperature structure of the cluster. 

\begin{figure*}[!htb]
  \begin{center}
    \begin{tabular}{cc}
      \includegraphics[width=2.5in,angle=-90]{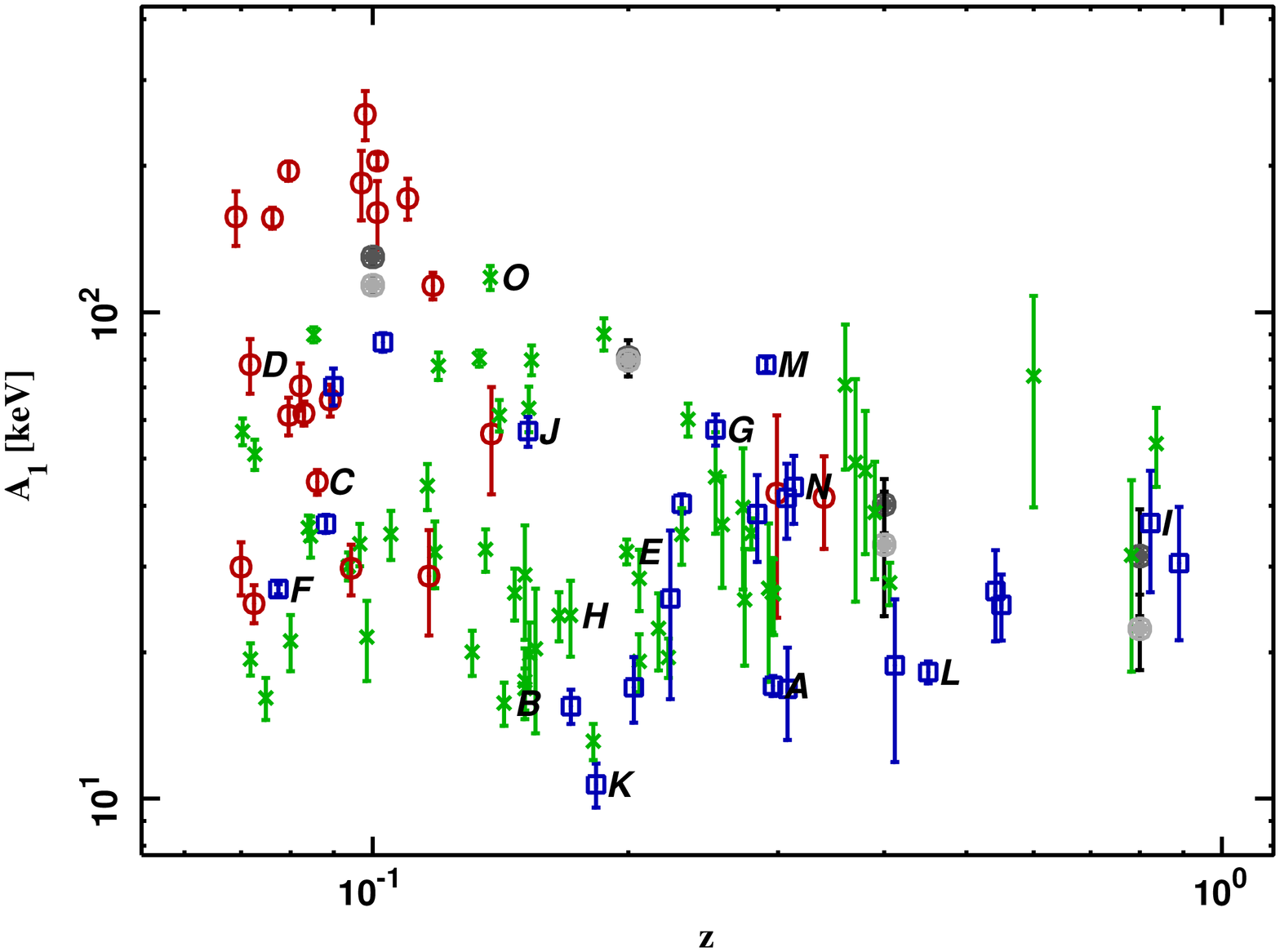} & 
      \includegraphics[width=2.5in,angle=-90]{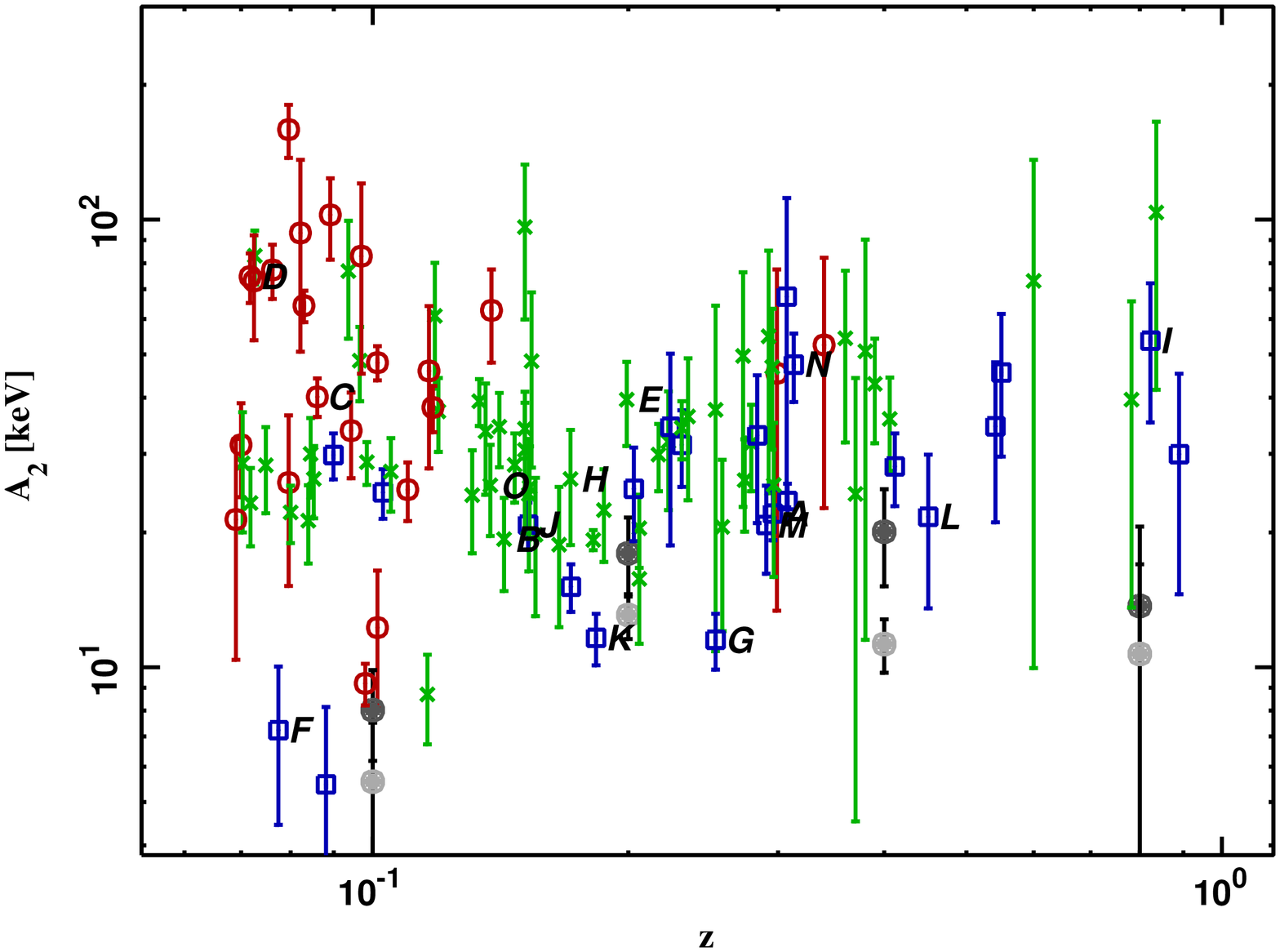} \\
    \end{tabular}
  \end{center}
  \caption{Temperature correlations for scales $250-500~$kpc ($A_1$) vs redshift (left panel) and $750-1000~$kpc ($A_2$) vs $z$ (right panel).  The different symbols represent luminosity (red circles 
$\ge 10^{43}~$erg s$^{-1}$, green stars $\ge 4~10^{44}~$erg s$^{-1}$, and blue squares $\ge 2~10^{45}~$erg s$^{-1}$). Simulated cool core clusters are shown as filled gray circles. Letters from A to O denote a few selected clusters (see Figure \ref{Lc}).
\label{Tcorr}}
\end{figure*}

The temperature correlation function $A(r_k)$ is integrated over $250$-$500~$kpc 
($A_1$) and over $750$-$1000~$kpc ($A_2$) and this is displayed in Figure \ref{Tcorr}, 
plotted against redshift. This shows the magnitude of temperature gradients 
over small and large scales respectively. 
These distance ranges are chosen since for the small scales, $250$-$500~$kpc, 
the cooling cores dominate the statistic. To distinguish these fluctuations from larger 
scale, merger related disturbances we compare it to the $750$-$1000~$kpc range 
where the core cannot be included (since we use a 500 kpc aperture). 

Strong cooling clusters such as A1835 and A2204 show large values of $A_1$ 
since the cool cores dominate the small scales. The same objects show low 
values of $A_2$ since these clusters are largely isothermal when the core is 
excluded. The same is true for our simulated cool core clusters, while the small 
scales cannot quite be resolved at high $z$.  
In contrast, disturbed clusters such as RXJ0658-55 or MS1054.4-0321 
show generally lower values of $A_1$ and 
larger values of $A_2$ whereas nearly isothermal clusters such as A1689 or A1413 
show low values for both. 
With the exception of a few low luminosity clusters at low redshift, large scale 
temperature structure ($A_2$) can be seen to increase slightly with redshift 
above $z=0.2$ when compared to the simulated clusters.
There is also an apparent drop in the temperature structure for smaller scales ($A_1$).

We note that the decrease in $A_1$ with redshift can be partly due to smoothing of the 
core caused by the PSF and loss of photons as seen in the evolution in the simulated 
clusters in Figure \ref{Tcorr}. 
However, it is unlikely that this effect is the cause 
of the increase in $A_2$ since temperature features are likely to be washed out 
by this effect. This is seen in the trend of $A_2$ for the simulated clusters.

\begin{figure*}[!htb]
  \begin{center}
    \begin{tabular}{c}
      \includegraphics[width=3in,angle=-90]{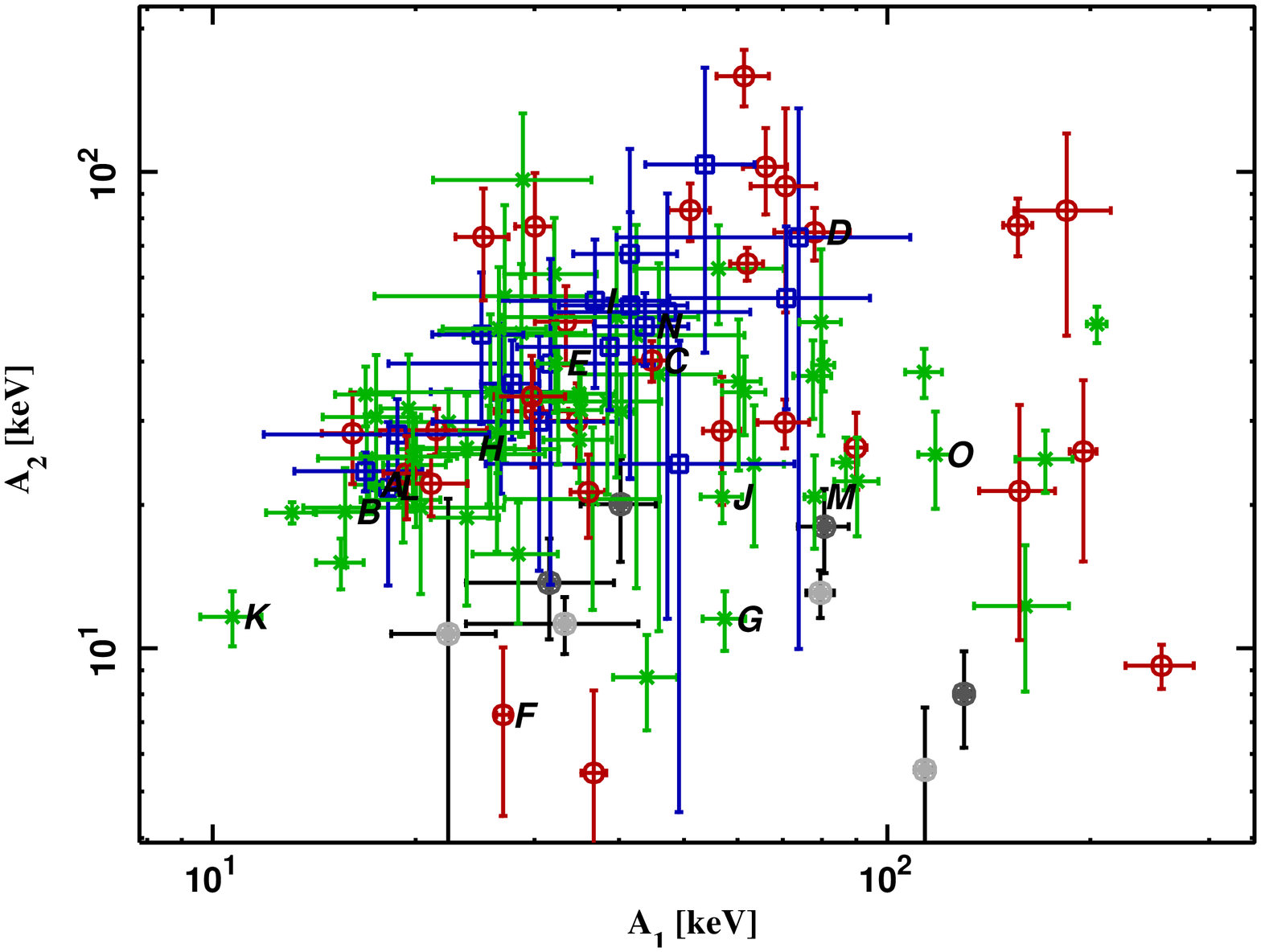} \\ 
    \end{tabular}
  \end{center}
  \caption{Temperature correlations for scales $750-1000~$kpc ($A_2$) plotted against $250-500~$kpc ($A_1$) scales.   Here the different symbols represent redshift (red circles $0.069\le z<0.1$, green stars $0.1\le z<0.3$, and blue squares $z\ge 0.3$). Simulated cool core clusters are shown as filled gray circles. Letters from A to O denote a few selected clusters (see Figure \ref{Lc}).
\label{Tcorr2}}
\end{figure*}

In Figure \ref{Tcorr2} we plot $A_2$ against $A_1$ to show the separation of 
``relaxed'' cool core clusters (lower right) from disturbed clusters (upper left).

\subsection{Notes on individual clusters}
\label{clustercomments}
In this section we describe breifly a few selected clusters in our sample, 
comparing our findings with previous results.
The luminosity and temperature maps of these clusters are presented in 
Figures \ref{A1068} to  \ref{ZW3146}.

\subsubsection{Abell 1068}
\begin{figure*}[!htb]
  \begin{center}
    \begin{tabular}{cc}
      \includegraphics[width=2in]{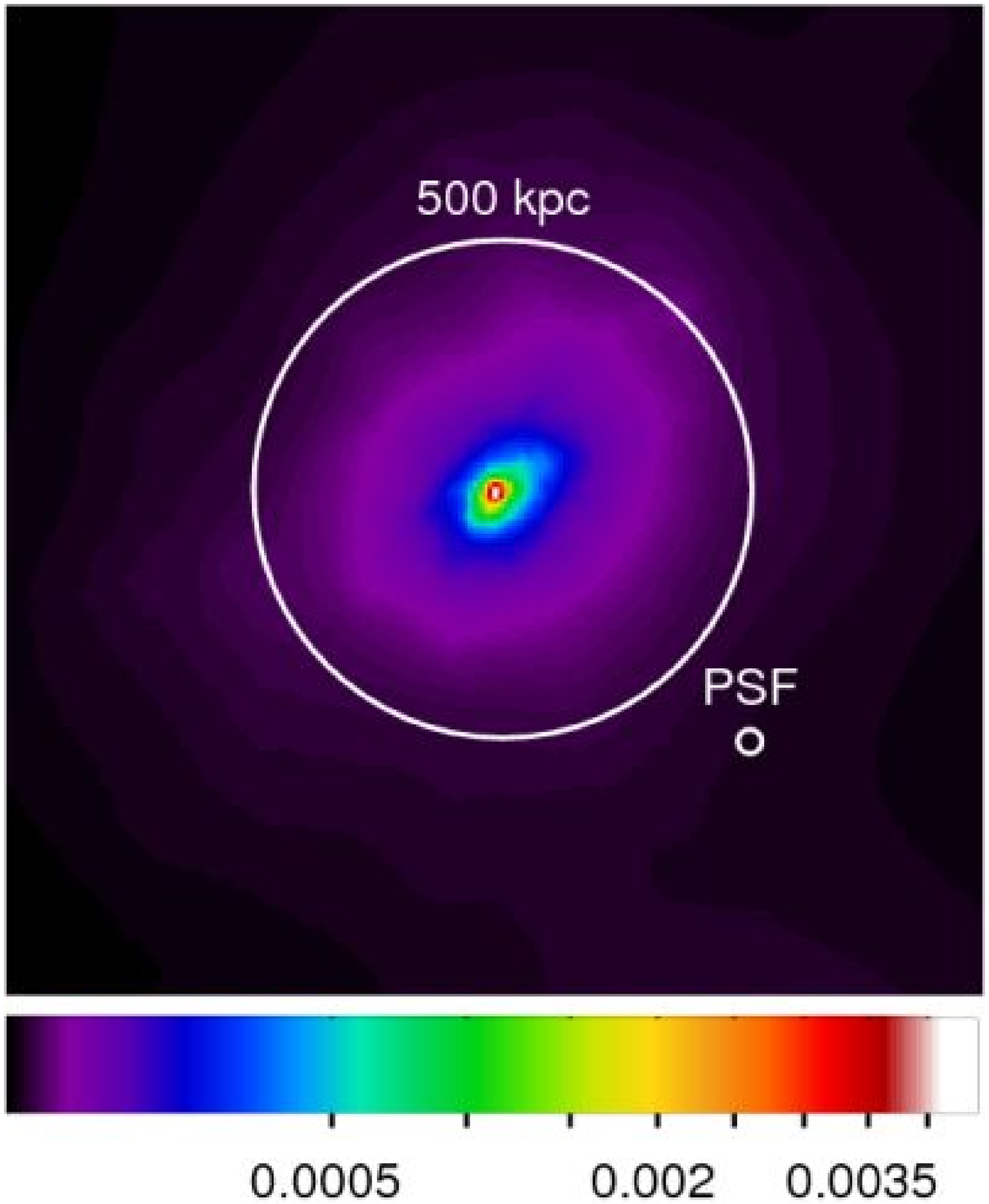} & 
      \includegraphics[width=2in]{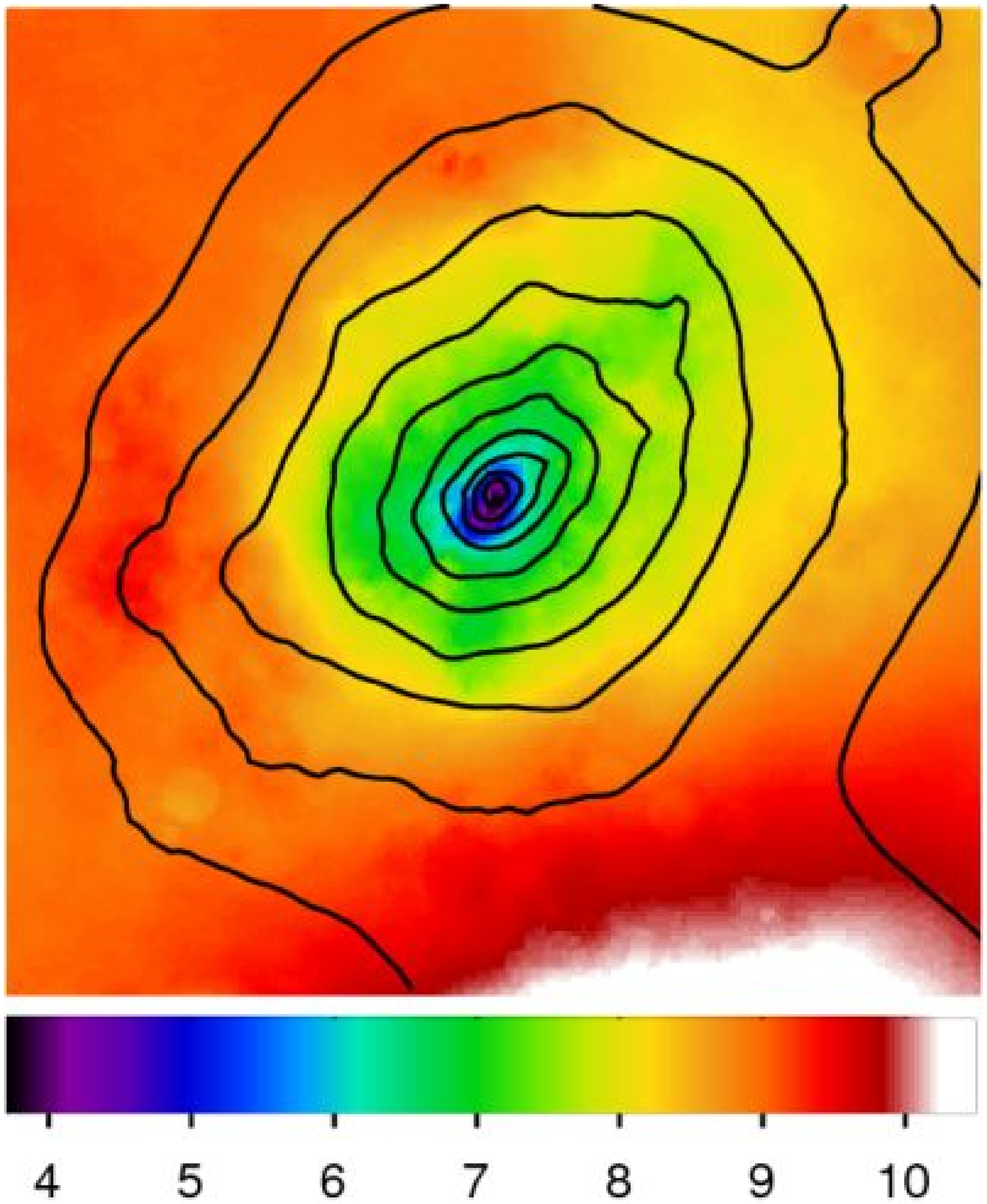} \\
    \end{tabular}
  \end{center}
  \caption{$2 \times 2$ Mpc images of the bolometric luminosity, in units of $10^{44}~$erg s$^{-1}$ $(\arcsec)^{-2}$ (left panel), and temperature, in keV (right panel), for Abell 1068. The size of the 500 kpc region used for power ratio analysis and two-point temperature correlation analysis is shown along with the size of the PSF (10$\arcsec$ radius) \label{A1068}.}
\end{figure*}

A high resolution temperature map of Abell 1068, using Chandra data, 
was previously presented in \citet{wise04} 
who find an large degree of substructuring within the central 
$80~\arcsec \times 80~\arcsec$ with temperatures varying between 2.5 and 6 keV. 
We find that this agrees well with our map in Figure \ref{A1068}, 
although the spatial resolution is not as high as in the Chandra map.
The size of the $80~\arcsec \times 80~\arcsec$ region 
can be estimated using 4 times the PSF width in Figure \ref{A1068}. 
In a radial temperature profile derived from the XMM-Newton data, \citet{snowden08} 
find a central temperature of $3.1~$keV increasing to $5.3~$keV around $0.4~r_{500}$ 
and dropping to $2.2~$keV around $0.8~r_{500}$ ($\sim 800~$kpc). 
In our map we find no indication of such a drop in $T$. We note, however, that 
due to the low cluster flux at these radii, the error on the temperature 
in the map is large ($\sim 3~$keV). There may also be some associated systematic 
effects (see Section \ref{systematic}). 
The average temperature of the cluster that we find from a simple spectral analysis 
is $3.81^{+0.08}_{-0.07}~$keV.

\subsubsection{Abell 1413}
Early studies of the temperature profile of Abell 1413 using XMM-Newton data 
find a central temperature of approximately $7~$keV declining to $6~$keV out 
to 1 Mpc radius \citep{pratt02}. \citet{snowden08} find a similar trend with a 
central temperature of $7.7~$keV declining to $5.7~$keV out to $\sim 1~$Mpc.
Our temperature map (Figure \ref{A1413}) exhibits a large degree of small scale substructure 
in this otherwise smooth surface brightness distribution 
with temperatures varying between $7$ and $9~$keV. 
We do not see a clear temperature drop with increasing radius but rather a large degree of 
asymmetry. We find an average spectral temperature of 
$7.30^{+0.19}_{-0.19}~$keV.

\subsubsection{Abell 1689}
The temperature map of Abell 1689 (Figure \ref{A1689}) is in good agreement with our earlier 
analysis of the XMM-Newton data \citep{andersson04,andersson07}. 
\citet{snowden08} find a slightly higher temperature of 11.6 keV 
around $0.1~r_{500}$ ($\sim 200~$kpc) whereas at other radii the results 
are consistent with ours with fluctuations from $8$ to $10~$keV.
The average spectral temperature is $9.07^{+0.17}_{-0.12}~$keV.

\subsubsection{Abell 1750}
The XMM-Newton data for Abell 1750, consisting of two subclusters, 
was analyzed by \citet{belsole04} who 
find temperatures varying between $3$ and $6~$keV within the double-cluster.  
The temperature structure overall is in good agreement with our temperature 
map while we find temperatures varying from $4.5$ to $8.5~$keV. 
This is likely related to the bias described in Section \ref{tempbias}. 
We find that the larger subcluster (pictured center in Fig. \ref{A1750}) 
is slightly hotter ($5.5~$keV) compared to the northern subcluster 
($4.5~$keV) with the gas in between having a temperature of $8~$keV. 
\citet{belsole04} find these temperatures to be $2.8$, $3.9$ and $5.1~$ keV 
respectively. 
In a simple spectral fit we find the average temperature of A1750 to be 
$4.45^{+0.13}_{-0.13}~$keV.

\subsubsection{Abell 1758}
Abell 1758 is a complicated system consisting of a northern and a southern component, 
separated by 2 Mpc in the plane of the sky, each of which, in turn, consists of two 
separate subclusters. 
We have limited our analysis to the northern component A1758N (See Fig. \ref{A1758}).
The Chandra and XMM-Newton data of A1758 have been analyzed earlier by \citet{david04} 
who conclude that A1758N is in the late stages of a large impact parameter merger 
of two 7 keV clusters. The hardness ratio map presented in \citet{david04} is 
qualitatively in good agreement with our temperature map. 
For their spectral fits in four separate regions they find temperatures of $7.2$, 
$6.6$, $7.2$ and $9.8~$ keV for the ``northwest wake'', ``southeast wake'', 
``core'' and ``halo'' using the XMM-Newton data. The values for the same regions 
in our temperature map are systematically approximately 1 keV higher. 
In our overall spectral fit we find a temperature of $9.16^{+0.39}_{-0.43}~$keV.

\subsubsection{Abell 1775}
Abell 1775 is a nearby ($z=0.072$), low mass cluster with high velocity dispersion, 
containing a pair of massive galaxies at the center (Figure \ref{A1775}). 
Studying the velocity distribution 
of the galaxies in this cluster, \citet{oegerle95} conclude that it, in fact, 
consists of two smaller subclusters that are in the process of merging. 
In our temperature map, we can see an arc of the coldest gas extending to the 
east of the cluster center, possibly a remnant of a subcluster core. 
In this low mass cluster, the surface brightness becomes low outside 500 kpc 
and the temperature map outside of this radius is not reliable. 
Within 500 kpc we find the gas temperature varying in the 4 to 7 keV range, 
biased upward by effects described in Section \ref{tempbias}. Adapting the 
prior range of the temperature distribution in this case to better accommodate 
for low temperature gas would give a more accurate result. 
\citet{snowden08} find a central temperature of $4.3~$keV decreasing to $3.5~$keV 
around $0.4~r_{500}$ ($\sim 600~$kpc). 
In our overall spectral fit we find a temperature of $3.70^{+0.10}_{-0.06}~$keV.

\subsubsection{Abell 1835}
Abell 1835 is a massive cluster with a pronounced cool core and a 
relaxed morphology. \citet{majerowicz02} find a central temperature of 
$4.1~$keV increasing to $8.5~$keV near 1 Mpc using XMM-Newton data. 
\citet{snowden08} find a central temperature of $6.1~$keV, increasing to 
$10.5~$keV just outside the cluster core and decreasing to $7.7~$keV 
out to 1 Mpc using the same data. This is in good agreement with our 
temperature map shown in Figure \ref{A1835}. 
In a simple spectral fit we find the average temperature of A1835 to be 
$7.14^{+0.10}_{-0.11}~$keV.

\subsubsection{Abell 2029}
Using an early Chandra observation of A2029, \citet{lewis02} find a central 
temperature as low as $4~$keV within 10 kpc increasing to approximately 9 keV 
out to 300 kpc. Using the same data, \citet{clarke04} find a disturbed core 
within a 20 kpc radius with some spatial correlation to the central radio emission. 
In our temperature map in Figure \ref{A2029} we find no significant central 
substructuring on the scales we are probing with the XMM-Newton data. 
\citet{bourdin08} find a central temperature of $5.7~$keV and a large degree 
of temperature substructure with fluctuations between $7$ and $9.5~$keV in 
their temperature map out to $600~$ kpc, generated using the XMM-Newton data. 
While in good quantitative agreement with the radial trend in our temperature 
map, we find no sign of the hot or cold substructures in our map.
In our overall spectral fit we find a temperature of $6.88^{+0.11}_{-0.08}~$keV.

\subsubsection{Abell 2142}
\citet{markevitch00} first argued that A2142 contains a ``cold front'', a 
steep nonaxisymmetric X-ray flux gradient and a steep temperature gradient 
originally believed to be due to a low-entropy remnant of a recently merged 
subcluster. \citet{tittley05} argue that this ``cold front'' is due to 
oscillatory motions of the core after a merger event. 
Our temperature map (Figure \ref{A2142}) shows a central temperature around $7.5~$keV with a 
plume-like structure of $8.5~$keV gas extending north and a steep temperature gradient, 
aligned with the sharp trend in surface brightness to the north-west, 
increasing to $11.5~$keV. This agree well with the results of \citet{markevitch00}, 
with the exception of the central $5~$keV gas which we do not detect. 
We find an overall spectral temperature of A2142 to be $8.15^{+0.24}_{-0.28}~$keV.

\subsubsection{Abell 2204}
Abell 2204 is a massive cool core cluster with a regular morphology. 
\citet{sanders05} find the temperature of the core region to be $3.26\pm0.20~$keV. 
The temperature map of \citet{sanders05} shows a gradually increasing temperature 
to $\sim 10~$keV at 100 kpc, continuing up to $16~$keV around 400 kpc after which 
it drops again to around 7 keV. The trend is continuous, with the exception of 
an arc-shaped region of hot gas extending around the northern part of the cluster 
at 50 kpc where the temperature suddenly jumps to 14 keV. 
Our map in Figure \ref{A2204} shows a similar trend with a central temperature of 
$4.5~$keV gradually increasing to $10~$keV around 300 kpc where after it 
drops again to $7~$keV at 1 Mpc. We find no signs of the hotter gas around 50 kpc. 
\citet{reiprich08} also find a similar radial profile using data from Chandra, 
XMM-Newton and Suzaku telescopes with a sharp increase in temperature outside 
of the core and a gradual decline out to the virial radius where a temperature of 
4 keV is measured. 
In our overall spectral fit we find a temperature of $6.44^{+0.08}_{-0.09}~$keV, 
dominated by the emission from the core.

\subsubsection{Abell 520}
A detailed temperature map of A520, using a short Chandra observation, was first 
presented in \citet{markevitch03}. The map shows the southwestern cluster core 
having a temperature of approximately $5~$keV while the main body of the cluster 
exhibits an irregular morphology with temperatures varying from 6 to 14 keV. 
Using a longer observation, \citet{markevitch05} detect a temperature drop in 
the southwest quadrant, from $11.5^{+6.7}_{-3.1}~$keV  at $400-600~$kpc distance 
from the center to $4.8^{+1.2}_{-0.8}~$keV at $600-1000~$kpc.
We do not find any evidence for such a temperature drop. This is most likely due 
to the fact that the surface brightness is low in this quadrant at this large radius.
Directly south of the main cluster, we see a region of possibly shocked gas at 
$11~$keV and this appears to be the source of the hottest emission in this 
cluster (see Figure \ref{A520}).
In the overall spectral fit we find this cluster has an average temperature of 
$8.45^{+0.33}_{-0.26}~$keV.

\subsubsection{MS0451.6-0305}
The temperature map in Figure \ref{MS0451.6-0305} reveals a low-temperature (7 keV) 
structure east of the main cluster (10 keV). It is possible that this component 
represents a subcluster in the process of merging with the main cluster, or 
perhaps a less massive cluster in projection.
This feature can be seen in the maps of surface brightness presented in 
\citet{donahue03}, where the soft-band image is seen to extend further east 
than the hard-band map.
\citet{donahue03} further find a central temperature ($r<31\arcsec$) of $10.3^{+2.2}_{-1.5}~$keV 
and an outer temperature ($31\arcsec<r<85\arcsec$) of $9.8^{+2.5}_{-1.7}~$keV.
In our overall spectral fit we find a temperature of $10.02^{+0.80}_{-0.60}~$keV 
in good agreement with their results.

\subsubsection{MS1054.4-0321}
MS1054.4-0321 has been studied previously with Chandra and XMM-Newton. 
\citet{gioia04} find an overall temperature of $7.2^{+0.6}_{-0.7}~$keV using 
the XMM data while \citet{jee05} find an overall temperature of $8.9^{+1.0}_{-0.8}~$keV 
using the Chandra data. \citet{jee05} also find the temperature of the western peak 
to be $7.5^{+1.4}_{-1.2}~$keV, the central peak to be $10.7^{+2.1}_{-1.7}~$keV 
and an overall positive gradient in their temperature map from west to east. 
We confirm the presence of a temperature gradient, seen rising from 9 to 11 keV 
in Figure \ref{MS1054.4-0321}. 
In our overall spectral fit we find a temperature of $9.20^{+1.26}_{-1.03}~$keV 
in good agreement with the Chandra results.

\subsubsection{MS2137-23}
MS2137-23 is a strong gravitational lensing cluster with a well defined central 
cool core. In the temperature map in Figure \ref{MS2137-23} we see the temperature 
increase from 5 to 8 keV within 500 kpc radius. The map exhibits a low degree of 
asymmetry which appears to increase with radius. 
In a deprojected temperature profile using a Chandra observation, \citet{arabadjis04} 
find a central ($\sim < 50~$kpc) temperature of 4 keV rising to around 7 keV at 
150 kpc radius and decreasing again at larger radii. An offset datapoint at 250 kpc radius 
shows a temperature of $12.5\pm 4.5~$keV.
Our overall spectral fit shows a temperature of $4.67^{+0.17}_{-0.19}~$keV, clearly 
showing the dominance of the cool core emission. 

\subsubsection{RXJ0658-55}
For a detailed discussion on the temperature structure of RXJ0658-55, see 
\citet{andersson07}. Our maps of luminosity and temperature are shown in 
Figure \ref{RXJ0658-55}.

\subsubsection{RXJ1347-1145}
The temperature map in Figure \ref{RXJ1347-1145} shows a central temperature of 
$9~$keV increasing to $11.5~$keV within 500 kpc and decreasing again towards 1000 kpc 
radius. The temperature map further shows a lack of the hottest emission in the 
north quadrant of the cluster. 
We compare our results with the radial temperature profile and map of \citet{gitti04}, 
derived from the XMM-Newton data. 
The temperature profile shows an overall good agreement with our results, however, the 
temperature map of \citet{gitti04} identifies a region of hot ($\sim 17~$keV) emission 
approximately 150 kpc south-east if the core. This region appears in our map as a region of 
$11.5~$keV emission aligned with the elongation of the surface brightness to the SE. 
Our map also shows less substructure overall.
We find this massive cluster has an overall spectral temperature of 
$11.44^{+0.26}_{-0.29}~$keV.
  
\subsubsection{ZW3146}
 The temperature map of the cool core cluster ZW3146 (Figure \ref{ZW3146}) 
reveals a central temperature of $5.5~$keV increasing to approximately 
$9~$keV at 500 kpc. At larger radii the temperature appears to increase 
further, however, here the measurement is uncertain.
Using XMM-Newton data, \citet{kausch07} find a temperature gradient from 4.7 
to 7 keV up to $1~$Mpc with most of the colder emission being localized within 
$100~$kpc. Their temperature map shows fluctuations between $4.5$ and $9$ keV 
in within 200 kpc. 
We do not see these fluctuations and conclude that the cluster has a 
relatively relaxed morphology with a smooth temperature gradient. The surface 
brightness contours have tighter spacing to the south-east which may indicate 
recent core motion relative to the cluster as a whole.
Our overall spectral fit shows a temperature of $6.21^{+0.14}_{-0.10}~$keV.

\section{Systematic effects}
\label{systematic}
\subsection{Selection bias}
Since the cluster sample was selected using all the public XMM data available 
at the time, 
it does not consist of a representative subset of clusters in general in the 
universe.  However, many studies, and observation proposals, are based 
on either the most regular known clusters or the most complex. Hence, the sample 
should contain both these extremes of the cluster population which is 
what is compared in this work.

\subsection{Background modeling}
The X-ray background is modeled as opposed to subtracted which 
is the usual procedure in standard spectral modeling of extended X-ray sources. This introduces 
an uncertainty different from the unsubtracted background flux which is common 
when dealing with combinations of several sources of background with different spatial and 
spectral distributions. Here, the distortions of different backgrounds are modeled 
correctly, but instead, there is the possibility of cluster particles mimicking 
the background flux.  This effect is small since the backgrounds are present 
across the entire field which is not the case for the majority of the cluster 
emission since all clusters are modeled out to at least 1 Mpc where the cluster 
emission tends to be faint.  
The low cluster flux at this radius does, however, impact on the precision 
of the temperature measurements at large radii. 
The nature of this uncertainty is seen in some temperature maps of median temperature 
as it reverts to the mean of the prior, which in this case is 7.75 keV. 

The spectral shape of the instrumental background also means that it is not easily 
modeled by cluster particles. The particle background has a nearly flat spectral shape 
across the entire energy range whereas the X-ray photons are modulated by the 
energy dependent effective area of the X-ray mirrors. This makes it statistically 
favorable to model these events using the background model which is not affected 
by mirrors as opposed to using the SPI particles. 

In order to estimate the effect of the background modeling on the temperature at 
large radii we compare the temperature correlation results, $A_1$ and $A_2$, 
for the simulated regular clusters with and without background. 
These are shown as filled circles in dark and light gray respectively in 
Figures \ref{Tcorr} and \ref{Tcorr2}. The inclusion of background leads to 
an overestimation of approximately 25\% in $A_1$ and 35\% in $A_2$. 
This is comparable to the 1-$\sigma$ statistical errors and since the 
effect appears to be relatively independent of redshift we conclude 
that it does not affect the evolution of the temperature correlation.

\subsection{Point-sources}
Point sources can significantly alter maps of cluster luminosity and temperature. 
In the cases where we can identify the point sources from the original event file 
we remove the flux from these sources using the technique described above (Section \ref{mapsec}, Figure \ref{psrmfig}). 
When a point source cannot be resolved it can have a large impact 
on the temperature maps even if the effect on the luminosity map is not 
visible. Ideally, Chandra data of the same cluster should be used in combination 
to identify the positions of point sources and to verify the accuracy of temperature 
features. Chandra data is however not available for all clusters in the sample 
and this type of analysis is beyond the scope of this work. 

\subsection{Temperature bias}
\label{tempbias}
As mentioned earlier, the average temperature from the SPI modeling becomes 
slightly higher than when standard analysis routines are applied. 
This effect stems from the fact that we are using multiple particles 
in our modeling.
Since the particles are extended and hundreds of particles are required 
to model a single cluster there will be multiple overlapping cluster 
particles at any given position. 
Each particle has a separate temperature and set of elemental abundances and 
this means that the model is inherently multi-phase. There are many combinations 
of thermal spectra that can be constructed to agree with the observed data and the 
final model consists of a wide range of phases at each spatial position. 
This range is very much dependent on the prior range for the parameter in 
question. Here, these ranges are visualized by making use of the median of the 
range at each spatial coordinate and this value again depends on the prior range. 
Since the measured differences of thermal spectra in XMM become smaller with increasing 
temperature the median is naturally biased towards the higher part of the prior. 
This may lead to a potential luminosity bias in the bolometric correction 
since the luminosities are calculated separately for each smoothed particle 
which will cause a higher luminosity if the temperature is higher. We estimate 
this offset to be less than $\sim 20$ \% based on a comparison when using 
the temperature from the ``standard'' analysis in the bolometric correction. 

\subsection{Modeling artifacts}
Since the modeling itself is based on particles, even if between 100 and 1000 
particles are required, there is a possibility for particle artifacts appearing 
in the luminosity and temperature maps. The cluster particles are constantly 
moving within the parameter space and interchanging positions with each other, 
and averaging over many Markov chain samples reduces this effect. 
However, since we are limited by CPU resources (see also Section \ref{uncertain}) 
and only 250 samples are used in the error calculation these artifacts may 
still show up in the maps as lumpy structures and this may have a small effect 
increasing the power ratios.
In comparing power ratios calculated from maps using 1500 samples 
with ones generated using 250 samples we establish that this effect 
corresponds to about a 20 \% increase for ratios ($P_2/P_0$ or $P_3/P_0$) 
between $10^{-8}$ and $10^{-7}$ and less for larger values. 
However, this is not sufficient to explain the discrepancy 
in the comparison with the values from \citet{jeltema05}.

\subsection{Trends of the simulated clusters}
In many of the plots, the trends in the relevant parameters with $z$ 
for the simulated cool core clusters appear similar to the evolution for the sample 
as a whole. In most cases, the trend for the simulations only represents 
the minimum disturbance that can be measured and the similarity does not 
necessarily mean that the evolution is not real. In fact, we see from 
our simulations of irregular clusters that the trend for a cluster with 
high $P_2/P_0$ or $P_3/P_0$ is much more stable than that of the circularly 
symmetric clusters. 
Ideally, to make a more rigorous claim of evolution, clusters of many different 
morphologies should be simulated and the apparent evolution analyzed 
separately. This will be the subject of future work and we are limited here, 
again, by CPU constraints. Markov chains should also be run for several more 
iterations to limit model noise.

\section{Discussion and conclusions}
We have applied a novel technique for X-ray galaxy cluster modeling to a large 
sample of XMM observed galaxy clusters in order to identify clusters with 
cooling cores and to study the redshift dependence 
of cluster substructure and the luminosity-temperature relation. 
We also developed a new statistic to quantify thermal substructure and to 
identify and distinguish cooling core 
clusters from isothermal and thermally disturbed clusters. We note here that 
while this technique is important in identifying clusters for which mass measurement 
can be used for cosmological applications, in its present form 
it is not capable of estimating cluster masses. However, a straightforward 
extension of it, 
with additional assumptions, should allow the determination 
of three-dimensional density and temperature of the cluster, and thus the mass of 
both gas and dark matter content.  

The number of cooling core clusters appears to be lower at high $z$ with 
32 \% identified as cooling clusters at $z<0.1$, 36 \% at $0.1\le z <0.3$ and 
11 \% at $z\ge 0.3$. This suggests that the formation of cooling cores in 
clusters is a recent phenomenon. However, due to the low number of clusters 
and various selection effects these percentages are uncertain. 

\subsection{L-T relation} 
The derived luminosity-temperature relation for clusters clearly shows a 
large offset for cooling core clusters, as expected. 
This highlights the importance 
of applying a redshift dependent criterion when excluding these sources. 
In contrast to previous work we do not attempt to correct for cooling cores 
by applying some standardized central profile. Instead we fit the relation 
for cooling clusters separately \citep[cf.][]{allen98} and find a similar 
slope of $\sim 3$ but with a normalization $\sim 3$ times higher 
than for the non-cooling sample. 
The normalization of our $L_X$-$T$ relation for cool core clusters is 
$\sim 50 \%$ higher for our sample compared to the findings of \citet{allen98}. 
This is likely due to the use of ROSAT and Asca data to determine temperatures 
and luminosities, where cool core clusters would hard to model with the 
low spatial resolution and the limited energy band.

We also note that the scatter in the $L_X$-$T$ relation is greater 
for cooling core clusters. 
We detect weak evolution in the non-cooling $L_X - T$ relation 
$\propto (1+z)^{0.50\pm0.34}$ in good agreement with self similar predictions 
and with an average slope of $\alpha_{LT}=2.87\pm0.14$. 
This is in contrast to the findings of \citet{vikh02} who detect strong evolution 
$\propto (1+z)^{1.5 \pm 0.3}$.
Due to the limited observable volume available 
at low $z$, high luminosity clusters are observed at mostly high redshifts. 
This could create  a $\alpha_{LT}$-$\beta_{LT}$ bias. 
We note also that when studying $\beta_{LT}$ for the cluster sample with 
$P_3/P_0 > 10^{-7}$, the sample shows significant evolution. This may be 
an important effect to account for in large surveys relying on the 
$L_X$-$T$ relation to measure the mass function of clusters over a range 
in $z$. Clearly, the $L_X$-$T$ relation has to be measured in greater 
detail with larger samples of well exposed clusters in order to 
reach a consensus of its evolution.
To measure the mass function of clusters to constrain cosmological parameters 
the knowledge of this evolution is crucial. 

\subsection{Structure quantification}
The power ratio analysis was applied uniformly for the 101 clusters within 
a 500 kpc radius. We find no significant increase in $P_2/P_0$ with $z$ but 
detect an increase in $P_3/P_0$ implying a higher level of substructure 
at high redshift. Our sample of cooling core clusters shows significantly 
lower values of $P_2/P_0$ and $P_3/P_0$ than the non-cooling sample, 
which is not surprising since cooling clusters generally are the most dynamically 
relaxed clusters. 

\begin{figure*}[!htb]
  \begin{center}
    \begin{tabular}{cc}
      \includegraphics[width=2in]{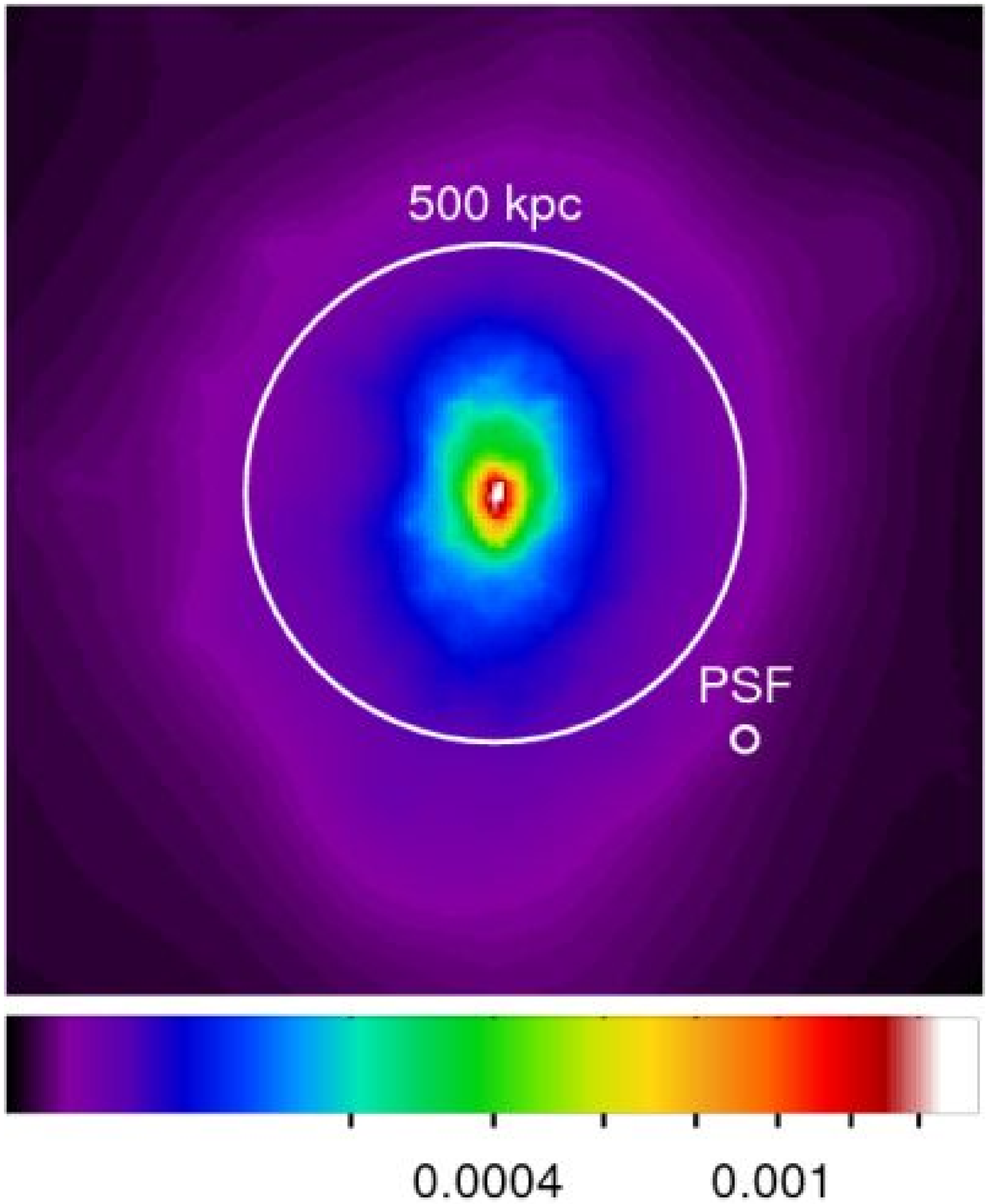} & 
      \includegraphics[width=2in]{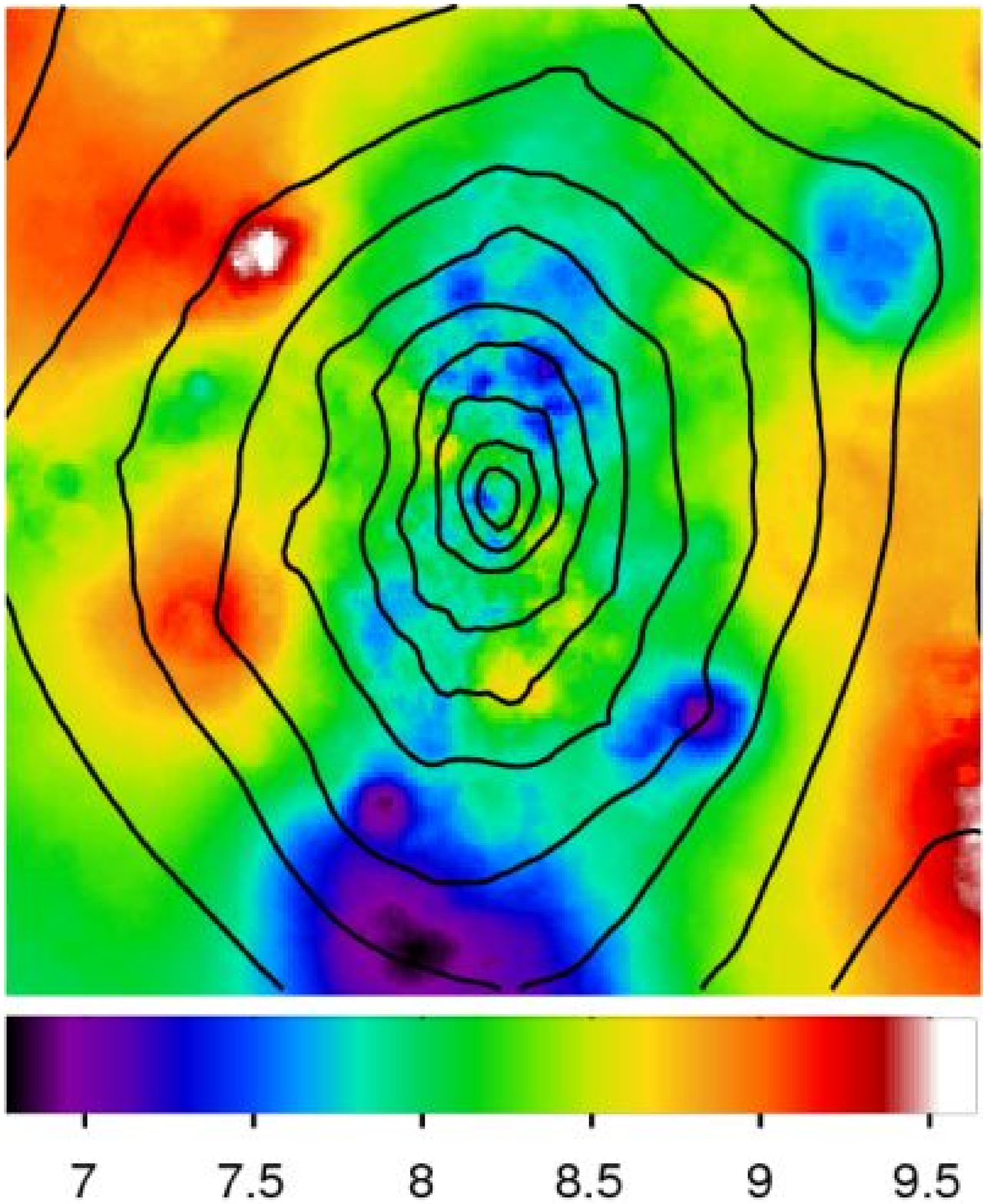} \\
    \end{tabular}
  \end{center}
  \caption{$2$ Mpc $ \times 2$ Mpc field showing luminosity (left) and temperature (right) maps of Abell 1413. \label{A1413}}
\end{figure*}

\begin{figure*}[!htb]
  \begin{center}
    \begin{tabular}{cc}
      \includegraphics[width=2in]{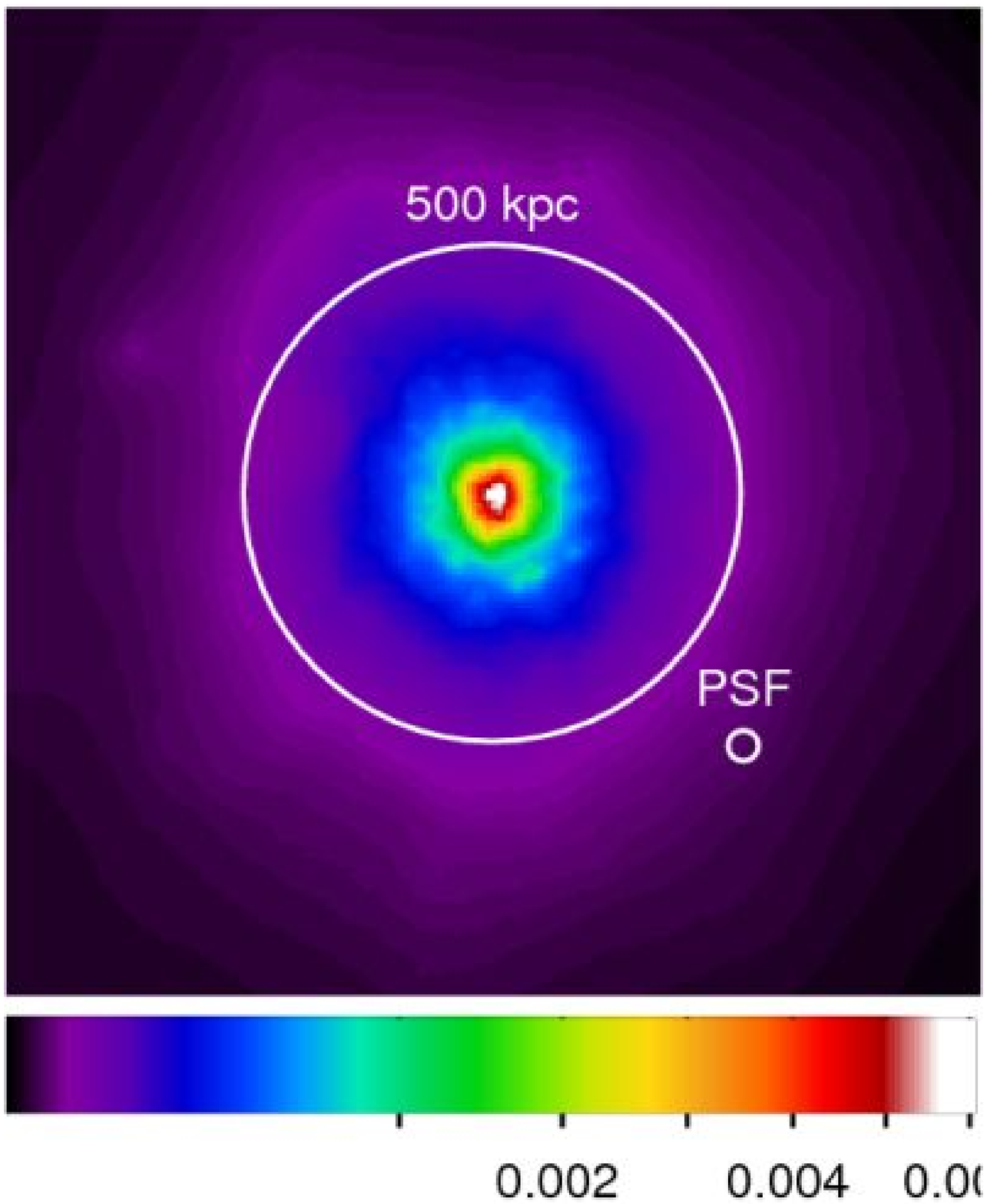} & 
      \includegraphics[width=2in]{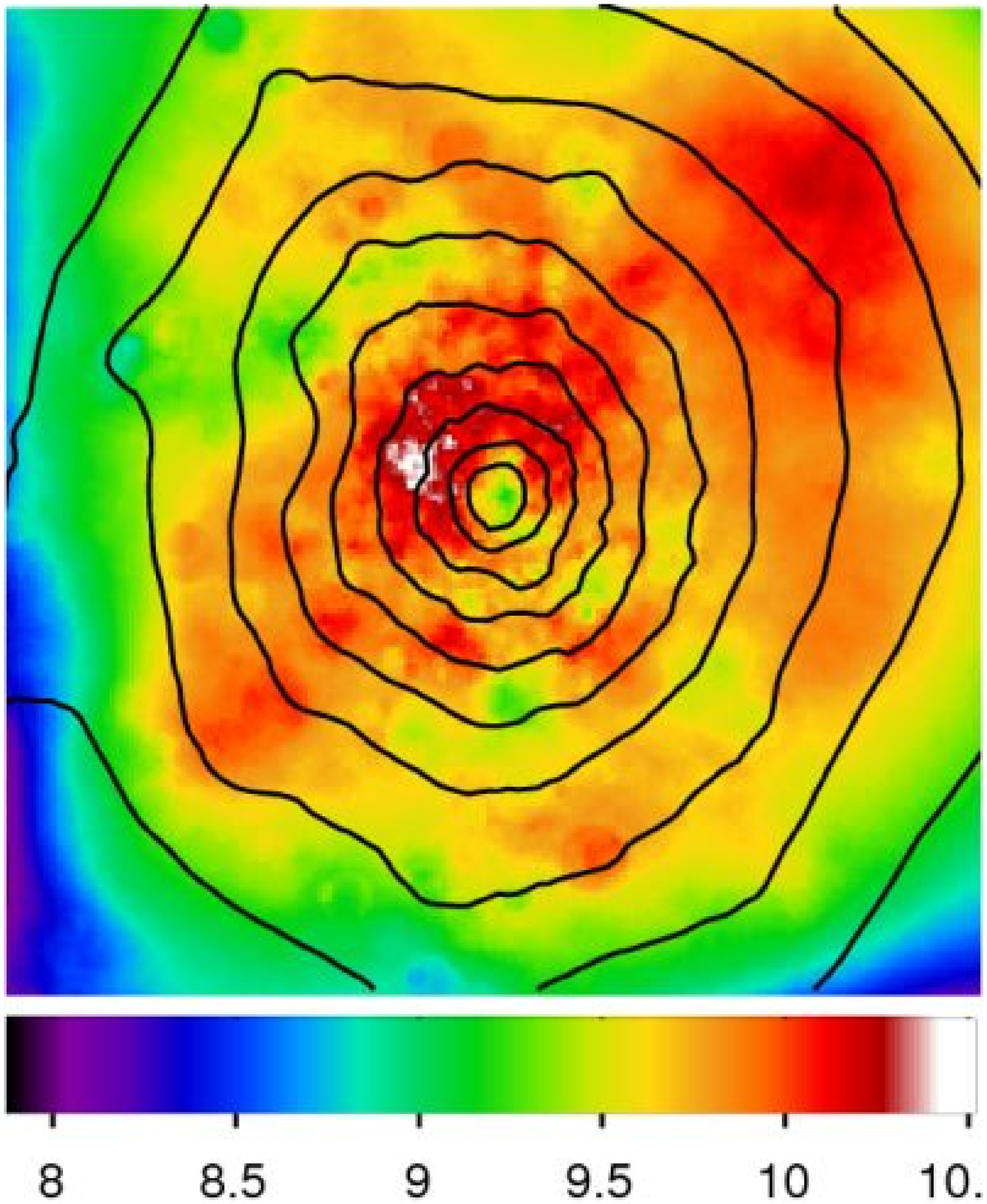} \\
    \end{tabular}
  \end{center}
  \caption{$2$ Mpc $ \times 2$ Mpc field showing luminosity (left) and temperature (right) maps of Abell 1689. \label{A1689}}
\end{figure*}

\begin{figure*}[!htb]
  \begin{center}
    \begin{tabular}{cc}
      \includegraphics[width=2in]{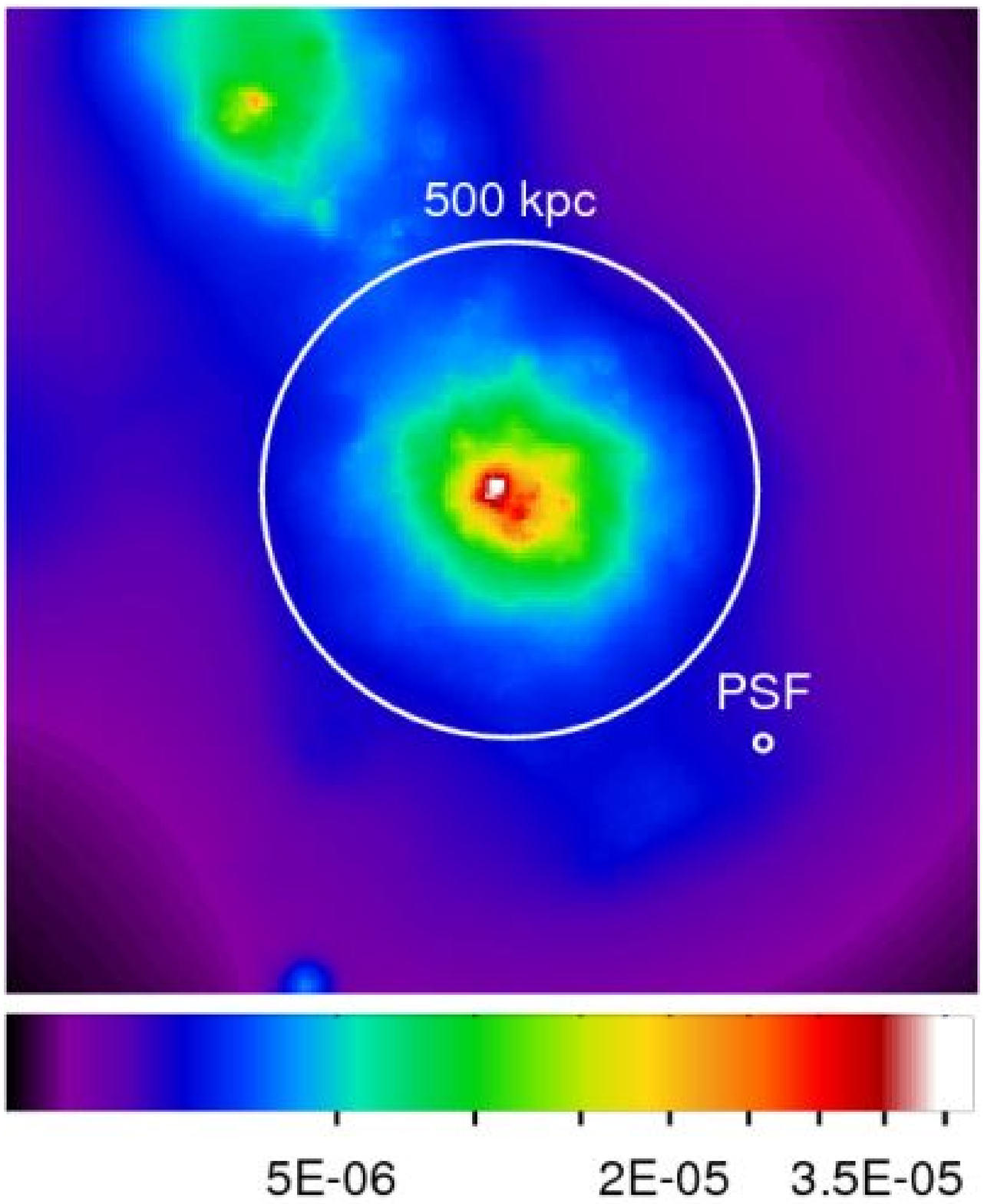} & 
      \includegraphics[width=2in]{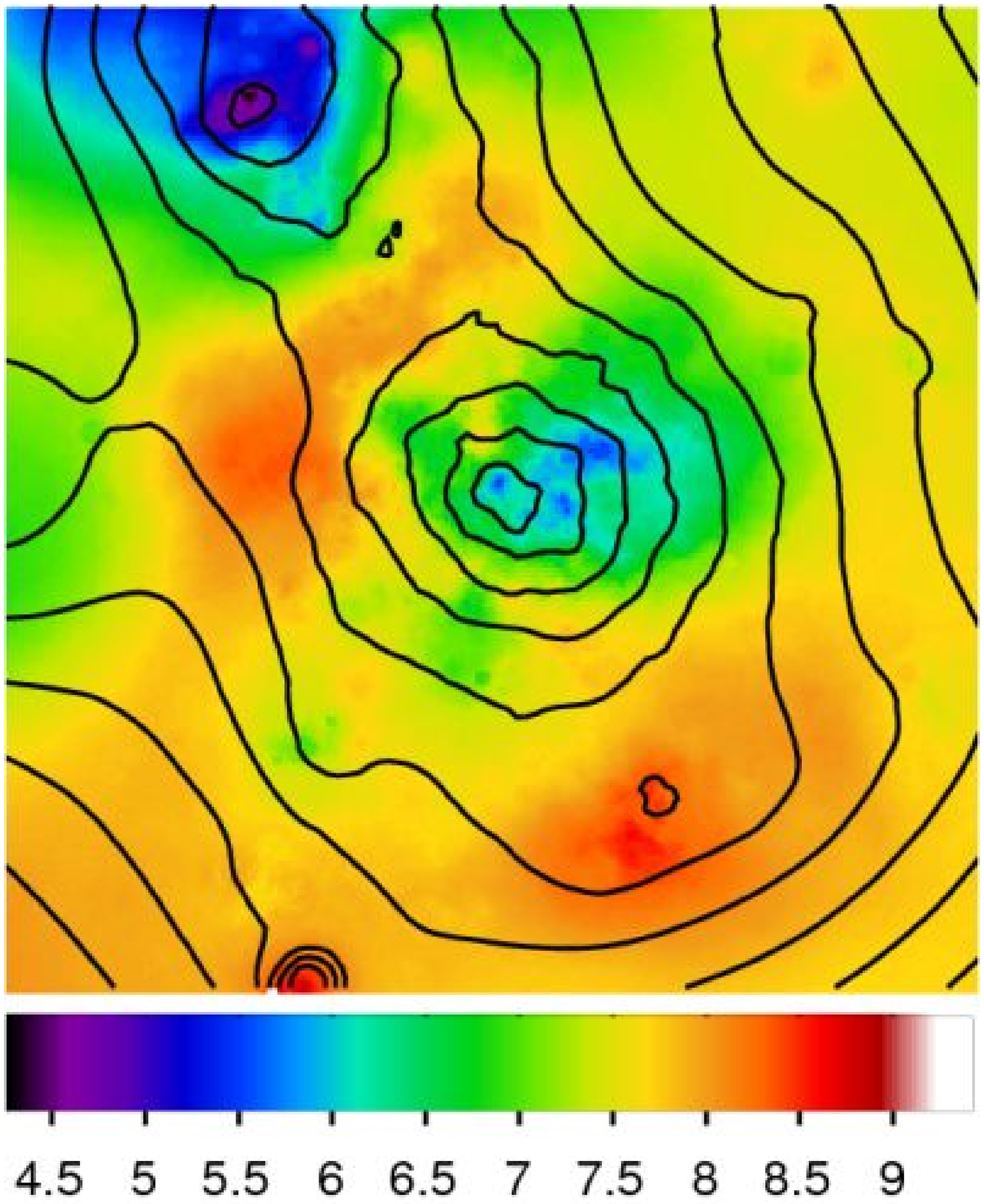} \\
    \end{tabular}
  \end{center}
  \caption{$2$ Mpc $ \times 2$ Mpc field showing luminosity (left) and temperature (right) maps of Abell 1750. \label{A1750}}
\end{figure*}

\begin{figure*}[!htb]
  \begin{center}
    \begin{tabular}{cc}
      \includegraphics[width=2in]{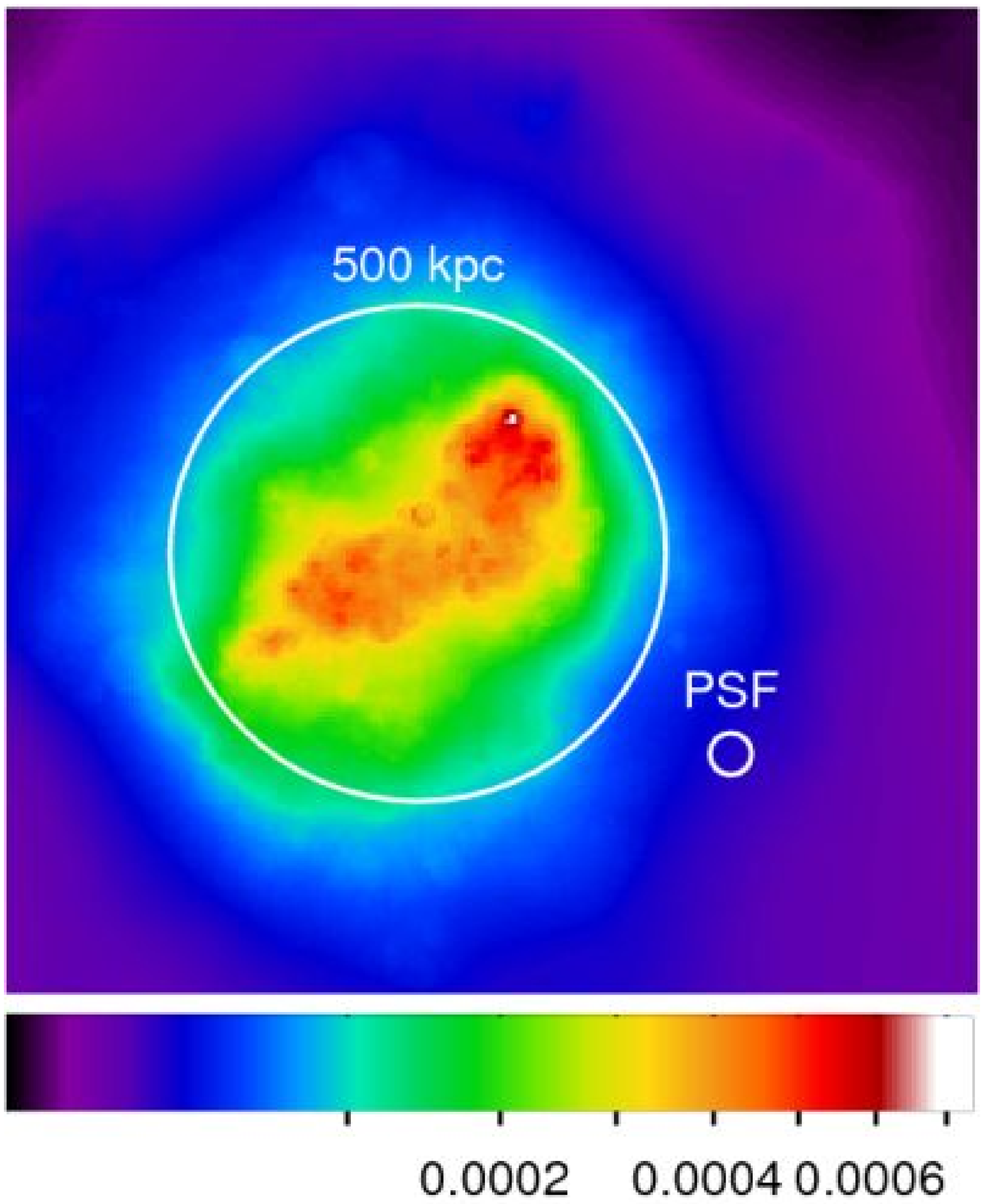} & 
      \includegraphics[width=2in]{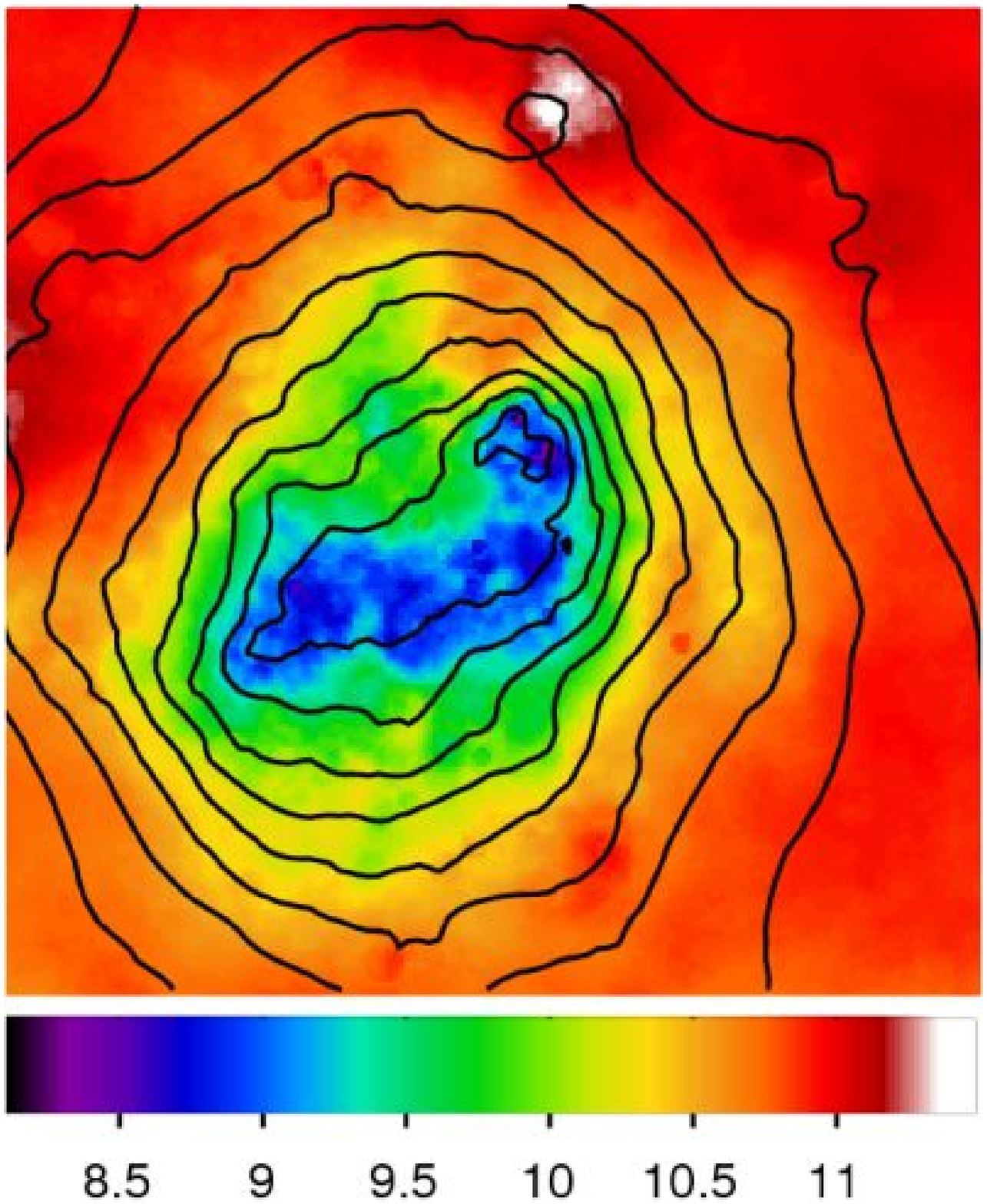} \\
    \end{tabular}
  \end{center}
  \caption{$2$ Mpc $ \times 2$ Mpc field showing luminosity (left) and temperature (right) maps of Abell 1758. \label{A1758}}
\end{figure*}

\begin{figure*}[!htb]
  \begin{center}
    \begin{tabular}{cc}
      \includegraphics[width=2in]{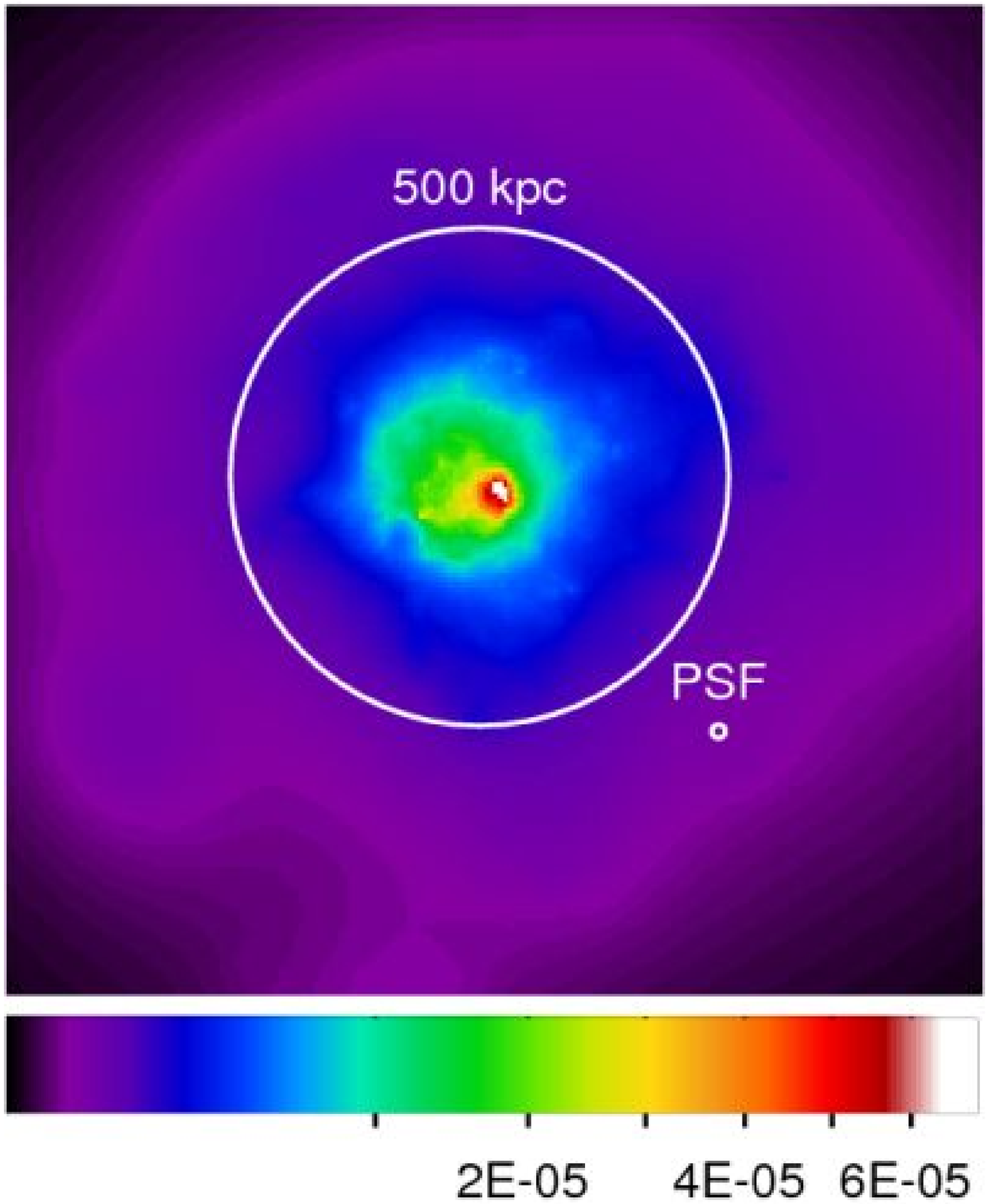} & 
      \includegraphics[width=2in]{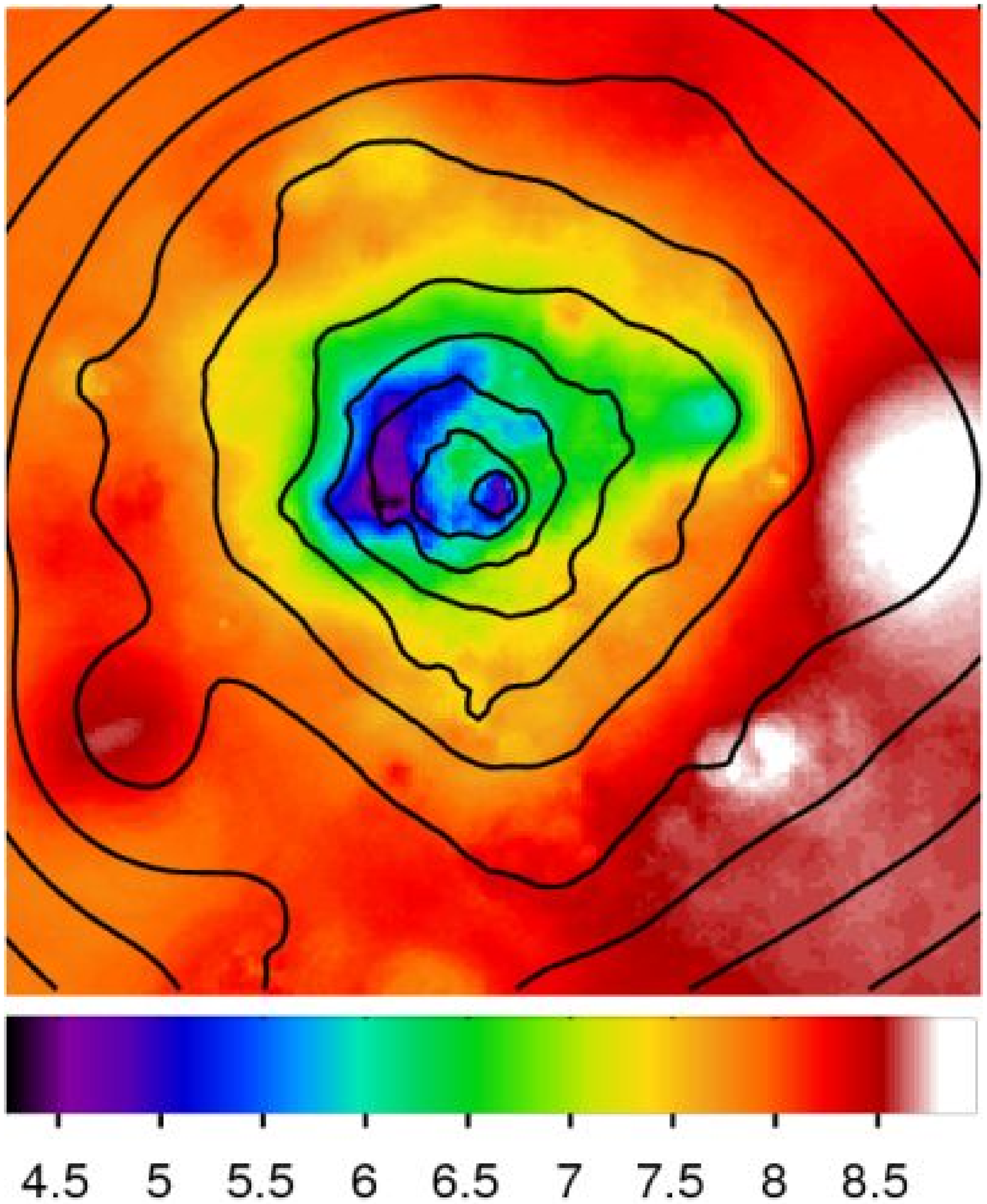} \\
    \end{tabular}
  \end{center}
  \caption{$2$ Mpc $ \times 2$ Mpc field showing luminosity (left) and temperature (right) maps of Abell 1775. \label{A1775}}
\end{figure*}

\begin{figure*}[!htb]
  \begin{center}
    \begin{tabular}{cc}
      \includegraphics[width=2in]{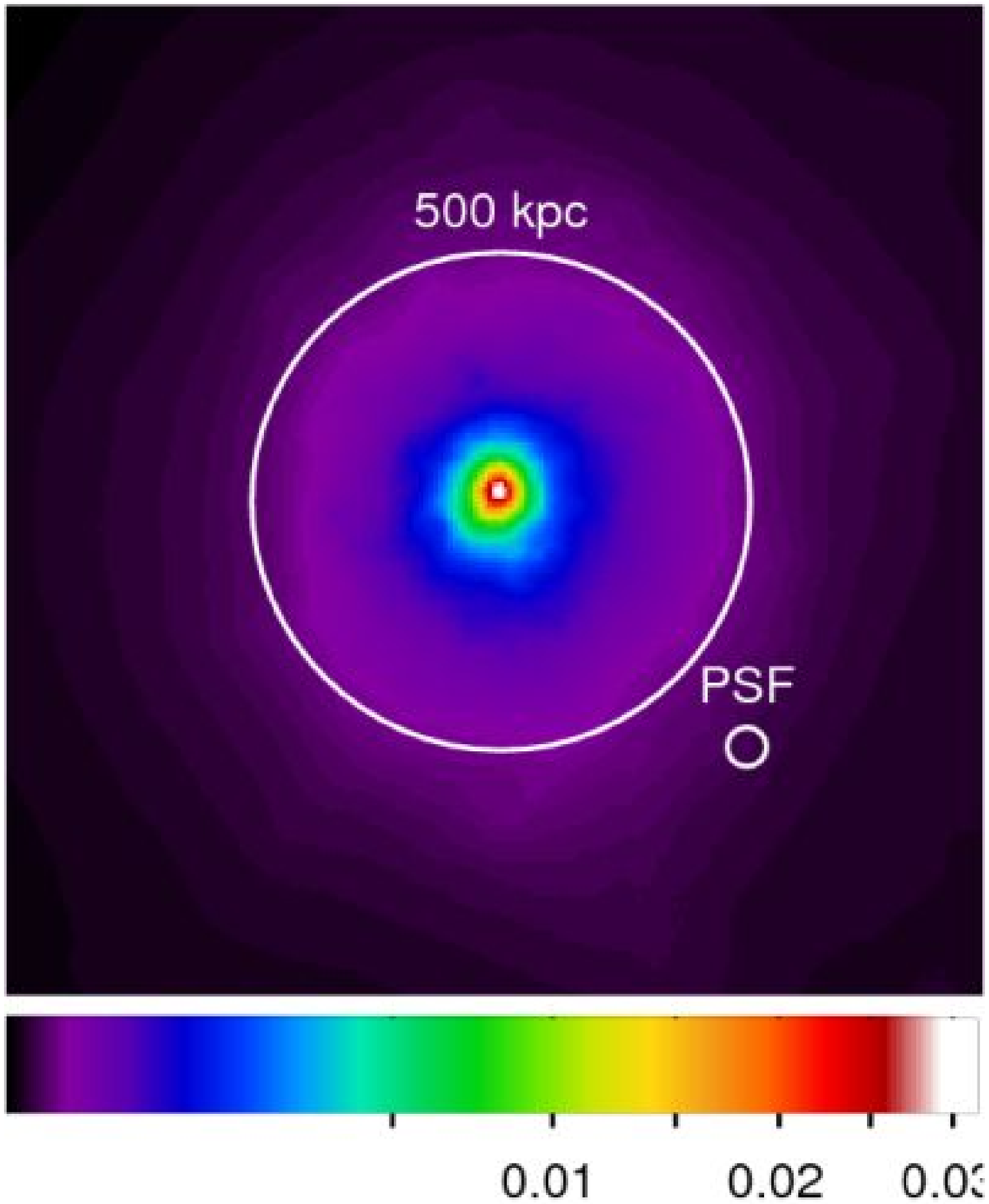} & 
      \includegraphics[width=2in]{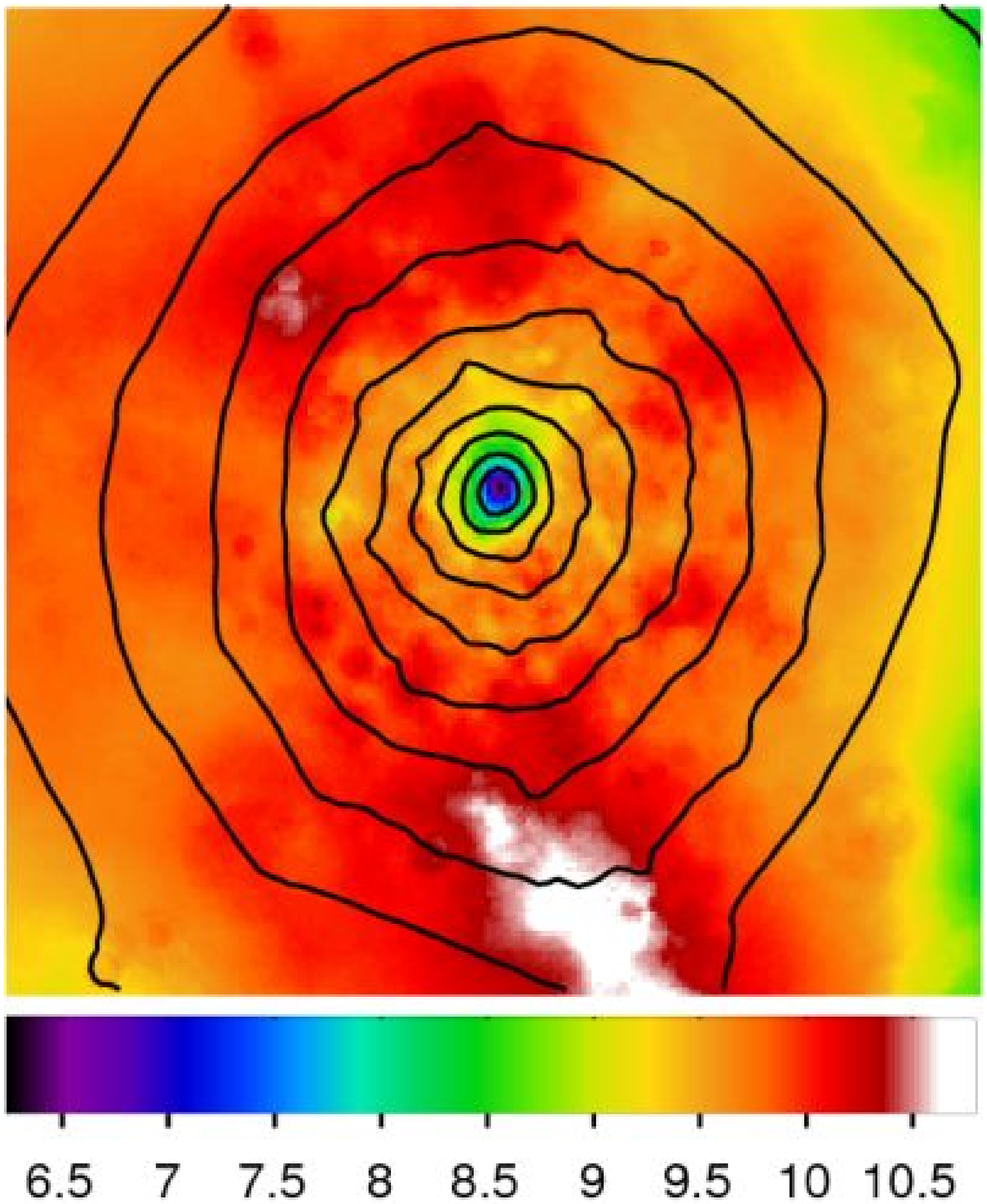} \\
    \end{tabular}
  \end{center}
  \caption{$2$ Mpc $ \times 2$ Mpc field showing luminosity (left) and temperature (right) maps of Abell 1835. \label{A1835}}
\end{figure*}

\begin{figure*}[!htb]
  \begin{center}
    \begin{tabular}{cc}
      \includegraphics[width=2in]{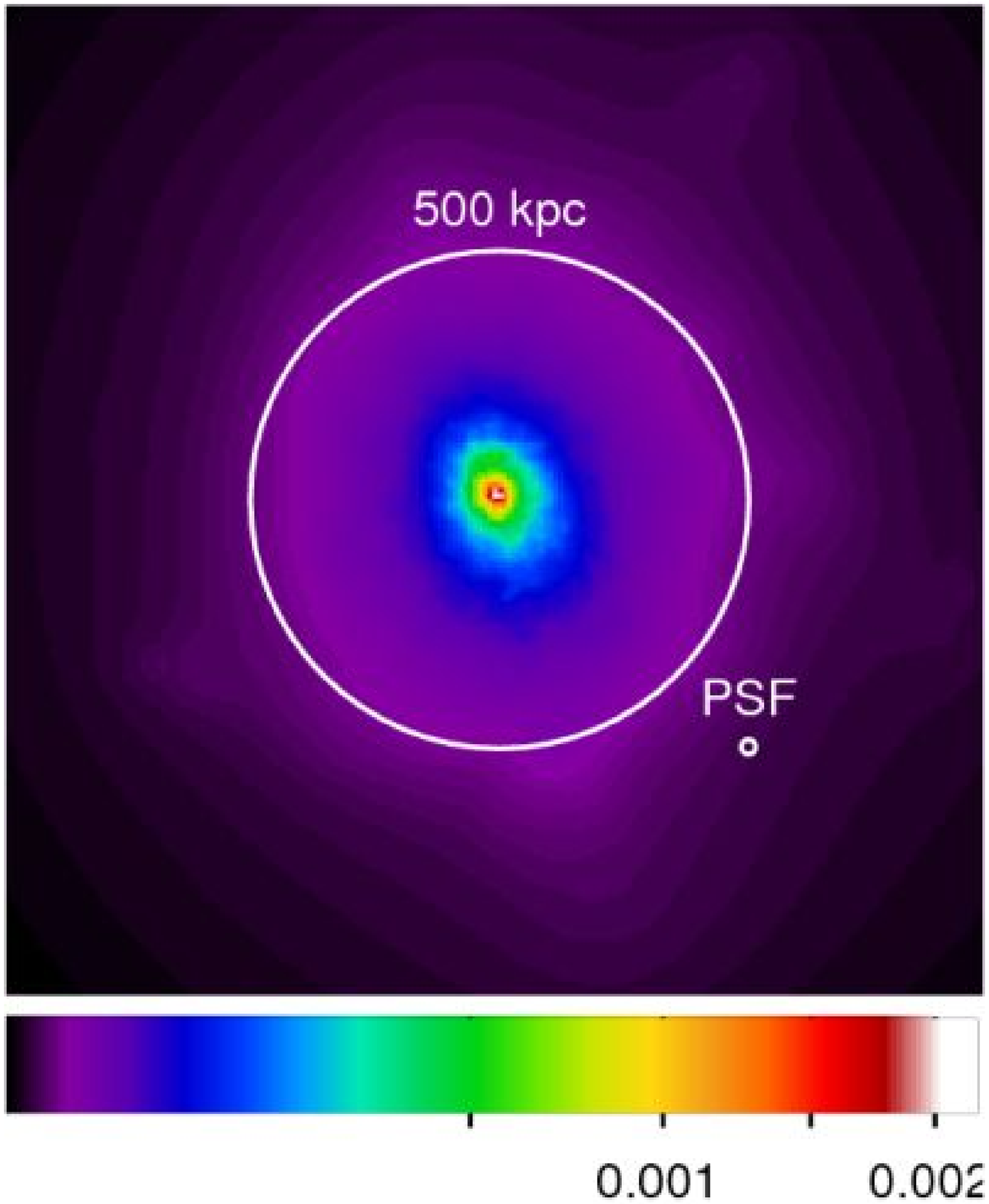} & 
      \includegraphics[width=2in]{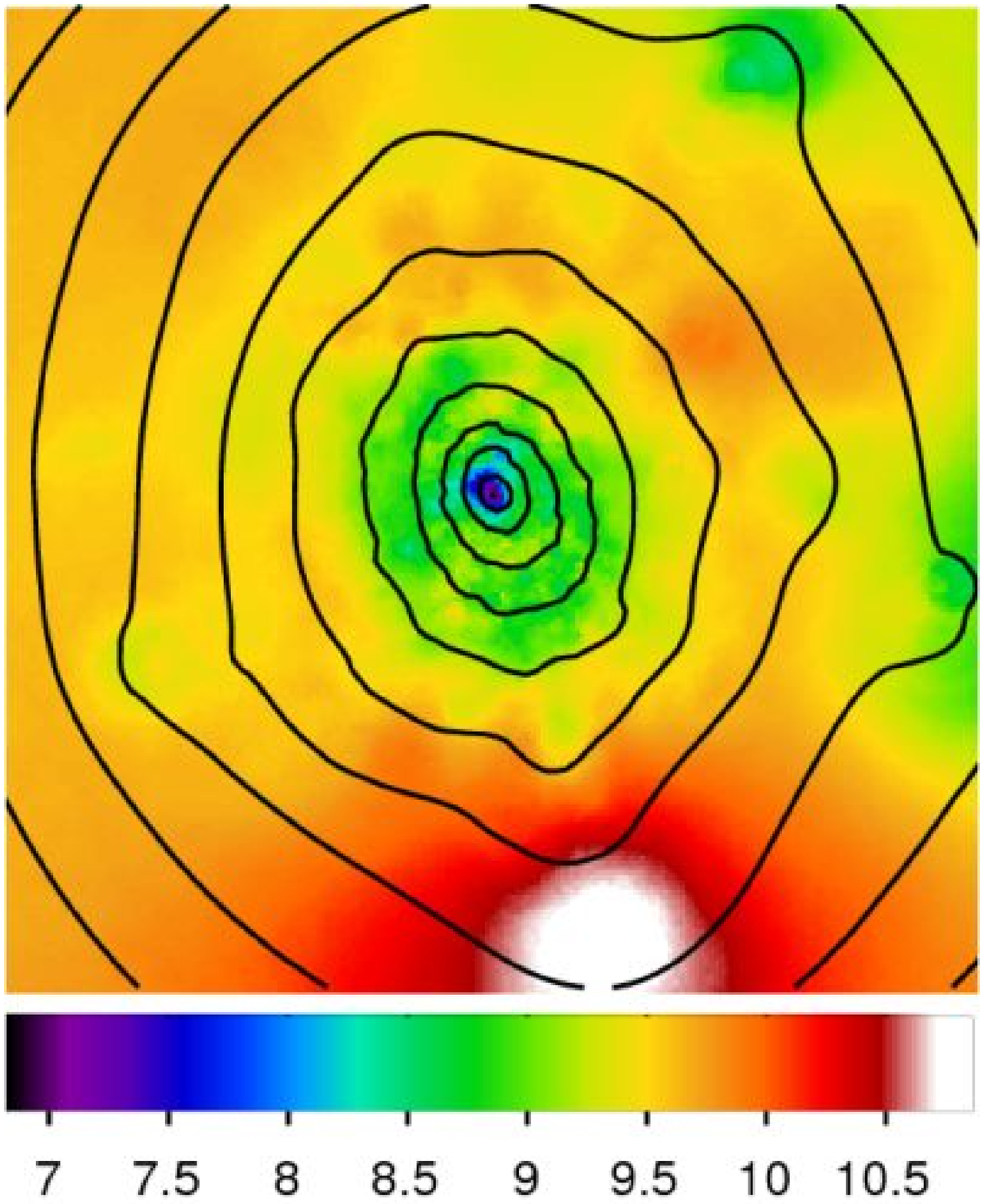} \\
    \end{tabular}
  \end{center}
  \caption{$2$ Mpc $ \times 2$ Mpc field showing luminosity (left) and temperature (right) maps of Abell 2029. \label{A2029}}
\end{figure*}

\begin{figure*}[!htb]
  \begin{center}
    \begin{tabular}{cc}
      \includegraphics[width=2in]{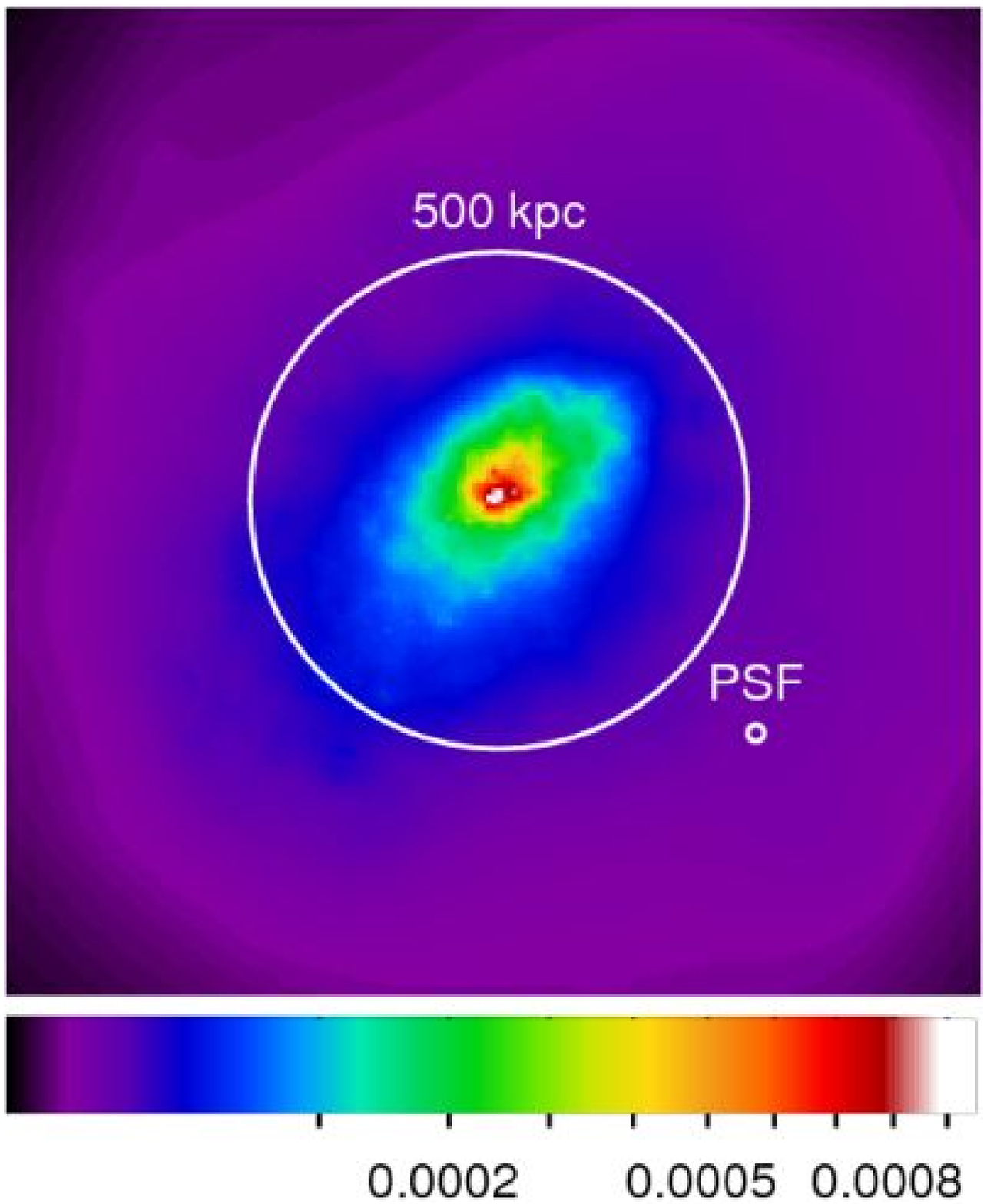} & 
      \includegraphics[width=2in]{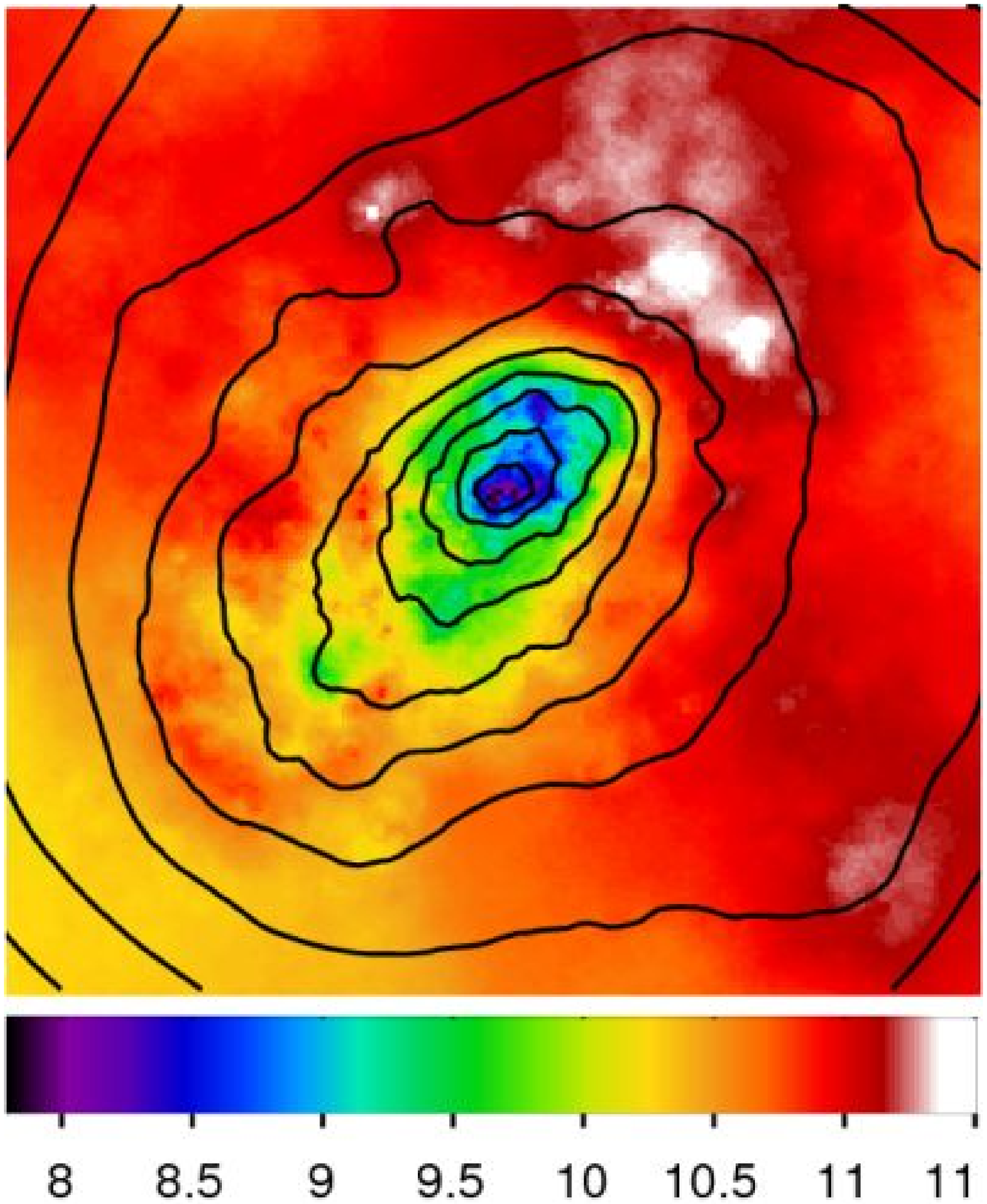} \\
    \end{tabular}
  \end{center}
  \caption{$2$ Mpc $ \times 2$ Mpc field showing luminosity (left) and temperature (right) maps of Abell 2142. \label{A2142}}
\end{figure*}

\begin{figure*}[!htb]
  \begin{center}
    \begin{tabular}{cc}
      \includegraphics[width=2in]{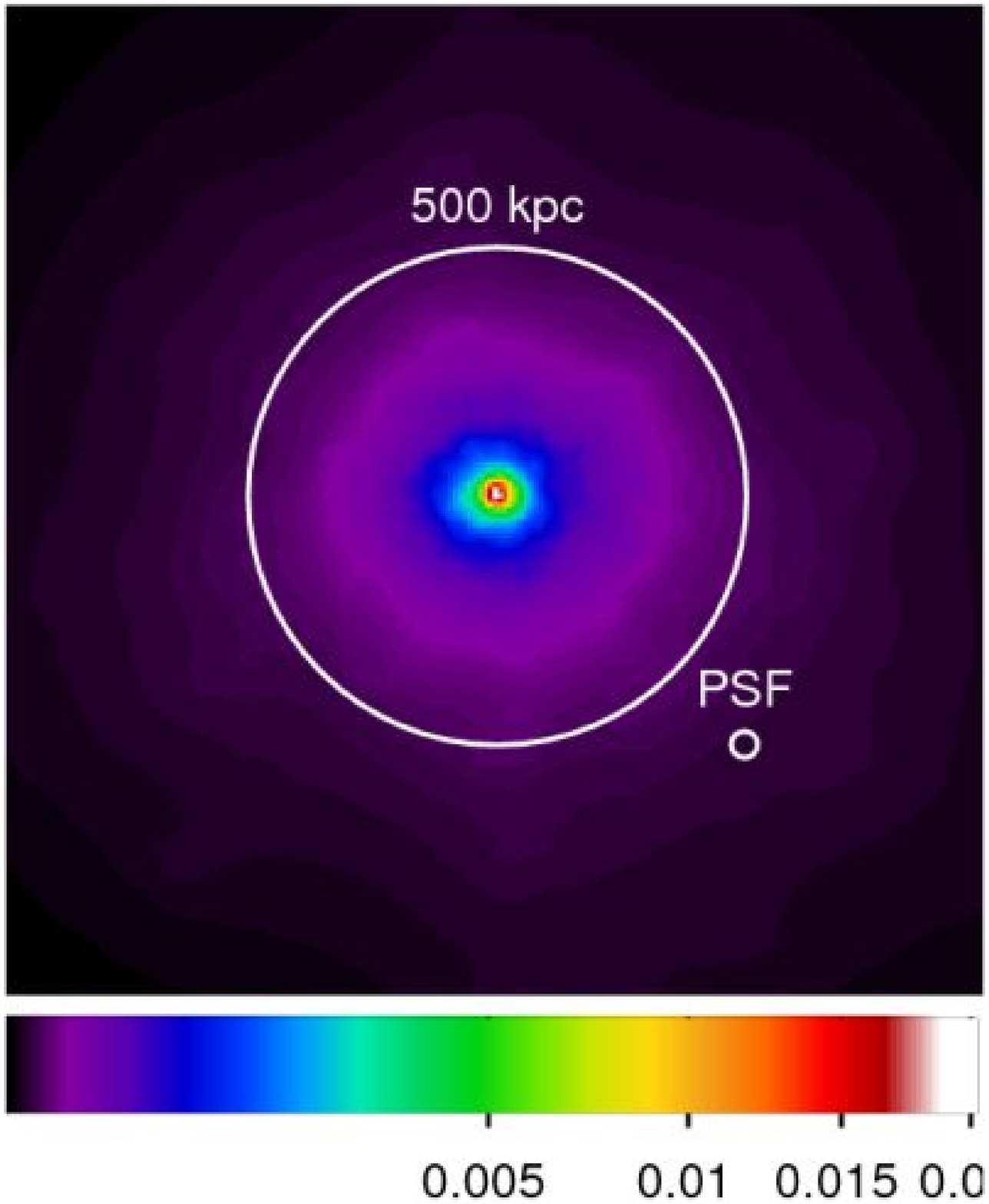} & 
      \includegraphics[width=2in]{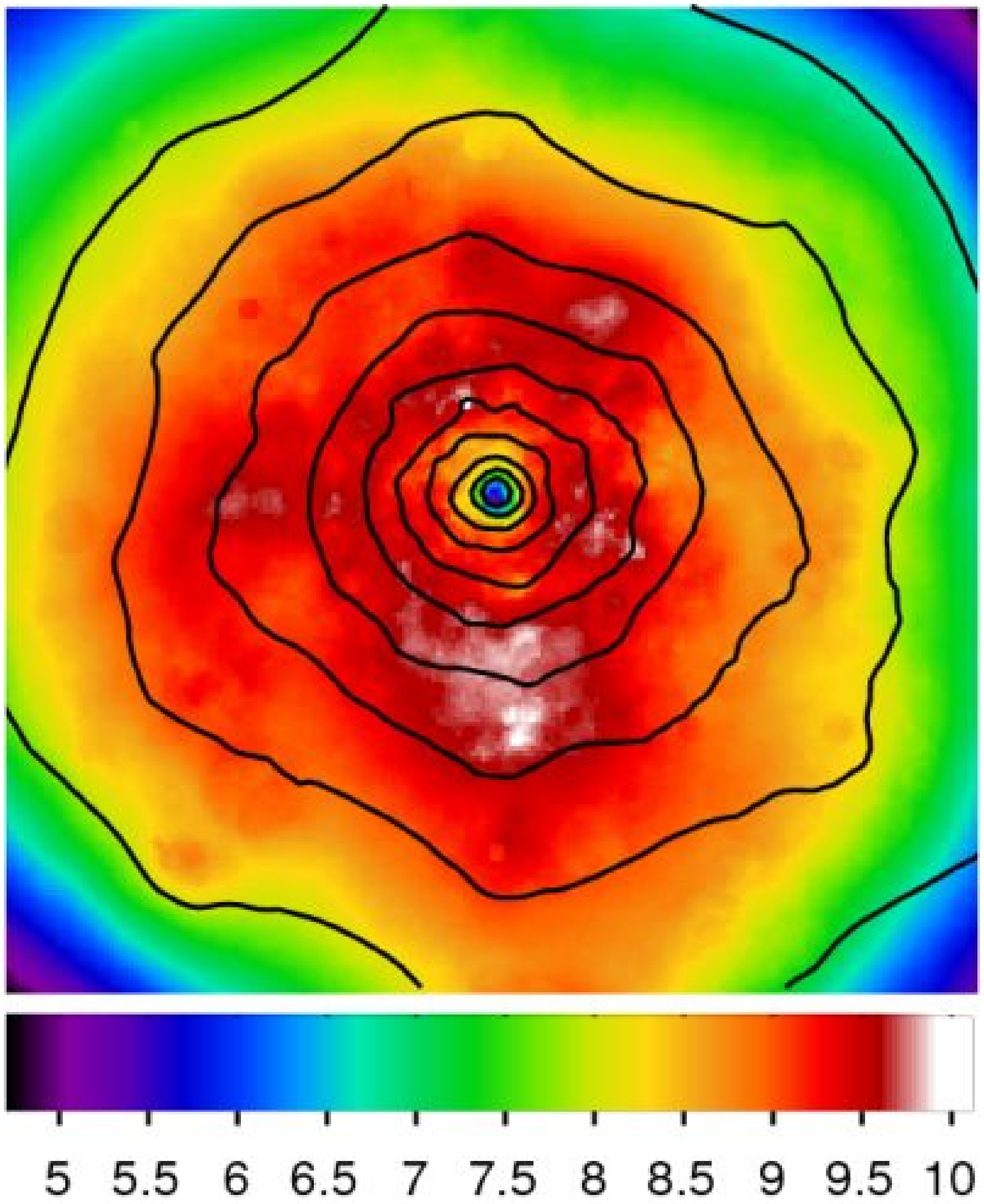} \\
    \end{tabular}
  \end{center}
  \caption{$2$ Mpc $ \times 2$ Mpc field showing luminosity (left) and temperature (right) maps of Abell 2204. \label{A2204}}
\end{figure*}

\begin{figure*}[!htb]
  \begin{center}
    \begin{tabular}{cc}
      \includegraphics[width=2in]{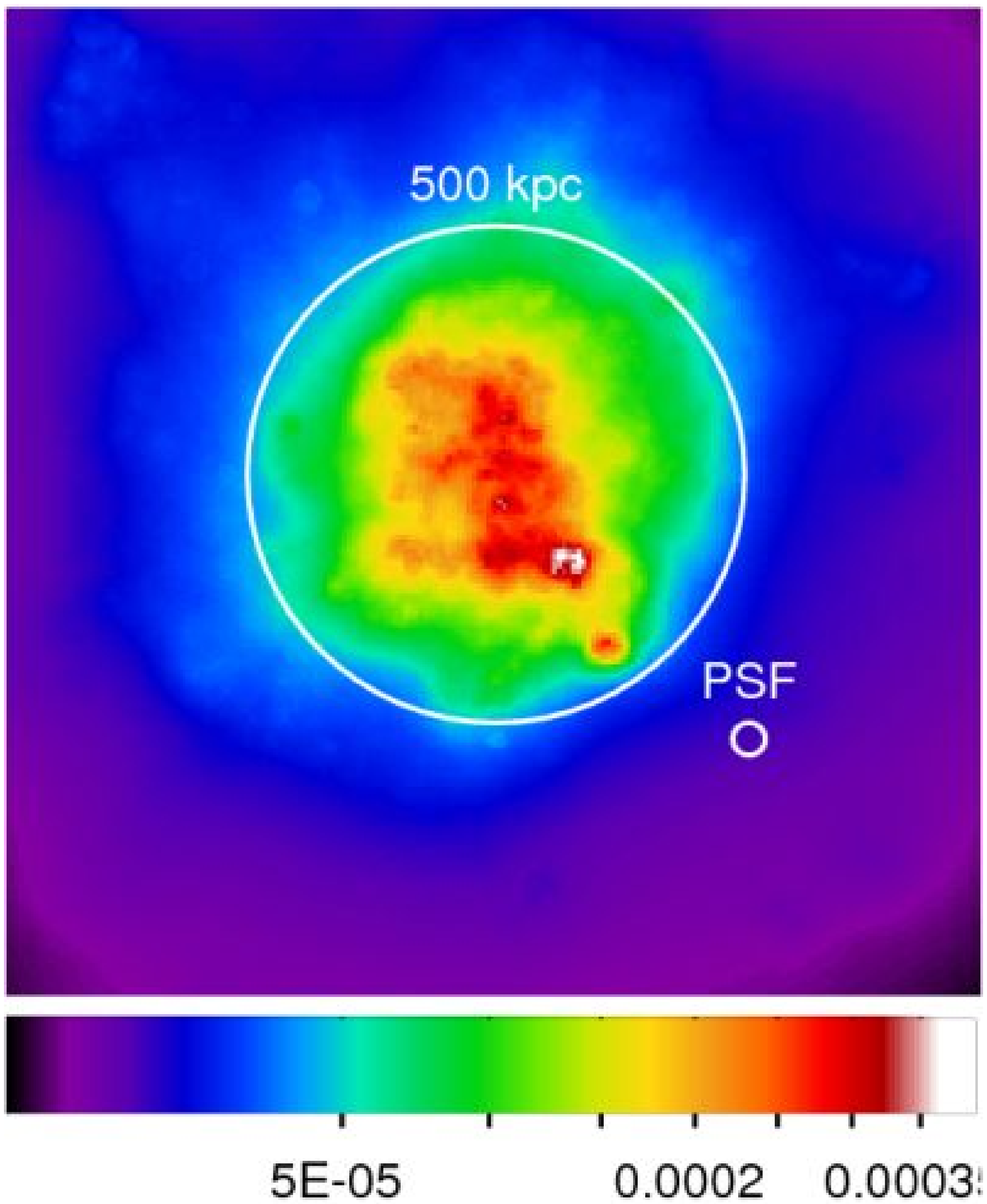} & 
      \includegraphics[width=2in]{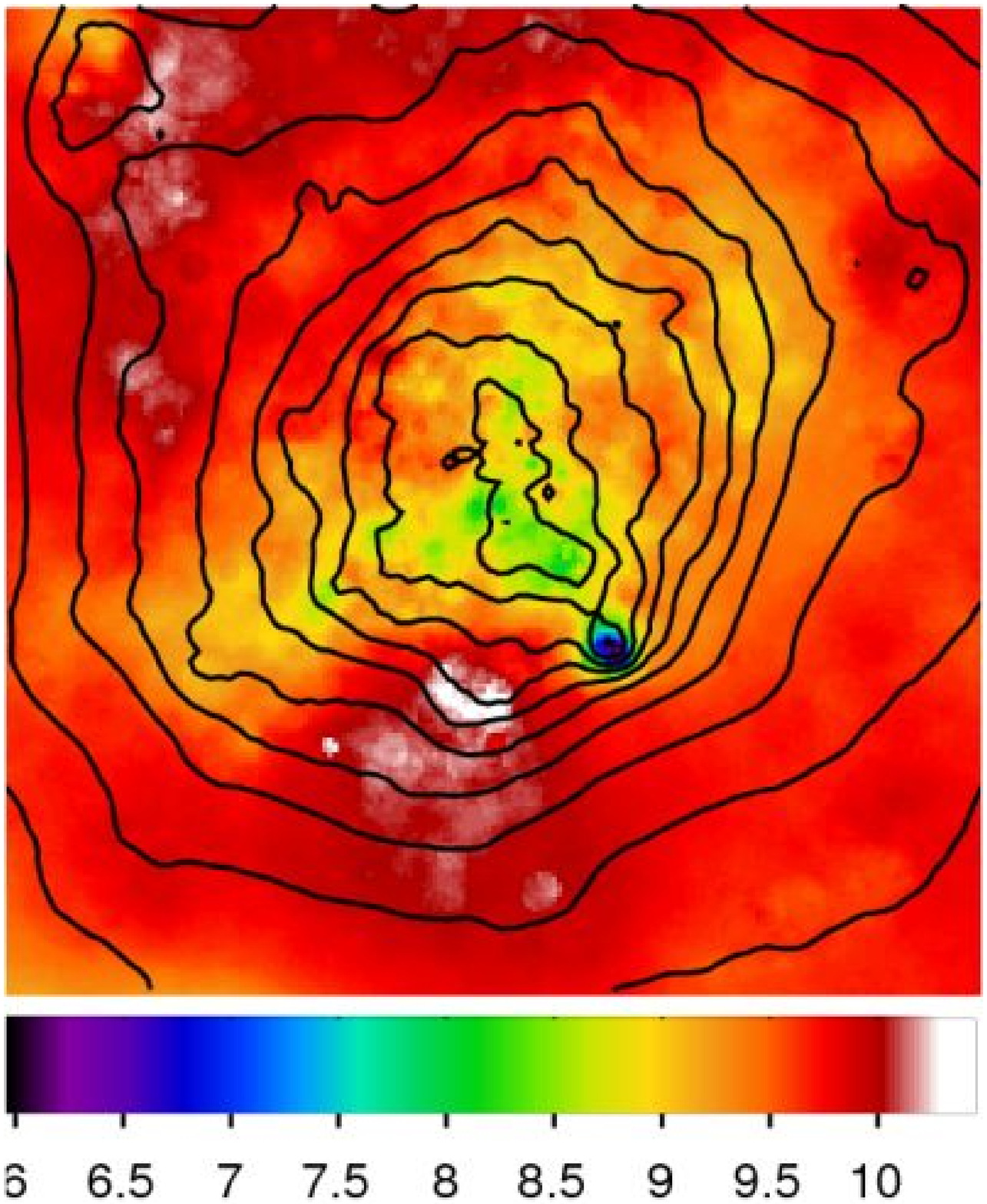} \\
    \end{tabular}
  \end{center}
  \caption{$2$ Mpc $ \times 2$ Mpc field showing luminosity (left) and temperature (right) maps of Abell 520. \label{A520}}
\end{figure*}

\begin{figure*}[!htb]
  \begin{center}
    \begin{tabular}{cc}
      \includegraphics[width=2in]{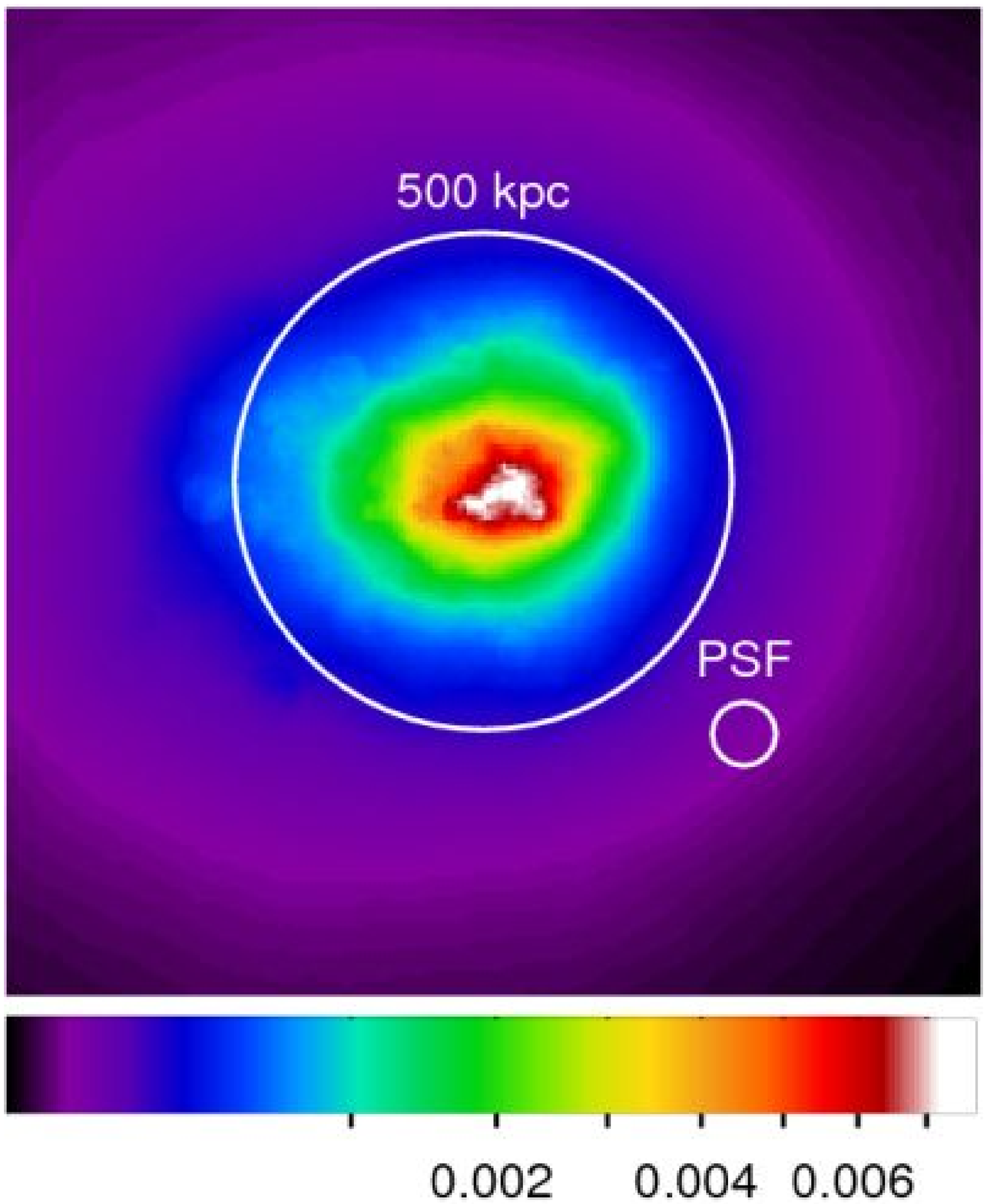} & 
      \includegraphics[width=2in]{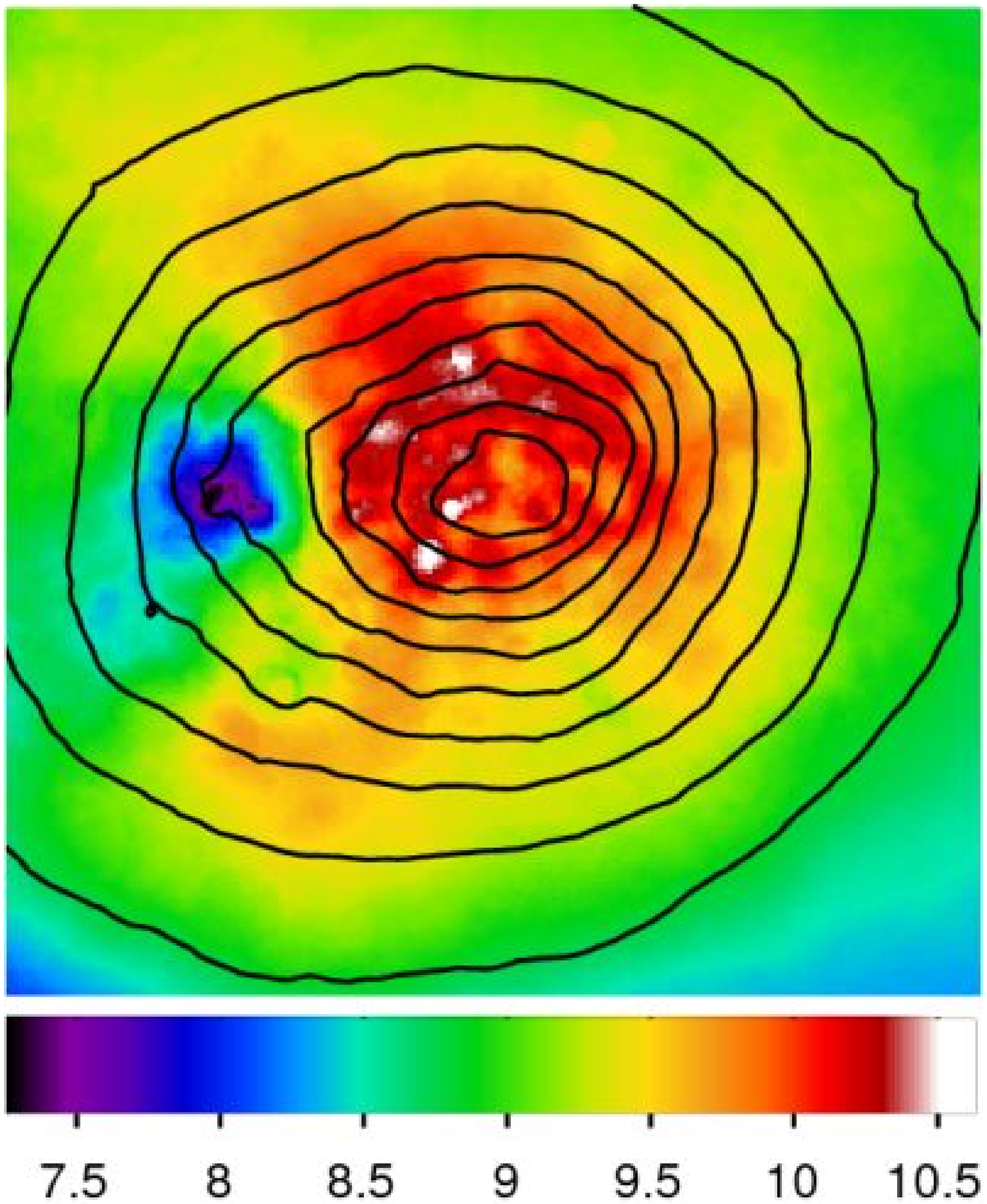} \\
    \end{tabular}
  \end{center}
  \caption{$2$ Mpc $ \times 2$ Mpc field showing luminosity (left) and temperature (right) maps of MS0451.6-0305. \label{MS0451.6-0305}}
\end{figure*}

\begin{figure*}[!htb]
  \begin{center}
    \begin{tabular}{cc}
      \includegraphics[width=2in]{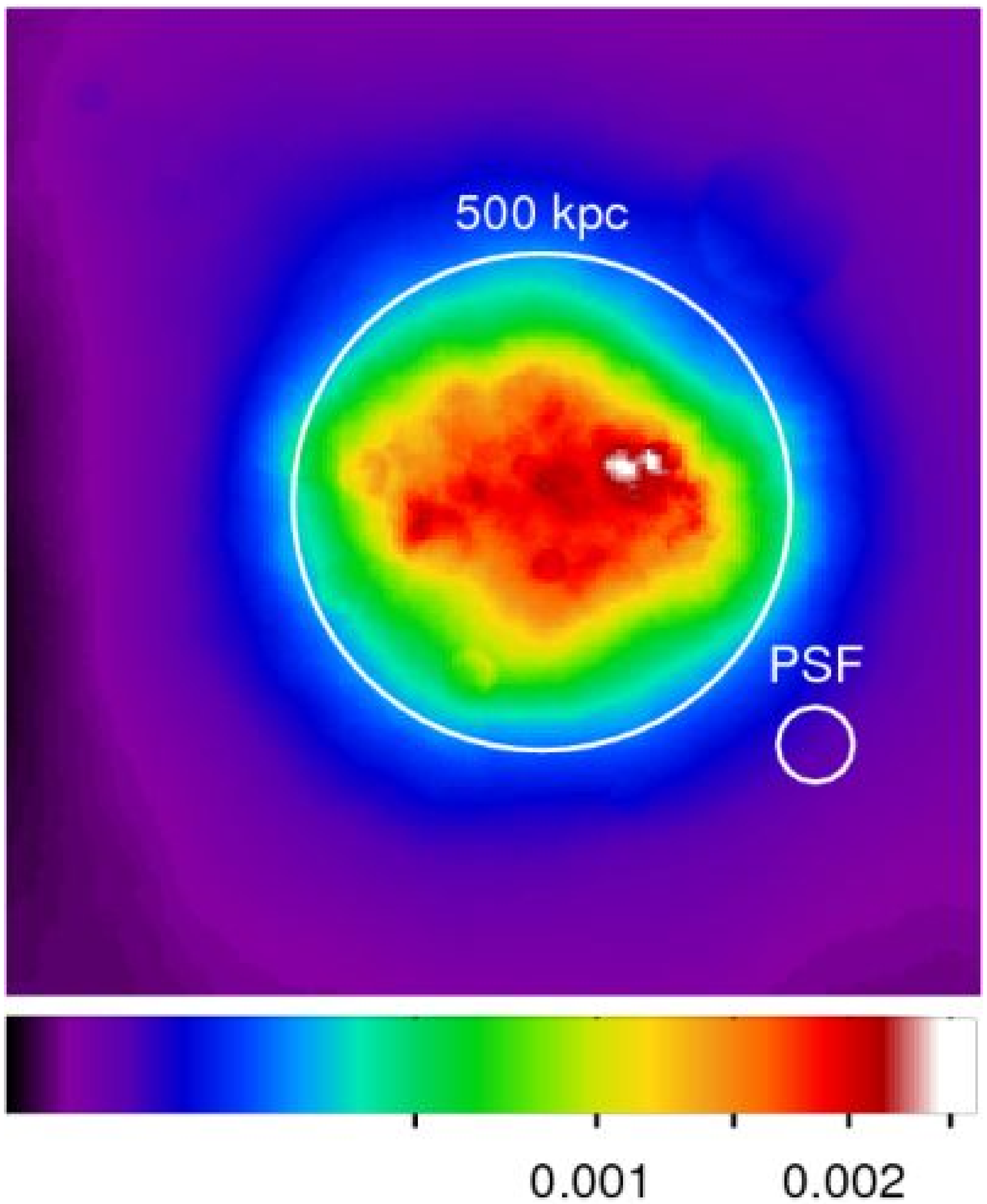} & 
      \includegraphics[width=2in]{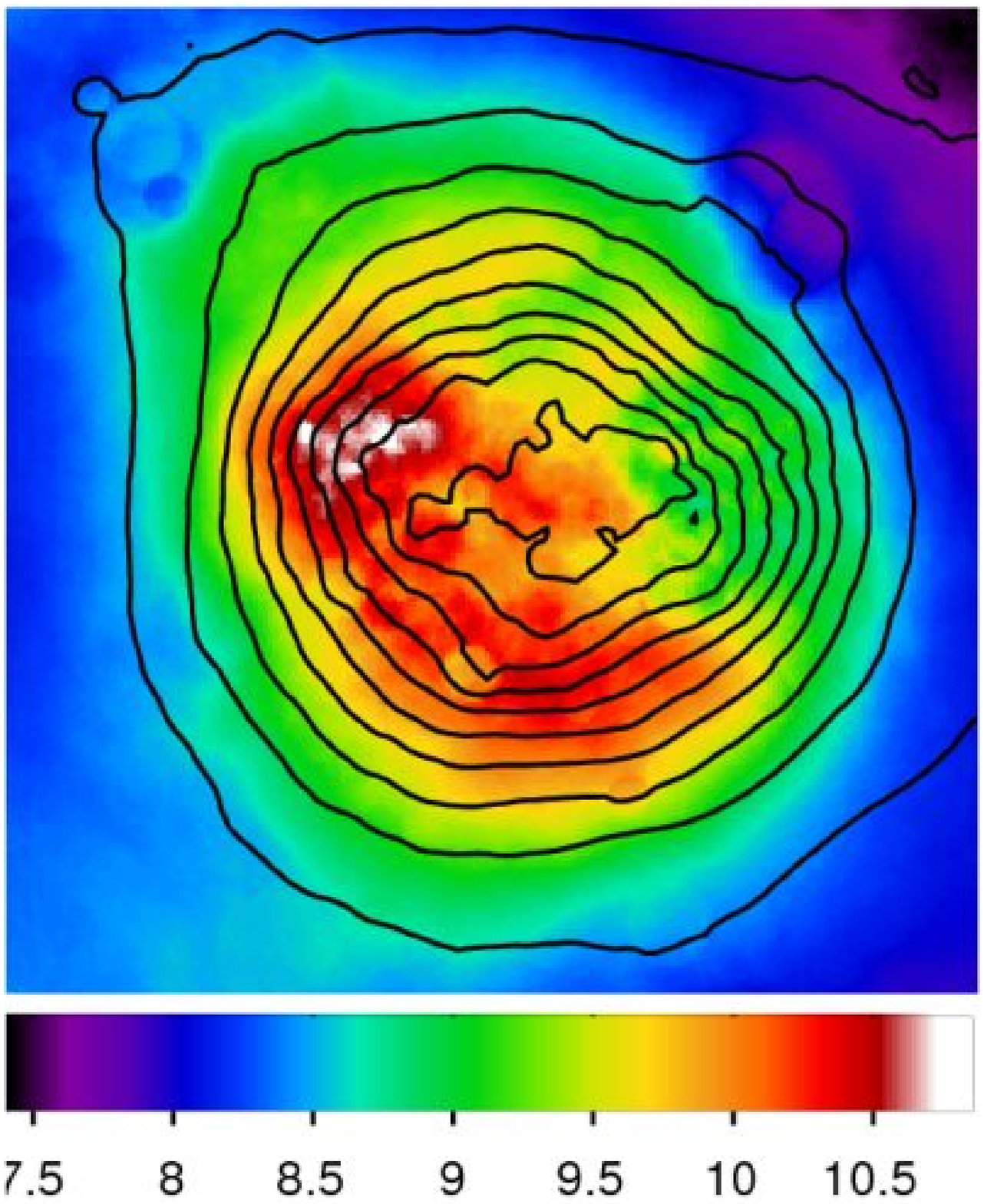} \\
    \end{tabular}
  \end{center}
  \caption{$2$ Mpc $ \times 2$ Mpc field showing luminosity (left) and temperature (right) maps of MS1054.4-0321. \label{MS1054.4-0321}}
\end{figure*}

\begin{figure*}[!htb]
  \begin{center}
    \begin{tabular}{cc}
      \includegraphics[width=2in]{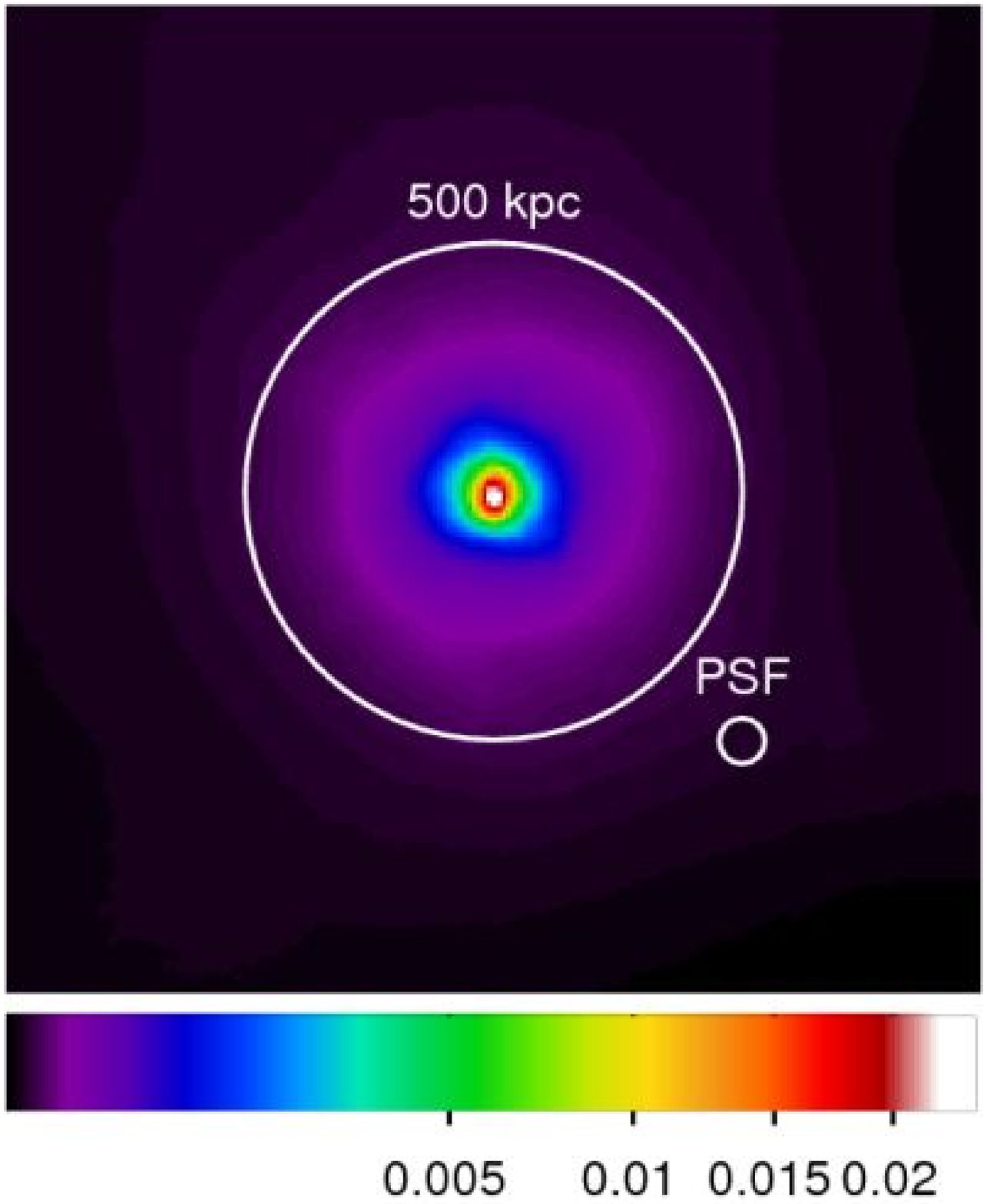} & 
      \includegraphics[width=2in]{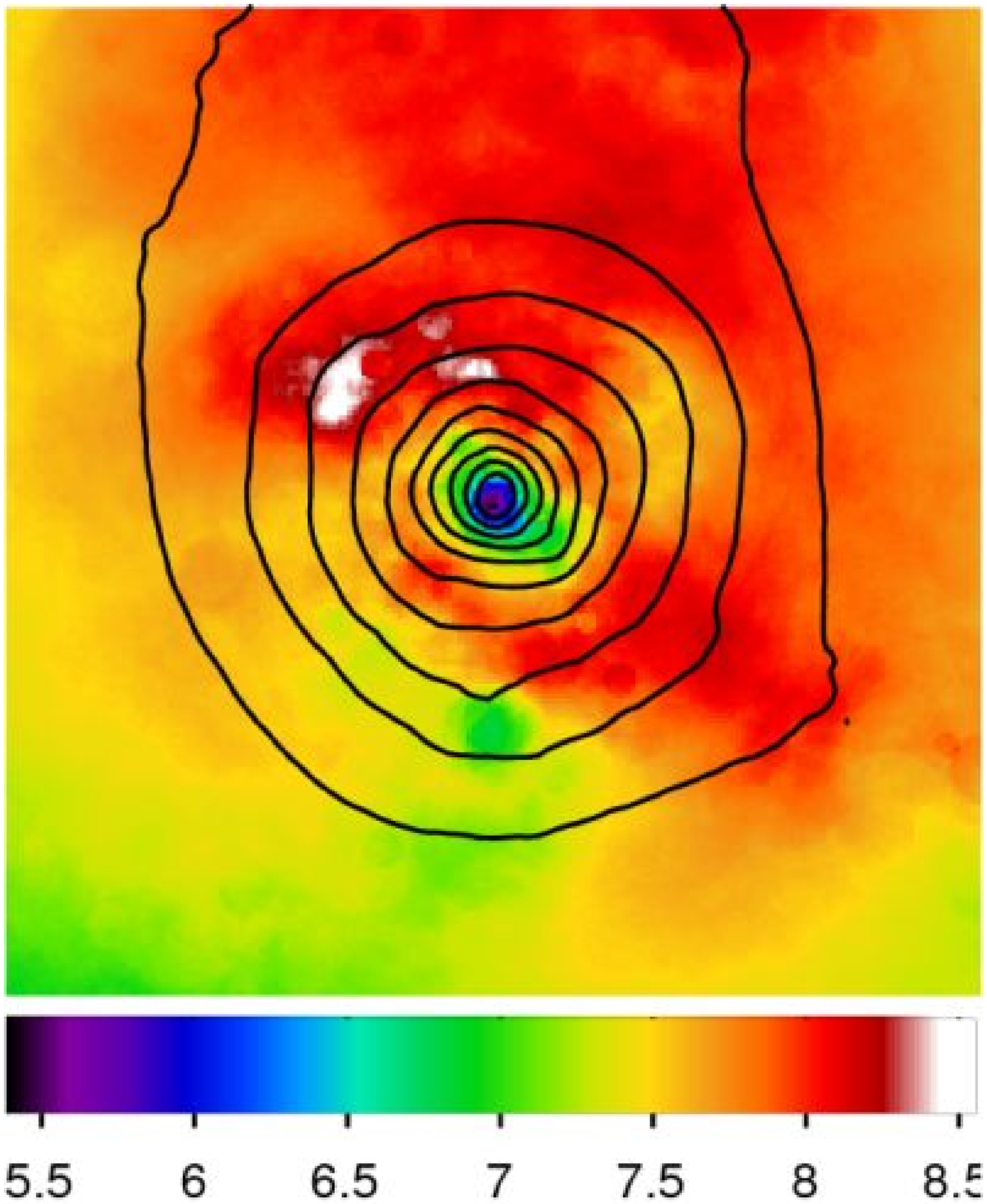} \\
    \end{tabular}
  \end{center}
  \caption{$2$ Mpc $ \times 2$ Mpc field showing luminosity (left) and temperature (right) maps of MS2137-23. \label{MS2137-23}}
\end{figure*}

\begin{figure*}[!htb]
  \begin{center}
    \begin{tabular}{cc}
      \includegraphics[width=2in]{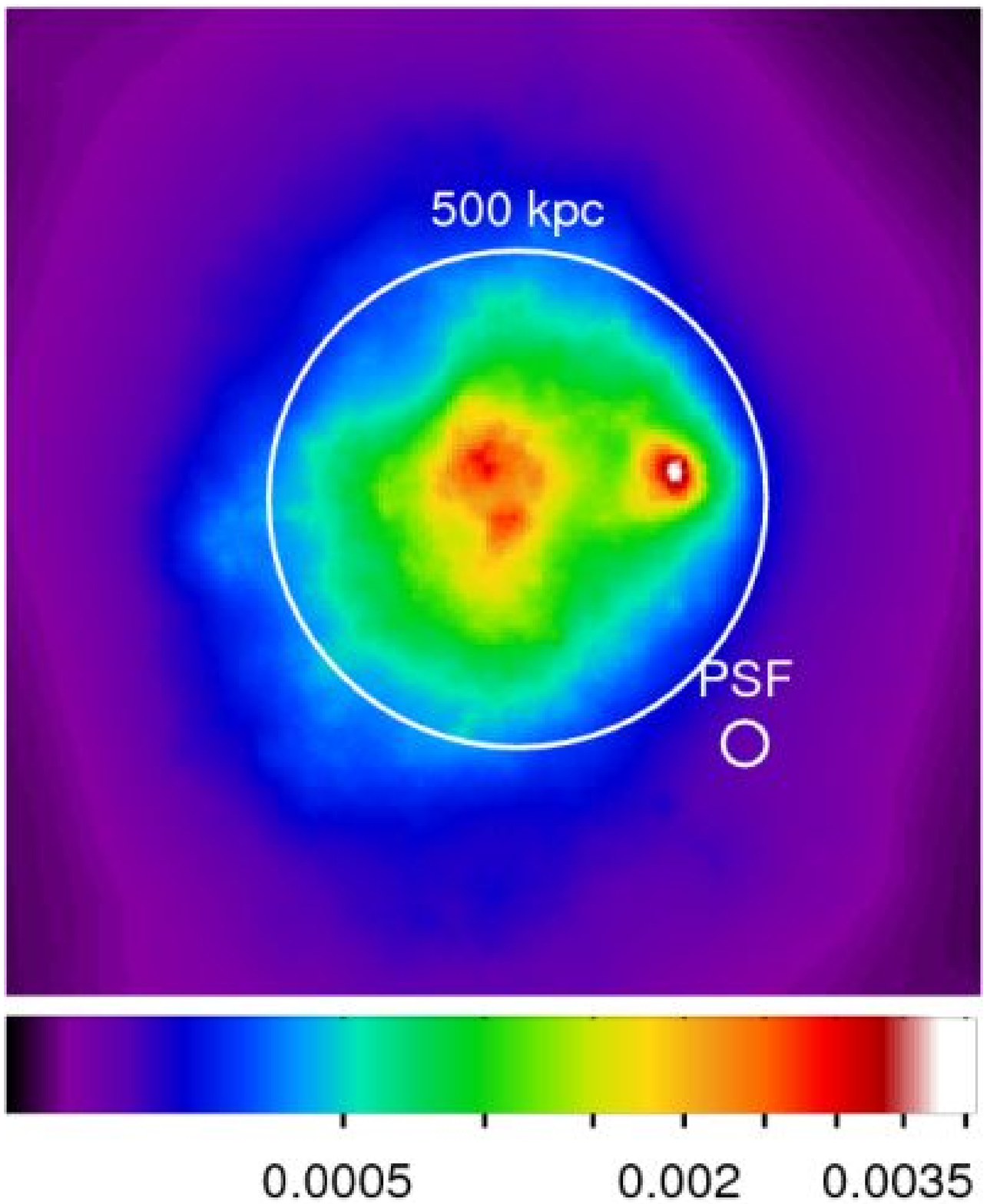} & 
      \includegraphics[width=2in]{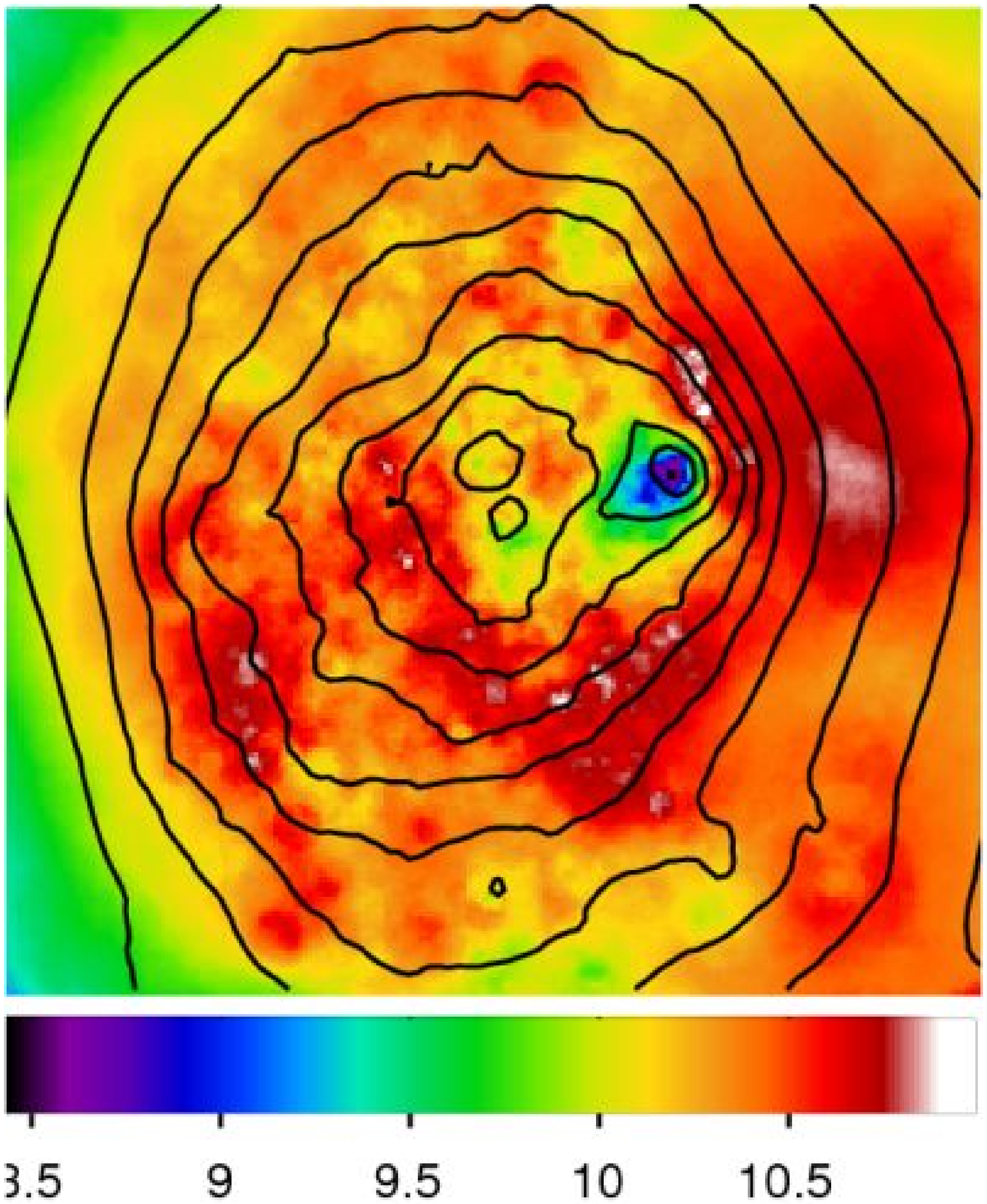} \\
    \end{tabular}
  \end{center}
  \caption{$2$ Mpc $ \times 2$ Mpc field showing luminosity (left) and temperature (right) maps of RXJ0658-55. \label{RXJ0658-55}}
\end{figure*}

\begin{figure*}[!htb]
  \begin{center}
    \begin{tabular}{cc}
      \includegraphics[width=2in]{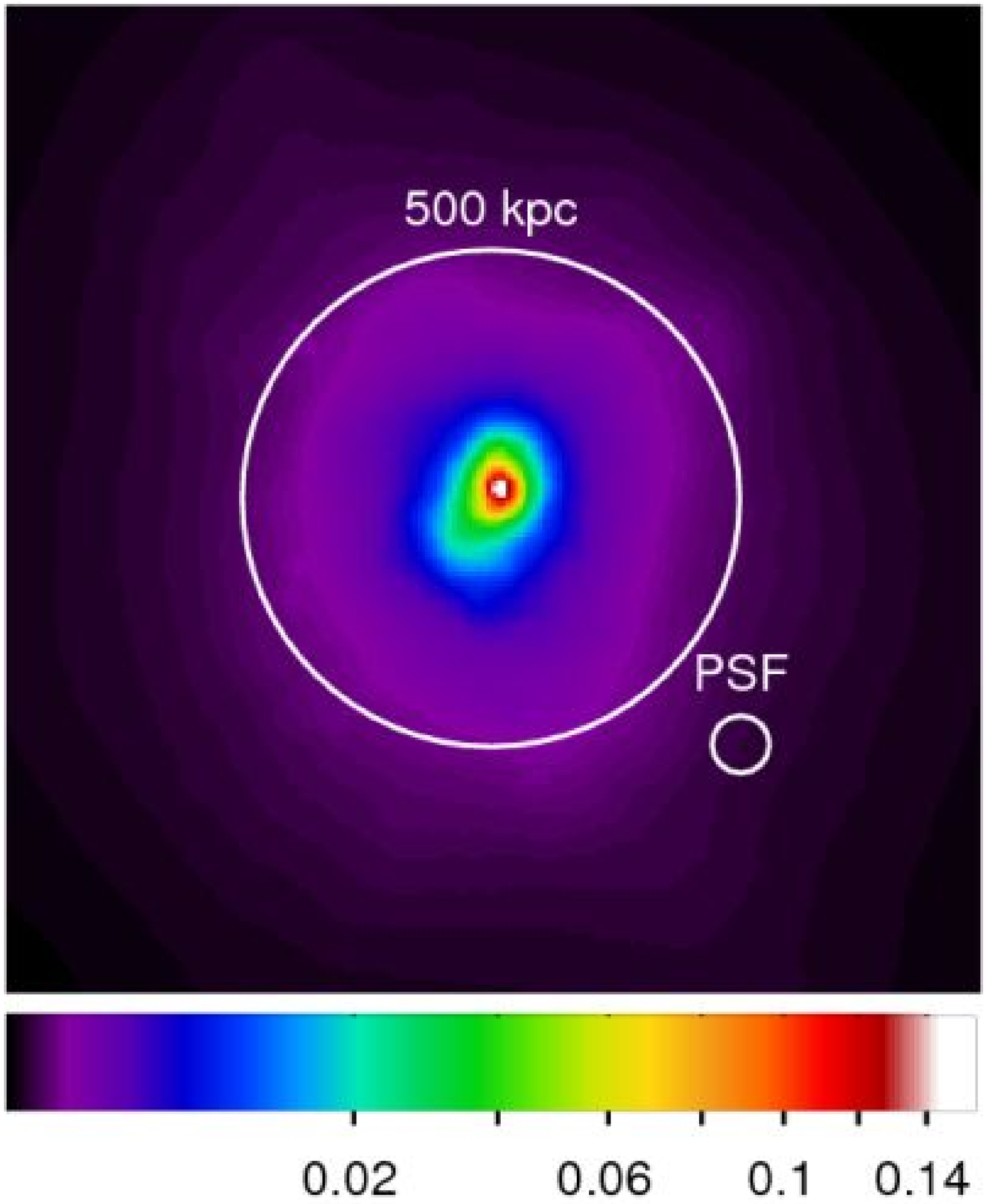} & 
      \includegraphics[width=2in]{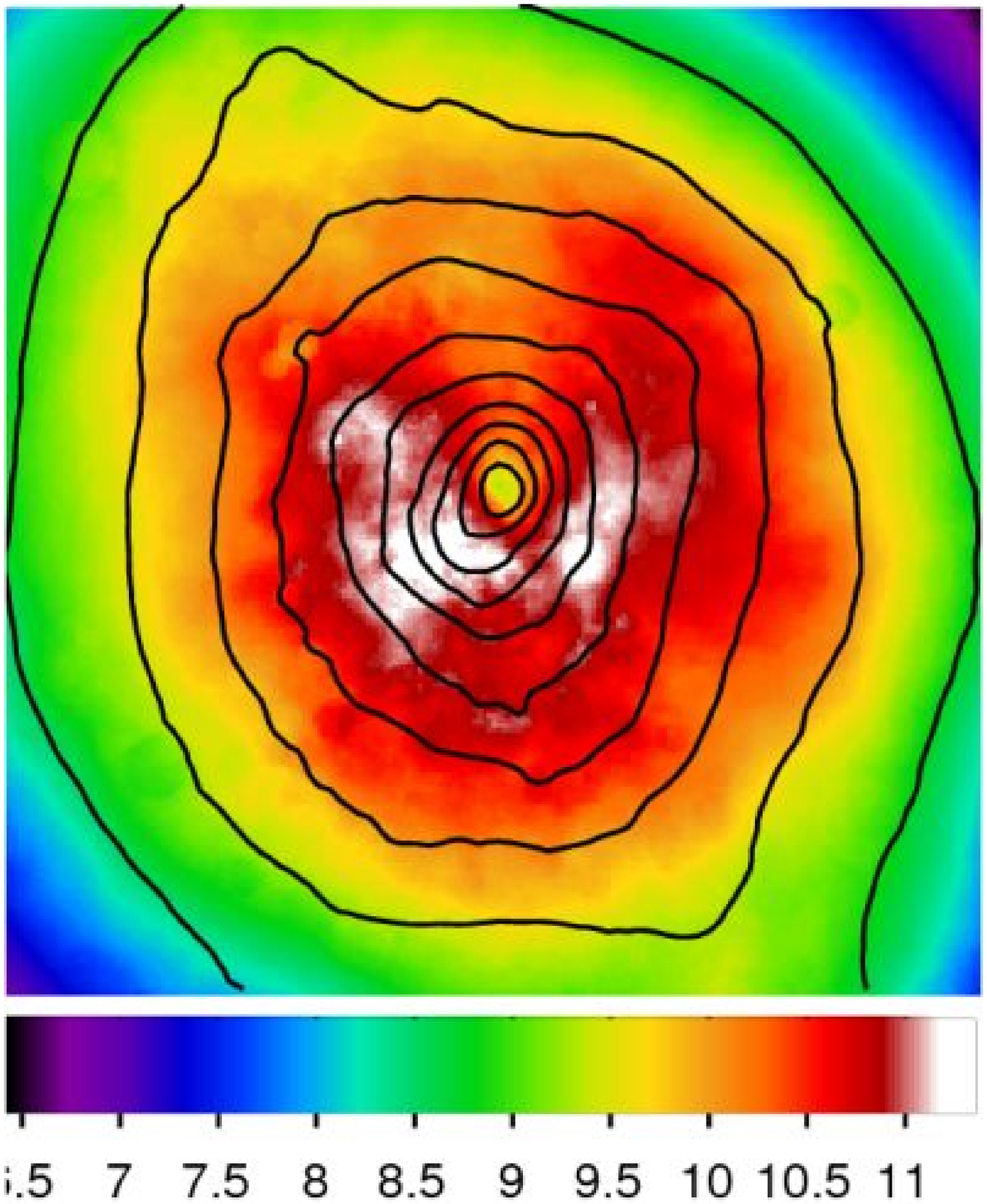} \\
    \end{tabular}
  \end{center}
  \caption{$2$ Mpc $ \times 2$ Mpc field showing luminosity (left) and temperature (right) maps of RXJ1347-1145. \label{RXJ1347-1145}}
\end{figure*}

\begin{figure*}[!htb]
  \begin{center}
    \begin{tabular}{cc}
      \includegraphics[width=2in]{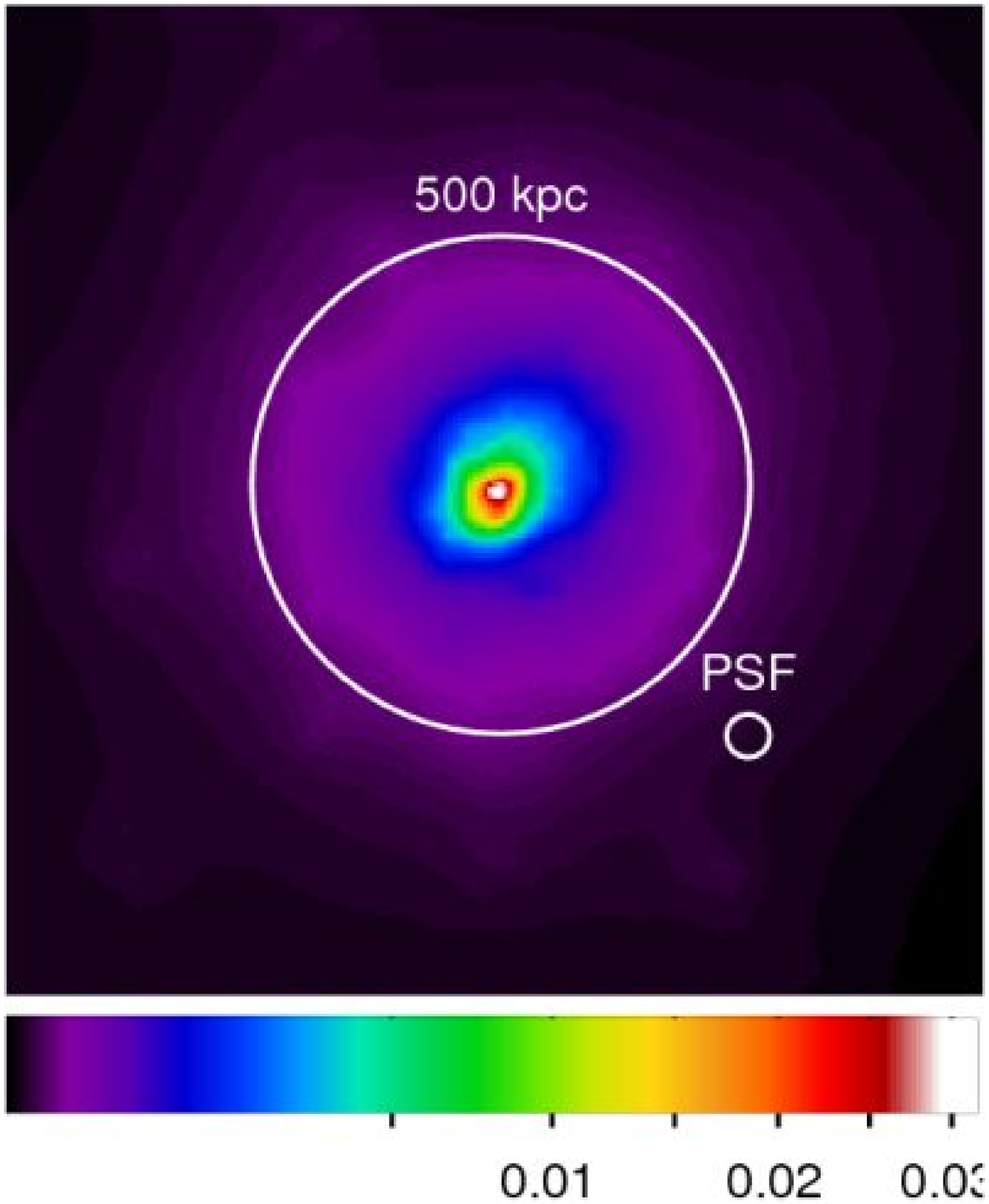} & 
      \includegraphics[width=2in]{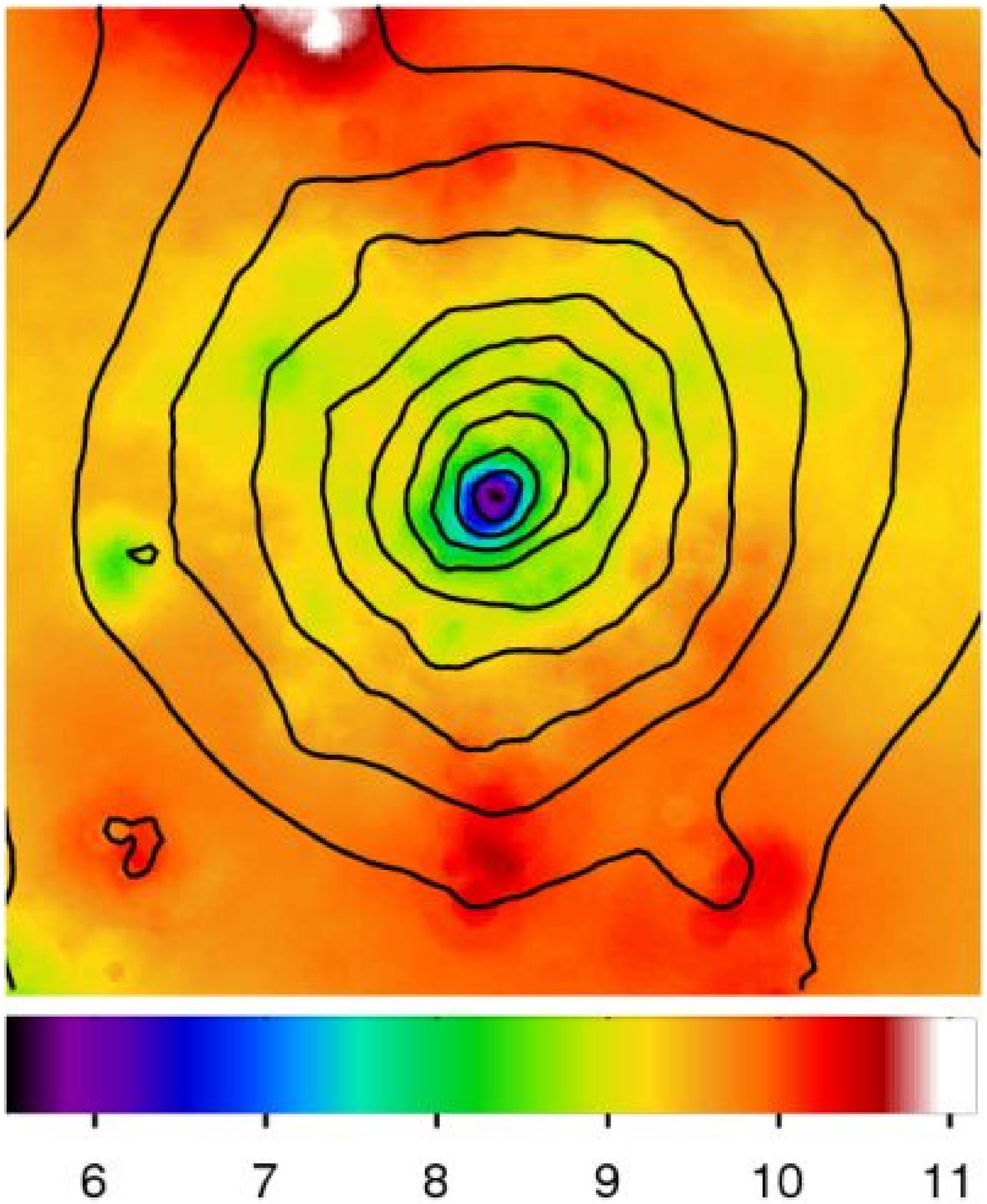} \\
    \end{tabular}
  \end{center}
  \caption{$2$ Mpc $ \times 2$ Mpc field showing luminosity (left) and temperature (right) maps of ZW3146. \label{ZW3146}}
\end{figure*}

Comparing this quantification of cluster structure with simulations 
of spherical clusters we show that all clusters in our sample are 
far from spherical, as determined by the value of $P_2/P_0$  
and should not be modeled as such. This is easily 
realized when viewing the luminosity maps in Figures \ref{A1068}-\ref{ZW3146} 
which exhibit a large degree of ellipticity. 
The large amount of features visible in the temperature maps confirm that 
a large number of processes govern the properties of the intra-cluster medium, 
that cluster merging can give rise to highly complex thermal structure and 
that temperature seldom is a simple function of radius. 

The SPI method inherently lends itself to a study of temperature structure 
of clusters, and to take advantage of it, we design a statistic, the 
``temperature two-point correlation'', 
capable of distinguishing between cooling clusters, isothermal clusters and 
clusters with significant temperature differences over large distances, 
such as merging clusters. 
The analysis using this function hints at an increase 
in the large scale thermal structure of clusters with 
redshift, indicating more frequent merging in a denser universe. 
This indication could be confirmed using a joint approach utilizing 
both XMM-Newton and Chandra datasets.


\section{Acknowledgments}
K.A. acknowledges financial support from the G\"{o}ran Gustavsson Foundation 
for Research in Natural Sciences  and Medicine.  This research was supported by 
NASA XMM-Newton observing grants NNX06AE39G and NNX07AE93G, and 
by the Department of Energy contract to SLAC no. DE-AC3-76SF00515.

\cleardoublepage


\end{document}